\documentclass{aa}

\usepackage[colorlinks,allcolors=blue]{hyperref}
\usepackage{graphicx}
\usepackage{txfonts}
\usepackage{upgreek}
\usepackage{tikz}
\tikzset{>=latex}
\usepackage{enumitem}
\usetikzlibrary{shapes,arrows}
\usetikzlibrary{decorations.pathmorphing,patterns}
\usetikzlibrary{patterns,snakes}
\usetikzlibrary{calc} 
\usetikzlibrary{arrows.meta,bending}
\usepackage{array}
\newcommand{\PreserveBackslash}[1]{\let\temp=\\#1\let\\=\temp}
\newcolumntype{C}[1]{>{\PreserveBackslash\centering}p{#1}}
\newcolumntype{R}[1]{>{\PreserveBackslash\raggedleft}p{#1}}
\newcolumntype{L}[1]{>{\PreserveBackslash\raggedright}p{#1}}

\makeatletter
\DeclareRobustCommand\bfseriesitshape{%
  \not@math@alphabet\itshapebfseries\relax
  \fontseries\bfdefault
  \fontshape\itdefault
  \selectfont
}

\pgfdeclaredecoration{ncoil}{coil}
{
  \state{coil}[switch if less than=%
    0.5\pgfdecorationsegmentlength+
    \pgfdecorationsegmentaspect\pgfdecorationsegmentamplitude+%
    \pgfdecorationsegmentaspect\pgfdecorationsegmentamplitude to last,
               width=+\pgfdecorationsegmentlength]
  {
    \pgfpathcurveto
    {\pgfpoint@oncoil{0    }{ 0.555}{1}}
    {\pgfpoint@oncoil{0.445}{ 1    }{2}}
    {\pgfpoint@oncoil{1    }{ 1    }{3}}
    \pgfpathcurveto
    {\pgfpoint@oncoil{1.555}{ 1    }{4}}
    {\pgfpoint@oncoil{2    }{ 0.555}{5}}
    {\pgfpoint@oncoil{2    }{ 0    }{6}}
    \pgfpathcurveto
    {\pgfpoint@oncoil{2    }{-0.555}{7}}
    {\pgfpoint@oncoil{1.555}{-1    }{8}}
    {\pgfpoint@oncoil{1    }{-1    }{9}}
    \pgfpathcurveto
    {\pgfpoint@oncoil{0.445}{-1    }{10}}
    {\pgfpoint@oncoil{0    }{-0.555}{11}}
    {\pgfpoint@oncoil{0    }{ 0    }{12}}
  }
  \state{last}[next state=final]
  {
    \pgfpathcurveto
    {\pgfpoint@oncoil{0    }{ 0.555}{1}}
    {\pgfpoint@oncoil{0.445}{ 1    }{2}}
    {\pgfpoint@oncoil{1    }{ 1    }{3}}
    \pgfpathcurveto
    {\pgfpoint@oncoil{1.555}{ 1    }{4}}
    {\pgfpoint@oncoil{2    }{ 0.555}{5}}
    {\pgfpoint@oncoil{2    }{ 0    }{6}}
  }
  \state{final}{}
}

\def\pgfpoint@oncoil#1#2#3{%
  \pgf@x=#1\pgfdecorationsegmentamplitude%
  \pgf@x=\pgfdecorationsegmentaspect\pgf@x%
  \pgf@y=#2\pgfdecorationsegmentamplitude%
  \pgf@xa=0.083333333333\pgfdecorationsegmentlength%
  \advance\pgf@x by#3\pgf@xa%
}

\pgfkeys{/eye/.cd,
  x/.code           = {\def\eye@x{#1}},
  y/.code           = {\def\eye@y{#1}},
  rotation/.code    = {\def\eye@rot{#1}},
  radius/.code      = {\def\eye@rad{#1}}
  }

\newcommand{\eye}[1][]{
\pgfkeys{/eye/.cd,
  x         = 0,
  y         = 0,
  rotation  = 0,
  radius    = 1
  } 
\pgfqkeys{/eye}{#1}   
   \draw[rotate around={\eye@rot:(\eye@x,\eye@y)}] 
         (\eye@x,\eye@y) -- ++(-25:\eye@rad) 
         (\eye@x,\eye@y) -- ++(25:\eye@rad);
   \draw (\eye@x,\eye@y) ++(\eye@rot+25:\eye@rad) arc (\eye@rot+25:\eye@rot-25:\eye@rad);
   \draw[fill=gray] (\eye@x,\eye@y)++(\eye@rot-17.5:\eye@rad) arc (\eye@rot+55+180:\eye@rot-55+180:0.38*\eye@rad);
   \draw[fill=black] (\eye@x,\eye@y) ++(\eye@rot+13:\eye@rad) arc (\eye@rot+13:\eye@rot-13:\eye@rad);
   \draw[fill=black] (\eye@x,\eye@y) ++(\eye@rot+13:\eye@rad) arc (\eye@rot-13+180:\eye@rot+13+180:\eye@rad);
   \draw (\eye@x,\eye@y) ++ (\eye@rot+25:\eye@rad) arc (-162.5:-50:0.15*\eye@rad);
   \draw (\eye@x,\eye@y) ++ (\eye@rot-25:\eye@rad) arc (180+162.5:180+50:0.15*\eye@rad);
}

\makeatother    
    
\DeclareTextFontCommand{\textbfit}{\bfseriesitshape}
\setlist[itemize]{label=\labelitemiii}
\begin{document}

   \title{The complex multi-component outflow of the Seyfert galaxy NGC~7130\thanks{Based on observations made at the European Southern Observatory using the Very Large Telescope under programmes \href{http://archive.eso.org/wdb/wdb/eso/sched_rep_arc/query?progid=60.A-9100(K)}{60.A-9100(K)} and \href{http://archive.eso.org/wdb/wdb/eso/sched_rep_arc/query?progid=60.A-9493(A)}{60.A-9493(A)}.}$^,$\thanks{The science-ready data cube can be accessed through the following link: \url{http://archive.eso.org/dataset/ADP.2020-12-09T12:34:28.554}.}}

  \author{ S.~Comer\'on\inst{1,2,3}, J.~H.~Knapen\inst{2,1}, C.~Ramos Almeida\inst{2,1}, and A.~E.~Watkins\inst{4}}
         
    \institute{Departamento de Astrof\'isica, Universidad de La Laguna, E-38200, La Laguna, Tenerife, Spain\\ \email{seb.comeron@gmail.com}
              \and Instituto de Astrof\'isica de Canarias E-38205, La Laguna, Tenerife, Spain
              \and Space Science and Astronomy, University of Oulu, P.O.~Box 3000, FI-90014 Oulu, Finland
              \and Astrophysics Research Institute, Liverpool John Moores University, IC2, Liverpool Science Park, 146 Brownlow Hill, Liverpool L3 5RF, UK
              }

 \titlerunning{The complex multi-component outflow of the Seyfert galaxy NGC~7130}
 \authorrunning{S.~Comer\'on et al.}

 \abstract{Active galactic nuclei (AGN) are a key ingredient for understanding galactic evolution, as their activity is coupled to the host galaxy properties through feedback processes. AGN-driven outflows are one of the manifestations of this feedback. The laser guide star adaptive optics mode for MUSE at the VLT now permits us to study the innermost tens of parsecs of nearby AGN in the optical. We present a detailed analysis of the ionised gas in the central regions of NGC~7130, which is an archetypical composite Seyfert and nuclear starburst galaxy at a distance of $64.8\,{\rm Mpc}$. We achieve an angular resolution of $0\farcs17$, corresponding to roughly 50\,pc. We performed a multi-component analysis of the main interstellar medium emission lines in the wavelength range of MUSE and identified nine kinematic components, six of which correspond to the AGN outflow. The outflow is biconic, oriented in an almost north--south direction, and has velocities of a few $100\,{\rm km\,s^{-1}}$ with respect to the disc of NGC~7130. The lobe length is at least 3\arcsec ($\sim900$\,pc). We decomposed the approaching side of the outflow into a broad and a narrow component with typical velocity dispersions below and above $\sim200\,{\rm km\,s^{-1}}$, respectively. The blueshifted narrow component has a sub-structure, in particular a collimated plume traced especially well by {[}O\,{\sc iii}{]}. The plume is aligned with the radio jet, indicating that it may be jet powered. The redshifted lobe is composed of two narrow components and a broad component. An additional redshifted component is seen outside the main north-south axis, about an arcsecond east of the nucleus. Line ratio diagnostics indicate that the outflow gas in the north--south axis is AGN powered, whereas the off-axis component has LINER properties. We hypothesise that this is because the radiation field that reaches off-axis clouds has been filtered by clumpy ionised clouds found between the central engine and the low-ionisation emitting region. If we account for all the outflow components (the blueshifted components), the ionised gas mass outflow rate is $\dot{M}=1.2\pm0.7\,M_{\odot}\,{\rm yr^{-1}}$ ($\dot{M}=0.44\pm0.27\,M_{\odot}\,{\rm yr^{-1}}$), and the kinetic power of the outflow is $\dot{E}_{\rm kin}=(2.7\pm2.0)\times10^{41}\,{\rm erg\,s^{-1}}$ ($\dot{E}_{\rm kin}=(7.0\pm4.8)\times10^{40}\,{\rm erg\,s^{-1}}$), which corresponds to $F_{\rm kin}=0.12\pm0.09\%$ ($F_{\rm kin}=0.032\pm0.022\%$) of the bolometric AGN power. The broad components, those with a velocity dispersion of $\sigma>200\,{\rm km\,s^{-1}}$, carry $\sim2/3$ ($\sim90\%$) of the mass outflow, and $\sim90\%$ ($\sim98\%$) of the kinetic power. The combination of high-angular-resolution integral field spectroscopy and a careful multi-component decomposition allows a uniquely detailed view of the outflow in NGC~7130, illustrating that AGN kinematics are more complex than those traditionally derived from less sophisticated data and analyses.}

   \keywords{Galaxies: active -- Galaxies: individual: NGC~7130 -- Galaxies: ISM -- Galaxies: nuclei -- Galaxies: Seyfert}

   \maketitle
%

\section{Introduction}

Active galactic nuclei (AGN) are compact luminous sources at the very centre of many giant galaxies. They are powered by the potential energy loss of material falling into a supermassive black hole \citep[SMBH;][]{Salpeter1964, LyndenBell1969}. AGN come in a multitude of varieties distinguishable by their spectral properties \citep[see e.g.~Table~1 in][]{Padovani2017}. AGN are interesting objects by themselves, and also because of their coupling with their host galaxies. An example of this is the fairly tight correlation between the SMBH mass and the velocity dispersion of the stellar spheroid \citep[the so-called $\mathcal{M}_{\rm BH}-\sigma_\star$ relation;][]{Ferrarese2000, Gebhardt2000}. AGN feedback is thought to be one of the mechanisms that limit the growth of massive galaxies \citep[e.g.][and references therein]{Harrison2017} and that contribute to the transformation of dark matter cusps into cores \citep[e.g.][]{Peirani2008}.

Active galactic nuclei feedback mechanisms include outflows \citep[for a review on AGN feedback, see][]{Morganti2017}. The first outflows were detected in ionised gas \citep[see the historical discussion in][]{Veilleux2005}, but they are nowadays known to be multi-phase and also carry H\,{\sc i} \citep{Morganti2005} and molecular gas \citep{Feruglio2010}. The kind of feature studied in this paper, ionised outflows, is sometimes complex and might require a multi-component approach to be accurately described \citep[e.g.][]{McElroy2015, Lena2015, Mingozzi2019}. Outflows are thought to be part of the self-regulation mechanism for the growth of the SMBH and to contribute to the quenching of star formation in the nuclear regions of the host galaxy \citep[for a recent review, see][]{Veilleux2020}.

The feeding of AGN is a matter of controversy, since it requires the existence of a mechanism for the gas to lose its angular momentum to reach a galaxy centre. Galaxy-galaxy interactions can generate torques to that effect \citep{Negroponte1983}. In non-interacting galaxies, and at scales larger than 1\,kpc, inwards gas transportation can be efficiently triggered by energy dissipation at shocks and gravitational torques associated with bars \citep{Schwarz1984, Athanassoula1992} and spiral arms \citep{Lubow1986, Kim2014}. Strong observational evidence of bars funnelling material towards galactic centres is provided by the detection of enhancements in the star formation, gas concentration, and central mass concentration in barred galaxies \citep[e.g.][]{Heckman1980, Hummel1981, Hawarden1986, Devereux1987, Sakamoto1999, Sheth2005, DiazGarcia2016, Lin2017, DiazGarcia2020}. Large-scale bars and spirals bring the gas to the inner Lindblad resonance \cite[ILR;][]{Schwarz1984} region, which is located at roughly one kiloparsec from the centre and sometimes traced by spectacular star-forming nuclear rings \citep{Knapen1995, Comeron2010}. Once near the ILR, it is unclear how the gas loses its remaining angular momentum to move further in, but it has been proposed that this can be achieved by shocks and gravitational torques in a `bar-within-bar' scenario \citep{Shlosman1989, Hunt2008, Querejeta2016} or in a nuclear spiral scenario \citep{Combes2014, Kim2017}. The same fuel that feeds the central engine can also ignite intense circumnuclear star formation episodes, or `nuclear starbursts'. Galaxies hosting both a Seyfert AGN and a nuclear starburst are referred to as `composite' \citep{Telesco1988}.

The study of the innermost parts of galaxies is crucial to understanding how AGN are fed \citep[inflows;][]{StorchiBergmann2019} and how they affect their surroundings \citep[through, e.g. outflows][]{RamosAlmeida2017, Hoenig2019}. Spectroscopic data are necessary to study both the kinematics and the physical conditions of the circumnuclear medium. Because of the relevant scales (a few hundred parsecs or less) sub-arcsecond angular resolution is required even for the closest galaxies. Hence, the advent of the laser guide star, GALACSI laser adaptive optics (AO) module \citep{Stuik2006}, in the Multi Unit Spectroscopic Explorer \citep[MUSE;][]{Bacon2010} integral field spectrograph at the VLT, provides a new tool to improve our understanding of galaxy-AGN interplay. In narrow field mode (MUSE-NFM), MUSE + GALACSI combine the wide MUSE wavelength range (4750 -- 9350\,\AA\ with a gap at 5780 -- 6050\,\AA\ to prevent contamination from the laser guide stars) and an extraordinary angular resolution that is nominally below $0\farcs1$ over the whole wavelength range. As a consequence, MUSE is able to obtain integral field data at an angular resolution comparable to that of the {\it Hubble} Space Telescope ({\it HST}) over a field of view of about $7\farcs5\times7\farcs5$. Such angular resolutions were achievable in the near-infrared \citep[see e.g. the works by][done with VLT SINFONI and Gemini NIFS data, respectively]{Davies2009, Riffel2009, Riffel2010}, but MUSE has expanded these capabilities to optical wavelengths, where the lines necessary to build, for example, Baldwin, Phillips, and Terlevich (BPT) diagnostics are found.

In \citet{Knapen2019}, we published the first ever MUSE-NFM AO-assisted study of the circumnuclear medium in an AGN-hosting galaxy, NGC~7130. Now, we build upon our previous work and provide a detailed analysis of the data in order to unveil the complex physics of the circumnuclear medium, including the outflow. In Sect.~\ref{NGC7130}, we summarise the properties of the target galaxy and the findings reported in the literature. In Sect.~\ref{processing}, we describe the data processing, including the reduction and the spectral analysis. In Sect.~\ref{results}, we describe our results, which are then discussed in Sect.~\ref{discussion}, where we also present a simple model to explain the observations. We summarise our findings in Sect.~\ref{conclusion}.

Throughout this paper, we assume the cosmology derived from the five-year WMAP mission combined with Type~Ia supernovae and baryonic acoustic oscillation data \citep{Hinshaw2009}, that is a Hubble-Lema\^itre constant of $H_0=70.5\,{\rm km\,s^{-1}\,Mpc^{-1}}$, a matter density parameter of $\Omega_{\rm m,0}=0.27$, and a cosmological constant density parameter of $\Omega_{\Lambda,0}=0.73$.

\section{Previous studies of NGC~7130}

\label{NGC7130}

The galaxy NGC~7130, also known as IC~5135, is a southern galaxy found at right ascension ${\rm RA}=21^{\rm h}48^{\rm m}19.\!\!^{\rm s}520$ and declination $\delta=-34^{\rm o}57^{\prime}04\farcs48$ (Epoch J2000.0) with a redshift $z=0.016151$, according to the NED\footnote{The NASA/IPAC Extragalactic Database (NED)
is operated by the Jet Propulsion Laboratory, California Institute of Technology, under contract with the National Aeronautics and Space Administration.}. It is a peculiar Sa galaxy \citep{Vaucouleurs1991} where infrared observations reveal a bar \citep{Mulchaey1997}. An inspection of the $HST$ images presented in \cite{Malkan1998} and \cite{EliasRosa2018} also shows the bar in optical, albeit partially obscured by conspicuous dust lanes. The bar is surrounded by a star-forming inner pseudo-ring \citep{Dopita2002, MunozMarin2007}. The proper, luminosity, and angular-diameter distances are $D_{\rm p}=64.8$\,Mpc, $D_{\rm L}=65.8$\,Mpc, and $D_{\rm A}=63.9$\,Mpc, respectively (based on the velocity with respect to the cosmic microwave background provided by the NED, $4586\pm23\,{\rm km\,s^{-1}}$). At that distance, one arcsecond corresponds to 310\,pc.

The infrared luminosity of NGC~7130 is ${\rm log}\,(L_{IR}/L_{\bigodot})=11.35$ \citep[][who assumed a value of $H_0=75\,{\rm km\,s^{-1}\,Mpc^{-1}}$]{Sanders2003} so it is classified as a luminous infrared galaxy (LIRG). In this kind of galaxy, the intense infrared emission is usually due to an intense star formation episode often linked to interactions between spiral galaxies \citep{Sanders1996}. NGC~7130 forms a pair with IC~5131 \citep{Sandage1994}, which is located at a distance of $12^{\prime}$, or 220\,kpc in projection. No obvious signs of interaction between the two galaxies are seen, but the distorted appearance of the outskirts of NGC~7130  \citep[already reported in][]{Vaucouleurs1964, Vaucouleurs1976} might indicate a past close encounter between them or with a smaller unidentified member of the group. The asymmetric velocity and velocity dispersion maps of the ionised gas in the galaxy \citep{Bellocchi2012} are further indicators of a likely recent interaction.

The galaxy NGC~7130 is nearly face-on, with an axial ratio $q=0.88$ and position angle ${\rm PA}=160\degr$ measured at the $K_{\rm s}$-band $20\,{\rm mag\,arcsec^{-2}}$ isophote \citep[from 2MASS;][]{Skrutskie2006}. Orientation parameters obtained from optical images are very similar \citep{Lauberts1989}.

The AGN of NGC~7130 was originally classified as a Seyfert~2 \citep{Phillips1983}. This was later refined to Seyfert~1.9 \citep{VeronCetty2010}, but see Sect.~\ref{discussion} for further details on the Seyfert type. The composite H\,{\sc ii} + Seyfert nature of the nucleus of NGC~7130 was independently found by \citet{Veron1981} and \citet{Phillips1983} and confirmed by \citet{Thuan1984} and \citet{Shields1990}, but \citet{Radovich1997} claimed that nuclear star formation is not required to explain the spectra of the inner kpc. The core of NGC~7130 has been found to emit in radio \citep{Norris1990}, and its optical spectrum has two kinematic components, narrow and broad \citep{Busko1988}, interpreted to be associated with H\,{\sc ii} regions and the AGN, respectively \citep{Shields1990}. The broad component is blueshifted and was later hypothesised to be associated with an outflow \citep{GonzalezDelgado1998, Bellocchi2012, Davies2014}.

High-resolution UV continuum images obtained by the $HST$ reveal a tiny circumnuclear ring $1^{\prime\prime}$ in size (major axis) and a few UV knots along the spiral arms associated with the bar \citep{GonzalezDelgado1998}. They also found a UV knot, presumably highly obscured, at the suspected location of the AGN engine. The ring is reminiscent of the ultra-compact nuclear rings (UCNRs) presented in \cite{Comeron2008}. The fact that the AGN of NGC~7130 is highly obscured was confirmed by the modelling of the spectral energy distribution \citep{Contini2002} and by X-ray observations \citep{Levenson2005}. The latter authors found that the AGN emits most of the hardest X-rays in NGC~7130 ($>2\,{\rm keV}$), but that two thirds of the total X-ray emission can be ascribed to an extended component associated with star-forming regions.

The centre of NGC~7130 has two dusty spiral arms within the bar. They coincide with molecular gas as traced by the CO(6--5) transition \citep{Zhao2016}. Extended CO emission, maybe partially correlated with the star-forming UCNR, is also found in the central $1^{\prime\prime}$. \citet{Zhao2016} found no traces of a molecular gas outflow using this CO transition (which traces very dense molecular gas). The spectral line energy distribution of CO in the central regions of NGC~7130 requires both star formation and X-Ray emission to be modelled \citep{Pozzi2017}.

In \citet{Knapen2019}, we presented the first optical high-angular resolution integral field study of the centre of NGC~7130. We found a tiny bipolar pattern in the velocity maps that we interpreted as a kinematically decoupled core $0\farcs2$ in radius. We confirmed the presence of an outflow whose line ratios indicate AGN ionisation. There, we assumed the location of the nucleus of NGC~7130 to be at the spaxel where a single-component fit of the ionised gas kinematics yields the largest velocity dispersion. This coincided with the centre of the small bipolar structure and with a bright knot located within the UCNR. Here, we assume the same position for the engine of the AGN.

\section{Observations and data processing}

\label{processing}

\subsection{Data obtention and reduction}

\label{obtention}

\begin{figure*}
   \centering
   \includegraphics[scale=0.30]{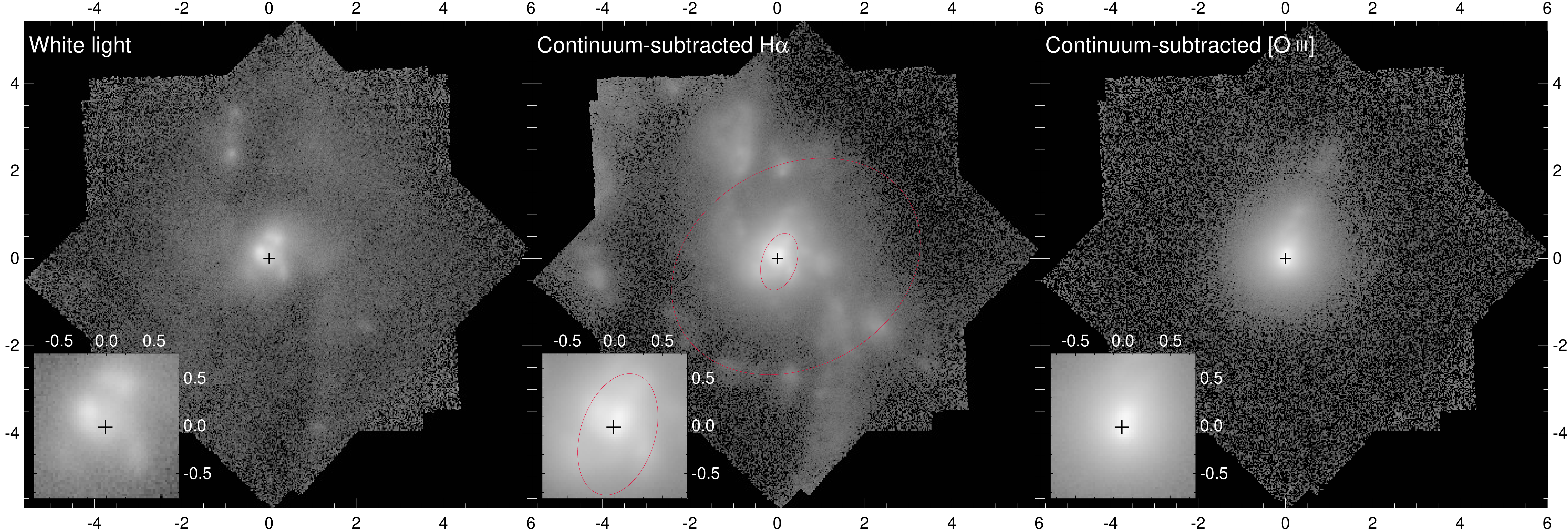}
   \caption{\label{images}{\it Left panel}: White-light image of the centre of NGC~7130 obtained from integrating the reduced MUSE data cube along the spectral direction. {\it Middle panel}: Continuum-subtracted H$\upalpha$ image obtained from the same data cube. The red ellipses indicate the outline of the nuclear rings (see Sect.~\ref{morphology}). {\it Right panel}: Continuum-subtracted {[}O\,{\sc iii}{]}\,$\uplambda$5007 image. The intensities are scaled logarithmically. In this and all subsequent maps, the plus sign indicates the inferred galaxy centre and the insets show an enlarged version of the centremost region of the galaxy. The axes are in arcseconds. North is up and east is left.}
\end{figure*}

We obtained MUSE-NFM AO data of the central region of NGC~7130 as part of the science verification programme for this observation mode. Our proposal aimed to obtain $4\times600\,{\rm s}$ on-target exposures intertwined with $180\,{\rm s}$ off-target exposures to model the sky. Ten 600\,s exposures were taken on the nights of September 15, 16, and 18 in 2018. Unfortunately, only in the two exposures taken on September 18 did the AO work well enough to bring the seeing below $0\farcs5$ \citep[full width at half maximum (FWHM) of $0\farcs16-0\farcs17$ as measured using the \texttt{imexamine} tool in \texttt{IRAF};][]{Tody1986}. After publishing our letter \citep{Knapen2019}, we checked the ESO archive for additional data. We were pleasantly surprised to find that eight 300\,s MUSE-NFM AO exposures of the same region had been taken during the commissioning of the instrument mode on June 19 and 21 2018. Three of these exposures have an angular resolution comparable to those used in \cite{Knapen2019} and are included in the present work, bringing the total exposure time to 2100\,s.

\setcitestyle{notesep={; }}
The raw MUSE data were processed using the standard MUSE pipeline \citep[version~2.8.1;][]{Weilbacher2012, Weilbacher2014} run under the version 2.9.1 of the \texttt{EsoReflex} environment \citep{Freudling2013}. The five exposures were manually aligned and then combined using the \texttt{muse\_exp\_combine} recipe. The processed data cube has an angular resolution of $0\farcs17$, comparable to the UVIS $HST$ images of the same region presented by \citet[][angular resolution $0\farcs11$]{EliasRosa2018}. The combined data cube has two extensions, namely one with the signal and another one with variances (error estimates). In Fig.~\ref{images}, we show a white-light and continuum-subtracted H$\upalpha$ and {[}O\,{\sc iii}{]}\,$\uplambda$5007 images (Sect.~\ref{binning}) produced from the final reduced data cube to illustrate the quality of the data.

\setcitestyle{notesep={, }}

\subsection{Use of \texttt{GIST} and \texttt{pyGandALF}}

We processed the data cube using a modified version of the Galaxy IFU Spectroscopy Tool\footnote{\url{https://abittner.gitlab.io/thegistpipeline/index.html}} \citep[\texttt{GIST}, version~2.0.0;][]{Bittner2019} and obtained the emission line properties using the Python implementation of \texttt{GandALF} \citep{Sarzi2006, FalconBarroso2006}, which is called \texttt{pyGandALF} \citep{Bittner2019}, included in \texttt{GIST}. \texttt{GIST} is an all-in-one Python pipeline that comprises many features to easily extract physical information and data cubes. The pipeline is complemented by \texttt{Mapviewer}, an extremely powerful interactive visualisation tool that has been key to understanding our data.

The \texttt{GIST} functions that we were particularly interested in were 1) the Voronoi binning of the data and 2) the extraction of stellar kinematics. How these are implemented in \texttt{GIST} and how these functions were modified for our purposes is explained in detail in Sects.~\ref{binning} and \ref{sstellar}. In Sects.~\ref{resolved}, \ref{kinematics}, and \ref{refinment}, we explain how we used \texttt{pyGandALF} to obtain the emission line properties.

\subsection{Voronoi binning}

\label{binning}

Voronoi binning was performed with the code written by \cite{Cappellari2003}. The user provides a signal and a noise value for each spaxel that are used to produce bins with a chosen signal-to-noise ratio (${\rm S/N}$). \texttt{GIST}'s original implementation allows for only a single Voronoi binning to be used both for the stellar and the gas emission. This was not convenient for our purposes because our data have a combination of poor ${\rm S/N}$ stellar emission and high ${\rm S/N}$ line emission. We thus modified the code to support two binnings.

The stellar emission binning was made using \texttt{GIST}'s original binning procedure, using the rest-frame wavelength range of $5500\,\text{\AA}-5680\,\text{\AA}$. We found the median signal and noise over the chosen wavelength range on a pixel-by-pixel basis before being fed into the binning code. We required a signal-to-noise ratio of ${\rm S/N}=50$ per stellar bin, which resulted in 111 bins.

The emission line or ionised gas binning was made based on a H$\upalpha$ continuum-subtracted image. We built the image by integrating the data cube flux in a 20\,\AA\ window centred on the restframe H$\upalpha$ wavelength of NGC~7130, and by subtracting the integral flux of a window of equal width but 50\,\AA\ redwards. The result is shown in Fig.~\ref{images}. The noise was estimated by quadratically summing the noise values from both the line and the continuum windows. For this binning, we required ${\rm S/N}=100$, which resulted in 2689 bins.

Both stellar and emission line binnings required a minimal single spaxel ${\rm S/N}$ to be considered for binning. We set this threshold as ${\rm S/N}=0.1$ for stars and ${\rm S/N}=0.5$ for ionised gas. Because some isolated spaxels or small spaxel clusters can be above the threshold without being connected to the main body of regions with a signal above the threshold, we modified \texttt{GIST} not to consider clusters smaller than 500\,spaxels for the binning.

For each binned spectrum we also computed a variance spectrum by summing the variances at each wavelength over all the individual spaxels in a bin. The procedure developed to produce H$\upalpha$ continuum-subtracted images was also used to obtain continuum-subtracted images in other lines, such as {[}O\,{\sc iii}{]}\,$\uplambda$5007 (Fig.~\ref{images}).

\subsection{Stellar kinematics}

\label{sstellar}

\texttt{GIST} recovers the stellar kinematics using \texttt{pPXF}. The latter code fits the spectra for each bin (stellar emission binning in this case) with a linear combination of spectral energy distribution templates convolved with the line-of-sight velocity distribution (LOSVD), which is what one ultimately desires to obtain. We only fitted the two lowest LOSVD momenta, namely the velocity, $V_\star$, and the velocity dispersion, $\sigma_\star$.

We used the templates from the E-Miles library\footnote{\url{http://miles.iac.es}} \citep{Vazdekis2016} with BaSTI isochrones \citep{Pietrinferni2004}, a Kroupa Universal stellar initial mass function \citep{Kroupa2001}, and the `base' abundances. E-Miles has a spectral resolution of 2.51\,\AA\ (FWHM) in the spectral range of interest. The spectral resolution of MUSE as a function of wavelength has been modelled to be \citep{Bacon2017}
\begin{equation}
 {\rm FWHM}(\lambda)=5.866\times10^{-8}\lambda^2-9.187\times10^{-4}\lambda+6.040
,\end{equation}
where $\lambda$ is expressed in \AA. The spectral resolution is worse than that of E-Miles on its blue side ($\lambda<6762\,\text{\AA}$) and better on its red side ($\lambda>6762\,\text{\AA}$). \texttt{GIST} supports a wavelength-dependent Gauss convolution of the templates so they match the instrumental spectral resolution. We have implemented a Gauss-convolution of the data so their spectral resolution matches that of the templates at wavelengths where the instrumental resolution is better than that of the templates.

Stellar kinematics were obtained using the wavelength range $4750\,\text{\AA}-7113\,\text{\AA}$ ($4675\,\text{\AA}-7000\,\text{\AA}$ in the restframe of NGC~7130). We ignored the reddest wavelengths because of the presence of many sky lines, and even though that region contains deep absorption lines that in principle can be used to characterise the stellar kinematics (the Ca\,{\sc ii} triplet), our experiments showed that using those regions caused the solutions to be noisier. The stellar continuum shape was modelled with an eight-degree additive Legendre polynomial.

The wavelength range affected by the AO laser guide stars ($5774\,\text{\AA}-6054\,\text{\AA}$) was masked. Sky lines with a peak flux larger than $6\times10^{-19}\,{\rm W\,m^{-2}\,\text{\AA}^{-1}\,arcsec^{-2}}$ in the sky line atlas by \citet{Hanuschik2003} were masked by windows with 20\,\AA\ in width. We also masked the ${\rm O}_2\,{\rm B}$ telluric feature at  $6862.1\,\text{\AA}-6964.6\,\text{\AA}$ \citep[wavelength range from][]{Buton2013}.

Based on our single-component fits in \citet{Knapen2019}, we know that emission lines from the galaxy display high velocity dispersion values of up to $400\,{\rm km\,s^{-1}}$, so we masked emission lines with windows of $3000\,{\rm km/s}$ in width. The MUSE observations are so deep that we detected more lines in the innermost arcsecond of NGC~7130 than previous observers, and we resorted to deep studies of another Seyfert galaxy, NGC~1068, for the identification of some of those lines \citep{Koski1978, Osterbrock1996, Kraemer2000}. The masked lines are listed in Table~\ref{masked_lines}.

\begin{table}
 \caption{List of emission lines masked for the stellar kinematics fit.}
 \label{masked_lines}
 \centering
 \begin{tabular}{l c | l c}
 \hline\hline
 Line & Rest wavelength&Line & Rest wavelength\\
     &  (\AA)& &  (\AA)\\
 \hline
  He\,{\sc ii}       &4686&  {[}O\,{\sc i}{]}   &6300\\
  {[}Ar\,{\sc iv}{]} &4711&  {[}O\,{\sc i}{]}   &6364\\
  {[}Ar\,{\sc iv}{]} &4740&  {[}N\,{\sc ii}{]}  &6548\\
  H$\upbeta$         &4861&  H$\upalpha$        &6563\\
  {[}O\,{\sc iii}{]} &4959&  {[}N\,{\sc ii}{]}  &6583\\
  {[}O\,{\sc iii}{]} &5007&  He\,{\sc i}        &6678\\
  {[}Fe\,{\sc vii}{]}&5159&  {[}S\,{\sc ii}{]}  &6716\\
  {[}N\,{\sc i}{]}   &5199&  {[}S\,{\sc ii}{]}  &6731\\
  {[}Fe\,{\sc vii}{]}&6087&  {[}Ar\,{\sc v}{]}  &7006\\ 
  \hline
 \end{tabular}
\end{table}

\begin{figure}
   \centering
   \includegraphics[scale=0.30]{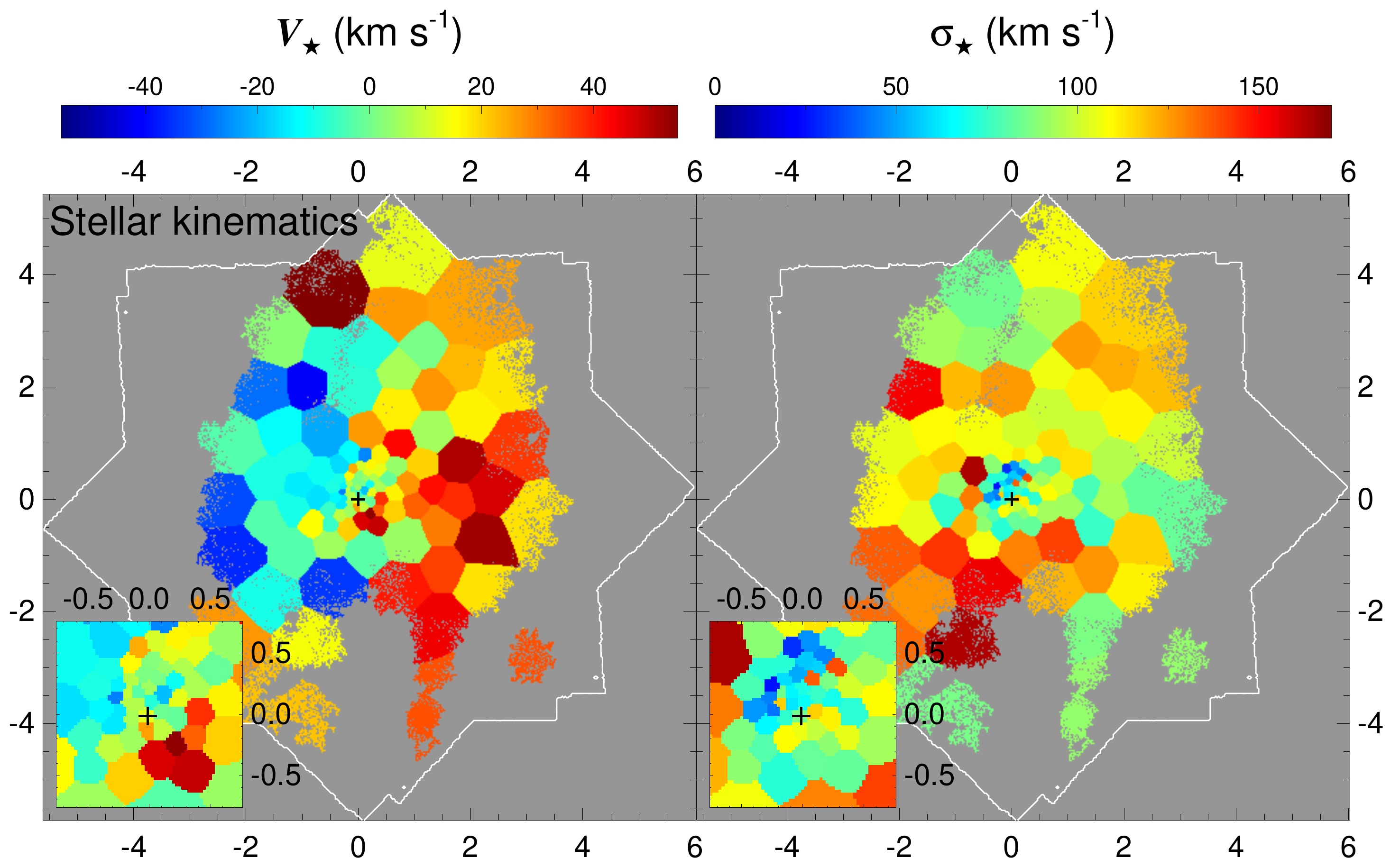}
   \caption{\label{stellar}{\it Left panel}: Velocity map of the stellar component. The zero in velocity corresponds to $z=0.016221$ ($cz=4863\,{\rm km\,s^{-1}}$; see Sect.~\ref{systemic}). {\it Right panel}: Velocity dispersion of the stellar component. In this and subsequent maps, the white contour delineates the field of view of the observations.}
\end{figure}

Our initial guesses for the fit were $V_\star=cz=4842\,{\rm km\,s^{-1}}$ (where $c$ is the speed of light in vacuum) and $\sigma_\star=30\,{\rm km\,s^{-1}}$, respectively. The stellar kinematic maps are shown in Fig.~\ref{stellar}. The velocity amplitude of the butterfly pattern is between 40 and 50\,${\rm km\,s^{-1}}$, which is similar to what is found for the ionised gas disc (Sect. \ref{orientation}). Hence, the circumnuclear stellar population is rotation-supported.

\subsection{One- to six-component fits to emission lines}

\label{resolved}

To study the ionised gas, we started by producing single-component fits of the gas using \texttt{GIST} \citep[akin to those made for][]{Knapen2019}. While examining them with \texttt{Mapviewer}, we discovered that a single-component description does not work well for many regions, especially in the innermost arcsecond. Hence, we wrote our own code that uses \texttt{pyGandALF} to perform multi-component fits with a series of criteria to decide the number of Gaussian components required for a given spectrum.

\begin{table}
 \caption{Fixed emission line flux ratios.}
 \label{ratios}
 \centering
 \begin{tabular}{c c}
 \hline\hline
 Ratio & Fixed value\\
 \hline
 {[}O\,{\sc iii}{]}\,$\uplambda$4959/{[}O\,{\sc iii}{]}\,$\uplambda$5007&0.335\\
 {[}O\,{\sc i}{]}\,$\uplambda$6364/{[}O\,{\sc i}{]}\,$\uplambda$6300    &0.330\\
 {[}N\,{\sc ii}{]}\,$\uplambda$6548/{[}N\,{\sc ii}{]}\,$\uplambda$6583  &0.327\\
  \hline
 \end{tabular}
 \tablefoot{Values estimated using the theoretical Einstein coefficients in \citet{Storey2000}.}
\end{table}

Increasing the number of gas components comes at the cost of increasing the number of free parameters, so we decided to reduce the complexity of the fit by tying the flux ratios of several doublets to their predicted values as calculated from the Einstein coefficients theoretically estimated by \citet{Storey2000}. Those flux ratios are shown in Table~\ref{ratios}. For all the lines in a component the kinematics were tied to those of H$\upalpha$.

We fitted the lines that are used in BPT diagnostics (Sect.~\ref{bpt}), namely H$\upbeta$, {[}O\,{\sc iii}{]}\,$\uplambda$4959, {[}O\,{\sc iii}{]}\,$\uplambda$5007, {[}O\,{\sc i}{]}\,$\uplambda$6300, {[}O\,{\sc i}{]}\,$\uplambda$6364, {[}N\,{\sc ii}{]}\,$\uplambda$6548, H$\upalpha$, {[}N\,{\sc ii}{]}\,$\uplambda$6583, {[}S\,{\sc ii}{]}\,$\uplambda$6716, and {[}S\,{\sc ii}{]}\,$\uplambda$6731. Additionally, we fitted the {[}S\,{\sc iii}{]}\,$\uplambda$9069 line, which, although in the infrared, is found in a window with no prominent sky lines. We only fitted windows with a width of $5000\,{\rm km\,s^{-1}}$ centred in the lines of interest. These windows contain two sky lines that were masked in the stellar kinematic fit, namely at $\uplambda=6364$\,\AA\ and $\uplambda=6864$\,\AA, which were left unmasked in here.

Since the ionised gas binning is much finer that the stellar one, the impact of stellar emission is very small and easily close to the noise level. We therefore ignored stellar emission altogether and simply assumed an underlying continuum modelled by a multiplicative eight-order Legendre polynomial (hence ignoring stellar lines). This is further justified by the fact that we only fit a narrow wavelength interval around the emission lines.

\begin{table}
 \setlength{\tabcolsep}{2.0pt}
 \caption{List of initial guesses for the emission line fits.}
 \label{guesses}
 \centering
 \begin{tabular}{c c | c c | c c | c c | c c | c c}
 \hline\hline
 $V_{1}$ & $\sigma_{1}$&$V_{2}$ & $\sigma_{2}$&$V_{3}$ & $\sigma_{3}$&$V_{4}$ & $\sigma_{4}$&$V_{5}$ & $\sigma_{5}$&$V_{6}$ & $\sigma_{6}$\\
 \multicolumn{2}{c|}{$\left({\rm km\,s^{-1}}\right)$}&\multicolumn{2}{c|}{$\left({\rm km\,s^{-1}}\right)$}&\multicolumn{2}{c|}{$\left({\rm km\,s^{-1}}\right)$}&\multicolumn{2}{c|}{$\left({\rm km\,s^{-1}}\right)$}&\multicolumn{2}{c|}{$\left({\rm km\,s^{-1}}\right)$}&\multicolumn{2}{c}{$\left({\rm km\,s^{-1}}\right)$}\\
  \hline
  \multicolumn{12}{c}{Single-component fits}\\
  \hline
 $0$  &  $80$  & --    & --    & --    & --    & --    & --    & --    & --    & --    & --  \\
  \hline
  \multicolumn{12}{c}{Two-component fits}\\
  \hline
 $0$  &  $80$  & $-300$& $200$ & --    & --    & --    & --    & --    & --    & --    & --  \\
 $0$  &  $80$  &  $300$& $200$ & --    & --    & --    & --    & --    & --    & --    & --  \\
  \hline
  \multicolumn{12}{c}{Three-component fits}\\
  \hline
 $0$  &  $80$  & $-350$& $200$ & $-200$& $200$ & --    & --    & --    & --    & --    & --  \\
 $0$  &  $80$  & $-350$& $200$ &  $200$& $200$ & --    & --    & --    & --    & --    & --  \\
 $0$  &  $80$  & $-350$& $200$ &  $400$& $200$ & --    & --    & --    & --    & --    & --  \\
 $0$  &  $80$  & $-200$& $200$ & $-100$& $70$  & --    & --    & --    & --    & --    & --  \\
 $0$  &  $80$  & $-200$& $200$ &  $200$& $200$ & --    & --    & --    & --    & --    & --  \\
 $0$  &  $80$  & $-200$& $200$ &  $400$& $200$ & --    & --    & --    & --    & --    & --  \\
 $0$  &  $80$  & $-200$& $200$ &  $400$& $200$ & --    & --    & --    & --    & --    & --  \\ 
  \hline
  \multicolumn{12}{c}{Four-component fits}\\
  \hline
 $0$  &  $80$  & $-350$& $200$ &  $200$& $100$ & $700$ & $100$ & --    & --    & --    & --  \\
 $0$  &  $80$  & $-350$& $200$ &  $400$& $100$ & $700$ & $100$ & --    & --    & --    & --  \\
 $0$  &  $80$  & $-200$& $200$ &  $200$& $100$ & $700$ & $100$ & --    & --    & --    & --  \\
 $0$  &  $80$  & $-200$& $200$ &  $400$& $100$ & $700$ & $100$ & --    & --    & --    & --  \\
 $0$  &  $80$  & $-150$& $200$ & $-100$&  $70$ & $300$ & $100$ & --    & --    & --    & --  \\
  \hline
  \multicolumn{12}{c}{Five-component fits}\\
  \hline  
 $0$  &  $80$  & $-380$& $200$ &  $200$& $100$ & $700$ & $100$ & $-150$& $100$ & --    & --  \\
 $0$  &  $80$  & $-380$& $200$ &  $300$& $100$ & $700$ & $100$ & $750$ & $250$ & --    & --  \\
 $0$  &  $80$  & $-200$& $200$ &  $200$& $100$ & $700$ & $100$ & $-150$& $100$ & --    & --  \\
 $0$  &  $80$  & $-200$& $200$ &  $300$& $100$ & $700$ & $100$ & $750$ & $250$ & --    & --  \\
  \hline
  \multicolumn{12}{c}{Six-component fits}\\
  \hline
 $0$  &  $80$  & $-380$& $200$ &  $200$& $100$ & $700$ & $100$ & $800$ & $250$ & $-150$&$100$\\
 $0$  &  $80$  & $-380$& $200$ &  $400$& $100$ & $700$ & $100$ & $800$ & $250$ & $-150$&$100$\\
 $0$  &  $80$  & $-200$& $200$ &  $200$& $100$ & $700$ & $100$ & $800$ & $250$ & $-150$&$100$\\
 $0$  &  $80$  & $-200$& $200$ &  $400$& $100$ & $700$ & $100$ & $800$ & $250$ & $-150$&$100$\\
  \hline  
  \end{tabular}
  \tablefoot{Velocities relative to the $cz$ of NGC~7130 assuming $z=0.016151$ ($cz=4842\,{\rm km\,s^{-1}}$).}
\end{table}

After a careful eyeball examination of hundreds of spectra, we decided that up to six Gaussian components per spectrum were required. To help the minimisation within \texttt{pyGandALF} to find the global minimum in multi-component fits, we established a set of initial guesses for the kinematics of the components (Table~\ref{guesses}). The need for this is illustrated in \cite{Ho2016}. The initial guesses were selected based on experience after manually fitting several hundred spectra representative of the whole data cube. For each fit, we calculated the chi-square:
\begin{equation}
 \chi_{j,n,k}^2=\sum_i\frac{\left(O^i_j-F_{j,n,k}^i\right)^2}{\left(s_j^i\right)^2}
,\end{equation}
where $O$ corresponds to the observed spectrum, $F$ corresponds to the fit, $s^2$ corresponds to the variances of the spectrum, $i$ is the running index over the unmasked elements of the spectrum, $n$ is the number of components included in the fit, and $k$ labels the fits with the same number of components but different initial guesses. For a bin $j$ and a given number of components, the best fit $F_{j,n}$ is defined to be the one that yields the smallest chi-square:
\begin{equation}
\chi^2_{j,n}=\min_{k}\left(\chi^2_{j,n,k}\right).
\end{equation}
We also calculated the chi-square for two restricted ranges in wavelength, namely restframe $4985\,\text{\AA}-5035\,\text{\AA}$ covering the brightest of the fitted {[}O\,{\sc iii}{]} lines (blue chi-square),
\begin{equation}
 \chi_{j,n}^{\rm b}{}^2=\sum_i\frac{\left(O^i_j-F_{j,n}^i\right)^2}{\left(s_j^i\right)^2}\,\,\,\,\,\,\,\,\,\,\,\,\forall i\,\mid\,4985\,\text{\AA}<\lambda_i<5035\,\text{\AA},
\end{equation}
and $6520\,\text{\AA}-6610\,\text{\AA}$ covering the complex of lines including the {[}N\,{\sc ii}{]} doublet and H$\upalpha$ (red chi-square),
\begin{equation}
 \chi_{j,n}^{\rm r}{}^2=\sum_i\frac{\left(O^i_j-F_{j,n}^i\right)^2}{\left(s_j^i\right)^2}\,\,\,\,\,\,\,\,\,\,\,\,\forall i\,\mid\,6520\,\text{\AA}<\lambda_i<6610\,\text{\AA}.
\end{equation}

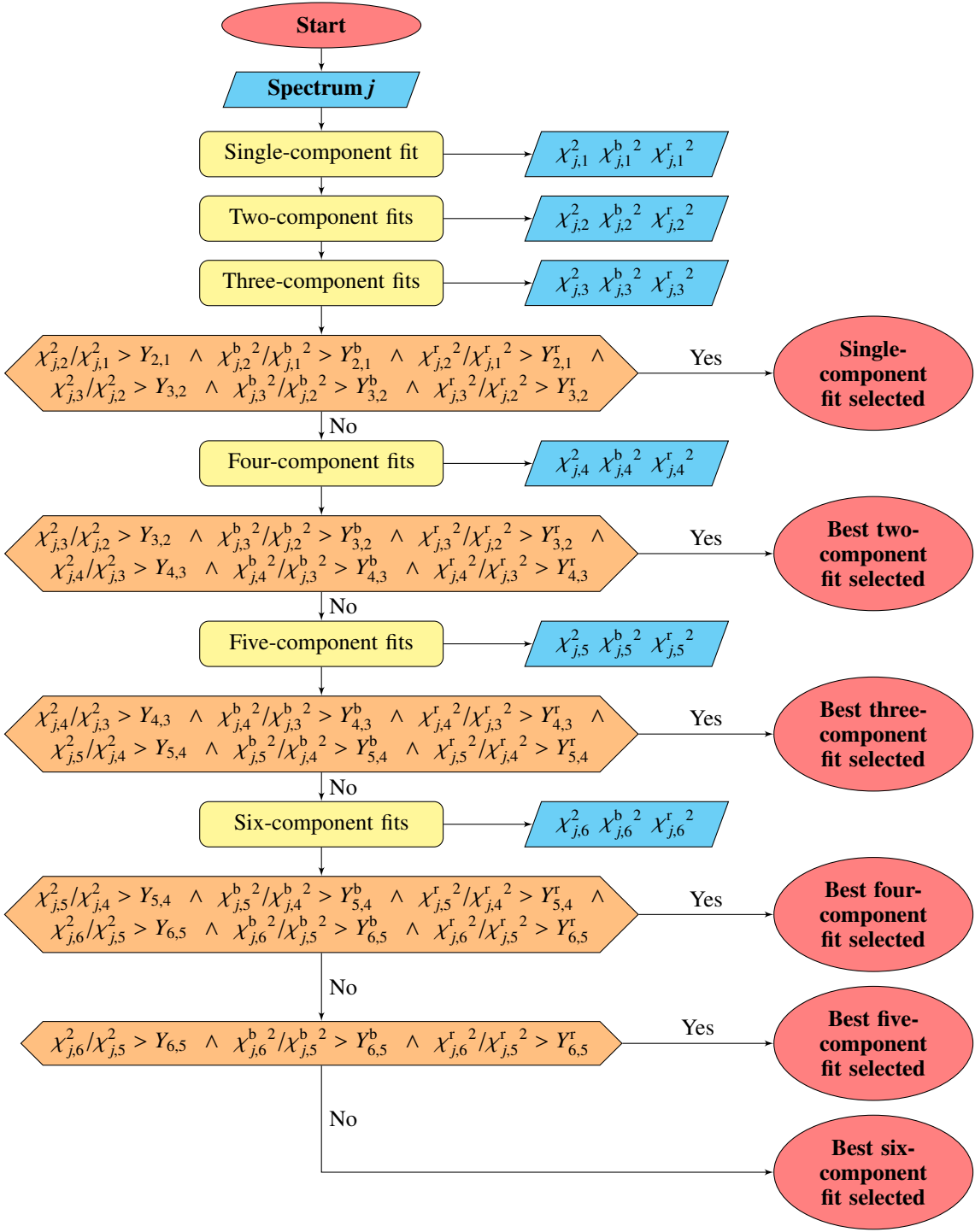
\begin{figure*}
\begin{center}
\begin{tikzpicture}[]
\tikzstyle{decision} = [chamfered rectangle, chamfered rectangle xsep=5cm, draw, fill=orange!50, 
    text width=25em, text badly centered, node distance=3.cm, inner sep=0pt]
\tikzstyle{action} = [rectangle, draw, fill=yellow!50, 
    text width=10em, text centered, rounded corners, minimum height=2em]
\tikzstyle{line} = [draw, -latex']
\tikzstyle{inout} = [draw, ellipse,fill=red!50, node distance=3cm,
    minimum height=2em, text badly centered, align=center, 
    text width=5.5em]
\tikzstyle{data} = [trapezium, trapezium left angle=70, trapezium right angle=110, text width=6.8em, draw, fill=cyan!50, text badly centered]
    
    \node [inout] (init) {\bf{Start}};
    \node [data,below of=init,node distance=1.0cm] (data) {\bf{Spectrum} \textbfit{j}};
    \node [action,below of=data,node distance=1.0cm] (single) {Single-component fit};
    \node [data,right of=single,node distance=4.7cm] (chi1) {$\chi^2_{j,1}$\,\,\,$\chi^{\rm b}_{j,1}{}^2$\,\,\,$\chi^{\rm r}_{j,1}{}^2$};
    \node [action,below of=single,node distance=1.0cm] (double) {Two-component fits};
    \node [data,right of=double,node distance=4.7cm] (chi2) {$\chi^2_{j,2}$\,\,\,$\chi^{\rm b}_{j,2}{}^2$\,\,\,$\chi^{\rm r}_{j,2}{}^2$};
    \node [action,below of=double,node distance=1.0cm] (triple) {Three-component fits};
    \node [data,right of=triple,node distance=4.7cm] (chi3) {$\chi^2_{j,3}$\,\,\,$\chi^{\rm b}_{j,3}{}^2$\,\,\,$\chi^{\rm r}_{j,3}{}^2$};
    \node [decision,below of=triple,node distance=1.4cm] (decision1) {$\chi_{j,2}^2/\chi_{j,1}^2 > Y_{2,1}\,\,\,\land\,\,\,\chi^{\rm b}_{j,2}{}^2/\chi^{\rm b}_{j,1}{}^2 > Y^{\rm b}_{2,1}\,\,\,\land\,\,\,\chi^{\rm r}_{j,2}{}^2/\chi^{\rm r}_{j,1}{}^2 > Y^{\rm r}_{2,1}\,\,\,\land\,\,\,\chi_{j,3}^2/\chi_{j,2}^2 > Y_{3,2}\,\,\,\land\,\,\,\chi^{\rm b}_{j,3}{}^2/\chi^{\rm b}_{j,2}{}^2 > Y^{\rm b}_{3,2}\,\,\,\land\,\,\,\chi^{\rm r}_{j,3}{}^2/\chi^{\rm r}_{j,2}{}^2 > Y^{\rm r}_{3,2}$ };
    \node [inout,right of=decision1,node distance=8.5cm] (end1) {\bf{Single-component fit selected}};
    \node [action,below of=decision1,node distance=1.4cm] (quadruple) {Four-component fits};
    \node [data,right of=quadruple,node distance=4.7cm] (chi4) {$\chi^2_{j,4}$\,\,\,$\chi^{\rm b}_{j,4}{}^2$\,\,\,$\chi^{\rm r}_{j,4}{}^2$};
    \node [decision,below of=quadruple,node distance=1.4cm] (decision2) {$\chi_{j,3}^2/\chi_{j,2}^2 > Y_{3,2}\,\,\,\land\,\,\,\chi^{\rm b}_{j,3}{}^2/\chi^{\rm b}_{j,2}{}^2 > Y^{\rm b}_{3,2}\,\,\,\land\,\,\,\chi^{\rm r}_{j,3}{}^2/\chi^{\rm r}_{j,2}{}^2 > Y^{\rm r}_{3,2}\,\,\,\land\,\,\,\chi_{j,4}^2/\chi_{j,3}^2 > Y_{4,3}\,\,\,\land\,\,\,\chi^{\rm b}_{j,4}{}^2/\chi^{\rm b}_{j,3}{}^2 > Y^{\rm b}_{4,3}\,\,\,\land\,\,\,\chi^{\rm r}_{j,4}{}^2/\chi^{\rm r}_{j,3}{}^2 > Y^{\rm r}_{4,3}$ };
    \node [inout,right of=decision2,node distance=8.5cm] (end2) {\bf{Best two-component fit selected}};
    \node [action,below of=decision2,node distance=1.4cm] (quintuple) {Five-component fits};
    \node [data,right of=quintuple,node distance=4.7cm] (chi5) {$\chi^2_{j,5}$\,\,\,$\chi^{\rm b}_{j,5}{}^2$\,\,\,$\chi^{\rm r}_{j,5}{}^2$};
    \node [decision,below of=quintuple,node distance=1.4cm] (decision3) {$\chi_{j,4}^2/\chi_{j,3}^2 > Y_{4,3}\,\,\,\land\,\,\,\chi^{\rm b}_{j,4}{}^2/\chi^{\rm b}_{j,3}{}^2 > Y^{\rm b}_{4,3}\,\,\,\land\,\,\,\chi^{\rm r}_{j,4}{}^2/\chi^{\rm r}_{j,3}{}^2 > Y^{\rm r}_{4,3}\,\,\,\land\,\,\,\chi_{j,5}^2/\chi_{j,4}^2 > Y_{5,4}\,\,\,\land\,\,\,\chi^{\rm b}_{j,5}{}^2/\chi^{\rm b}_{j,4}{}^2 > Y^{\rm b}_{5,4}\,\,\,\land\,\,\,\chi^{\rm r}_{j,5}{}^2/\chi^{\rm r}_{j,4}{}^2 > Y^{\rm r}_{5,4}$ };
    \node [inout,right of=decision3,node distance=8.5cm] (end3) {\bf{Best three-component fit selected}};
    \node [action,below of=decision3,node distance=1.4cm] (sextuple) {Six-component fits};
    \node [data,right of=sextuple,node distance=4.7cm] (chi6) {$\chi^2_{j,6}$\,\,\,$\chi^{\rm b}_{j,6}{}^2$\,\,\,$\chi^{\rm r}_{j,6}{}^2$};
    \node [decision,below of=sextuple,node distance=1.4cm] (decision4) {$\chi_{j,5}^2/\chi_{j,4}^2 > Y_{5,4}\,\,\,\land\,\,\,\chi^{\rm b}_{j,5}{}^2/\chi^{\rm b}_{j,4}{}^2 > Y^{\rm b}_{5,4}\,\,\,\land\,\,\,\chi^{\rm r}_{j,5}{}^2/\chi^{\rm r}_{j,4}{}^2 > Y^{\rm r}_{5,4}\,\,\,\land\,\,\,\chi_{j,6}^2/\chi_{j,5}^2 > Y_{6,5}\,\,\,\land\,\,\,\chi^{\rm b}_{j,6}{}^2/\chi^{\rm b}_{j,5}{}^2 > Y^{\rm b}_{6,5}\,\,\,\land\,\,\,\chi^{\rm r}_{j,6}{}^2/\chi^{\rm r}_{j,5}{}^2 > Y^{\rm r}_{6,5}$  };
    \node [inout,right of=decision4,node distance=8.5cm] (end4) {\bf{Best four-component fit selected}};
    \node [decision,below of=decision4,node distance=2.0cm] (decision5) {$\chi_{j,6}^2/\chi_{j,5}^2 > Y_{6,5}\,\,\,\land\,\,\,\chi^{\rm b}_{j,6}{}^2/\chi^{\rm b}_{j,5}{}^2 > Y^{\rm b}_{6,5}\,\,\,\land\,\,\,\chi^{\rm r}_{j,6}{}^2/\chi^{\rm r}_{j,5}{}^2 > Y^{\rm r}_{6,5}$  };
    \node [inout,right of=decision5,node distance=8.5cm] (end5) {\bf{Best five-component fit selected}};    
    \node [inout,below of=end5,node distance=2.0cm] (end6) {\bf{Best six-component fit selected}};

    \path[line] (init) -- (data);
    \path[line] (data) -- (single);
    \path[line] (single) -- (double);
    \path[line] (double) -- (triple);
    \path[line] (triple) -- (decision1);
    \path[line] (single) -- (chi1);
    \path[line] (double) -- (chi2);
    \path[line] (triple) -- (chi3);
    \path[line] (decision1) -- node [above]{Yes} (end1);
    \path[line] (decision1) -- node [right]{No} (quadruple);
    \path[line] (quadruple) -- (chi4);
    \path[line] (quadruple) -- (decision2);
    \path[line] (decision2) -- node [above]{Yes} (end2);
    \path[line] (decision2) -- node [right]{No} (quintuple);
    \path[line] (quintuple) -- (chi5);
    \path[line] (quintuple) -- (decision3);
    \path[line] (decision3) -- node [above]{Yes} (end3);
    \path[line] (decision3) -- node [right]{No} (sextuple);
    \path[line] (sextuple) -- (chi6);
    \path[line] (sextuple) -- (decision4);
    \path[line] (decision4) -- node [above]{Yes} (end4);
    \path[line] (decision4) -- node [right]{No} (decision5);
    \path[line] (decision5) -- node [above]{Yes} (end5);
    \path[line] (decision5) |- node [right, near start]{No} (end6);

\end{tikzpicture}
\end{center}
\caption{\label{flow} Flow diagram illustrating how our code decides the number of components that a bin $j$ requires to have its ionised gas lines fitted.}
\end{figure*}

We used the chi-square values to decide whether adding extra components to a fit improved it enough to justify the growth in complexity. Originally, this was done by calculating the ratio of the chi-square values for $n+1$ and $n$ components, $\chi_{j,n+1}^2/\chi^2_{j,n}$ and comparing it to a threshold ratio of chi-squared values $Y_{n+1,n}$ chosen so the automatic procedure would, in general, choose as many components as we would if fitting the spectra manually. This method was, for example, implemented by \citet{Davis2012} to choose between single- and two-component fits in IFU data of NGC~1266. However, occasionally, adding an extra component to a fit did not improve it significantly, but adding two caused a large leap in the quality of the fit. Therefore, our final implementation takes into account both $Y_{n+1,n}$ and $Y_{n+2,n+1}$ before considering whether extra components are necessary. Also, since some fit improvements only show in very narrow wavelength ranges that are not well described by the global chi-square value $\chi_{j,n}^2$, we also took into account the improvements in the blue and red restricted ranges as quantified by $\chi^{\rm b}_{j,n+1}{}^2/\chi^{\rm b}_{j,n}{}^2$ and $\chi^{\rm r}_{j,n+1}{}^2/\chi^{\rm r}_{j,n}{}^2$ through the thresholds $Y^{\rm b}_{n+1,n}$, $Y^{\rm b}_{n+2,n+1}$, $Y^{\rm r}_{n+1,n}$, and $Y^{\rm r}_{n+2,n+1}$. In Appendix.~\ref{example1}, we show examples of fits that justify considering these choices. Our complete multi-component fitting strategy is depicted in the flow diagram in Fig.~\ref{flow}.

\begin{table}
 \caption{Thresholds in the ratio of the chi-square values used to select the number of components in the emission line fits.}
 \label{thresholds}
 \centering
 \begin{tabular}{c c c c}
 \hline\hline
 Chi-square & \multicolumn{3}{c}{Value of the threshold}\\
 ratio      &  $N<200$& $200\leq N<1000$ &$N\geq1000$\\
 \hline
 $Y_{2,1}$ & 0.50 & 0.40 & 0.30\\
 $Y_{3,2}$ & 0.70 & 0.60 & 0.50\\
 $Y_{4,3}$ & 0.80 & 0.70 & 0.60\\
 $Y_{5,4}$ & 0.80 & 0.70 & 0.60\\
 $Y_{6,5}$ & 0.80 & 0.70 & 0.60\\
 \hline
 $Y^{\rm b}_{2,1}$ & 0.30 & 0.30 & 0.20\\
 $Y^{\rm b}_{3,2}$ & 0.50 & 0.40 & 0.30\\
 $Y^{\rm b}_{4,3}$ & 0.75 & 0.65 & 0.55\\
 $Y^{\rm b}_{5,4}$ & 0.75 & 0.65 & 0.55\\
 $Y^{\rm b}_{6,5}$ & 0.75 & 0.65 & 0.55\\
 \hline
 $Y^{\rm r}_{2,1}$ & 0.30 & 0.30 & 0.20\\
 $Y^{\rm r}_{3,2}$ & 0.55 & 0.45 & 0.35\\
 $Y^{\rm r}_{4,3}$ & 0.75 & 0.65 & 0.55\\
 $Y^{\rm r}_{5,4}$ & 0.75 & 0.65 & 0.55\\
 $Y^{\rm r}_{6,5}$ & 0.75 & 0.65 & 0.55\\
 \hline
 \end{tabular}
 \tablefoot{$N$ denotes the number of spaxels in a Voronoi bin.}
\end{table}

After comparing automatic fits with manual ones, we found that if we choose the same $Y_{n+1,n}$, $Y^{\rm b}_{n+1,n}$, and $Y^{\rm r}_{n+1,n}$ values for all bins, we would either be overfitting the bins with a large surface, or underfitting the bins with a small one. A possible cause for that is that the variances of the spectra calculated by the MUSE pipeline might be underestimated in the low-surface-brightness regime, which impacts the $\chi^2$ determinations. Therefore, we chose different thresholds for bins with fewer than $N=200$\,spaxels, bins between $N=200$\,spaxels and $N=999$\,spaxels, and bins with $N=1000$\,spaxels or more. The selected values are listed in Table~\ref{thresholds}. Although these values are to some degree arbitrary, they constitute a reasonable compromise between the need to describe as many visually identified components as possible and the necessity to avoid spurious components that would be overfitting the data.

\subsection{Identification of the kinematic components}

\label{kinematics}

We found nine distinct kinematic components based on their spatial location, their kinematics, and BPT ratios (see more on the latter criterion in Sect.~\ref{bpt}). One corresponds to the disc, five are narrow components ($\sigma\lesssim250\,{\rm km\,s^{-1}}$), and three are broad components ($\sigma\gtrsim250\,{\rm km\,s^{-1}}$). Since we were only fitting up to six of them per bin, not all are co-spatial. The kinematic properties of these components (as measured after the refining process explained in Sect.~\ref{refinment} and Fig.~\ref{flow2}) are summarised in Table~\ref{components}. The refined kinematic maps and the contours of the components are shown in Figs.~\ref{gas_kinematics} and \ref{locations}, respectively. The labels in the top-left corners of the velocity maps identify each component and are assigned a colour that is used consistently throughout this paper. Because the colour scale in Fig.~\ref{gas_kinematics} does not allow one to see the details in the kinematic maps of the disc component, we redisplay it in Fig.~\ref{disckin} with an adapted colour scale.

\begin{table*}
 \caption{Kinematics and morphology of the kinematically-defined components of the ionised gas.}
 \label{components}
 \centering
 \begin{tabular}{p{3.1cm} C{2.6cm} C{2.2cm} p{8.7cm}}
 \hline\hline
  Component& Typical $V$ & Typical $\sigma$ & Morphology\\
  &  $\left({\rm km\,s^{-1}}\right)$  &$\left({\rm km\,s^{-1}}\right)$ &\\
 \hline
 Disc  & $-60\lesssim V \lesssim 60$& $\sigma\lesssim100$& Butterfly pattern. The central $1^{\prime\prime}$ is blueshifted with respect to the systemic velocity of the galaxy as deduced from the kinematics at larger radii.\\
 \hline
 Blueshifted narrow & $-200\lesssim V\lesssim-50$&$50\lesssim\sigma\lesssim200$ & Runs from the galactic nucleus towards the north-west to the edge of the field of view at $\sim3^{\prime\prime}$ ($\sim900$\,pc).\vspace{1.mm}\\
 Zero-velocity narrow &$-100\lesssim V\lesssim0$&$100\lesssim\sigma\lesssim250$ & A patch of this component seems to correlate with the north-eastern spiral arm. The other patches correlate with the UCNR.\vspace{0.8mm}\\
 Crescent narrow & $0\lesssim V\lesssim 300$&$100\lesssim\sigma\lesssim250$& Redshifted component with an inward-pointing crescent shape $\sim0\farcs7$ ($\sim200$\,pc) west of the nucleus.\vspace{0.8mm}\\
 Redshifted narrow 1 & $100\lesssim V\lesssim300$&$50\lesssim\sigma\lesssim200$ & Found up to $\sim1^{\prime\prime}$ ($\sim300$\,pc) north and south of the nucleus.\vspace{0.8mm}\\
 Redshifted narrow 2 & $500\lesssim V\lesssim800$&$\sigma\lesssim200$& Found up to $\sim1^{\prime\prime}$ ($\sim300$\,pc) north and south of the nucleus.\\
 \hline
 Blueshifted broad & $-600\lesssim V\lesssim-100$&$250\lesssim\sigma\lesssim500$ &
 Centred in the nucleus and elongated in the south-east to north-west direction with a width of $3^{\prime\prime}-4^{\prime\prime}$ (900\,pc -- 1200\,pc).\vspace{0.8mm}\\
 Zero-velocity broad&$-200\lesssim V\lesssim200$ &$250\lesssim\sigma\lesssim600$ & Found around the nucleus with an extension towards the south.\vspace{0.8mm}\\
 Redshifted broad &$800\lesssim V\lesssim1300$ & $200\lesssim\sigma\lesssim500$& Found south of the nucleus in a blob $\sim0\farcs5$ ($\sim150$\,pc) across.\\
 \hline
 \end{tabular}
\end{table*}

The disc component has a butterfly pattern, a low velocity dispersion ($\sigma<100\,{\rm km\,s^{-1}}$), and BPT ratios compatible with star formation (Sect.~\ref{bpt}). All of the remaining kinematic components, except the zero-velocity narrow component, have line ratios compatible with AGN excitation. The locations of the spiral arms within the bar (Fig.~\ref{images}) have velocities that are slightly different to those of their surroundings: the north-eastern arm is blueshifted, and the south-western arm is redshifted. We argue that this is evidence for gas inflow through the arms (Sect.~\ref{discussion}).

The blueshifted broad component occupies a $3^{\prime\prime}-4^{\prime\prime}$ (900\,pc -- 1200\,pc) wide region that runs in the south-east to north-west direction. All of the other blue- and redshifted components are found within the area covered by this component.

The blueshifted narrow component nearly overlaps (in projection) with the blueshifted broad component in the region north-west of the nucleus. The region with the highest velocity relative to the disc of NGC~7130 (darker shades of blue in Fig.~\ref{gas_kinematics}, or $V\sim-200\,{\rm km\,s^{-1}}$) runs first to the north and then to the north-west. Other regions of this component, especially those to the north-east and to the west of the nucleus, are not as blueshifted and might be to some degree confused with the zero-velocity narrow component. The $\eta$ parameter, which describes the ionisation mechanism, was used to distinguish them (Sect.~\ref{bpt}).

The crescent narrow component is found $0\farcs7$ (200\,pc) west of the nucleus.  The most redshifted regions of the crescent narrow component have velocities comparable to those in the redshifted narrow component~1. We considered them as two separate components because we see a narrow gap between them (Fig.~\ref{locations}). Also, the ionisation mechanism of this component is different to that of other outflow components, as it has LINER line ratios rather than Seyfert ones (Sect.~\ref{bpt}).

The redshifted narrow components~1 and 2 overlap in projection throughout most of their extent. In regions around and south of the nucleus, they correspond to well-defined shoulders in {[}O\,{\sc iii}{]}. At $1^{\prime\prime}$ (300\,pc) north of the nucleus, these components are less well defined and might actually be describing the wings of the disc and the blueshifted components. The redshifted broad component is associated with the regions south of the nucleus where the redshifted narrow component~2 is located, and it could be interpreted as a non-Gaussian red wing.

The zero-velocity narrow component is seen in the inner arcsec in clumps north-east, north-west, and south-west of the nucleus (in locations overlapping with the UCNR, see Fig.~\ref{images}) as well as in the north-eastern arm. Thus, this component correlates with star formation (the line ratios indicate ionisation from star-forming regions; Sect.~\ref{bpt}). Because it is slightly blueshifted with respect to the disc ($V>-100\,{\rm km\,s^{-1}}$) it could correspond to star-formation driven outflows.

The zero-velocity broad component is found in regions where the redshifted narrow components are also present. As discussed in Appendix~\ref{examples}, although a visual examination of the spectra does not reveal an obvious need for this component, some regions require a low-amplitude very broad line for the redshifted narrow components to be properly characterised. Thus, this component might hold no physical meaning, and instead, in many bins, may describe the sum of the effects of non-Gaussian wings of the remaining components. In a few bins it fits with a single Gaussian what a human observer would distinguish as the blueshifted and the redshifted narrow components.

\subsection{Refinement of the fits}

\label{refinment}

\begin{figure}
\begin{center}
\begin{tikzpicture}[]
\tikzstyle{decision} = [chamfered rectangle, chamfered rectangle xsep=5cm, draw, fill=orange!50, 
    text width=16.em, text badly centered, node distance=3.cm, inner sep=0pt]
\tikzstyle{action} = [rectangle, draw, fill=yellow!50, 
    text width=10em, text centered, rounded corners, minimum height=2em]
\tikzstyle{line} = [draw, -latex']
\tikzstyle{inout} = [draw, ellipse,fill=red!50, node distance=3cm,
    minimum height=2em, text badly centered, align=center, 
    text width=5.5em]
\tikzstyle{data} = [trapezium, trapezium left angle=70, trapezium right angle=110, text width=6.8em, draw, fill=cyan!50, text badly centered]
\tikzstyle{line2} = [draw]
   
    \node [inout] (init) {\bf{Start}};
    \node [data,below of=init,node distance=0.9cm] (data) {\bf{Spectrum} \textbfit{j}};
    \node [decision,below of=data,node distance=1.0cm] (selection) {Is the size of the bin $j$ smaller than $N=200$\,spaxels?};
    \node [decision,below of=selection,node distance=2.6cm] (component) {Do at least $b$ out of the $a$ closest bins (including $j$) have a fitted component $n$?};
    \coordinate[left of=component,node distance=1cm](inter);
    \node [action,below of=inter,node distance=2.0cm] (add) {Initial guess for component $n$ set to the median of the previously fitted values among the $a$ closest bins.};
    \node [decision,below of=component,node distance=3.8cm] (all) {Have all the nine components been considered?};
    \coordinate[right of=all,node distance=4.5cm](inter2);
    \coordinate[right of=component,node distance=4.5cm](inter3);
    \node [action,below of=all,node distance=1.4cm] (fit) {Fit};
    \node [inout,below of=fit,node distance=1.0cm] (result) {\bf{Result}};

    \path[line] (init) -- (data);
    \path[line] (data) -- (selection);
    \path[line] ([xshift=-0.4cm]selection.south) -- node[left]{\begin{tabular}{r}Yes\\$a=9$\\$b=5$\\$n=1$\end{tabular}}([xshift=-0.4cm]component.north);
    \path[line] ([xshift=+0.4cm]selection.south) -- node[right]{\begin{tabular}{l}No\\$a=5$\\$b=3$\\$n=1$\end{tabular}}([xshift=+0.4cm]component.north);
    \path[line] ([xshift=-1.0cm]component.south) -- node[right]{Yes}(add);
    \path[line] (add) -- ([xshift=-1.0cm]all.north);
    \path[line] ([xshift=1.5cm]component.south) -- node[right]{No}([xshift=+1.5cm]all.north);
    \path[line2] (all) -- node[above]{No}(inter2);
    \path[line2] (inter2) -- node[left]{$n=n+1$}(inter3);
    \path[line] (inter3) -- (component);
    \path[line] (all) -- node[right]{Yes}(fit);
    \path[line] (fit) -- (result);

\end{tikzpicture}
\end{center}
\caption{\label{flow2} Flow diagram that illustrates the procedure for rerunning the fit of a spectrum $j$. This process was done three times after the initial run.}
\end{figure}
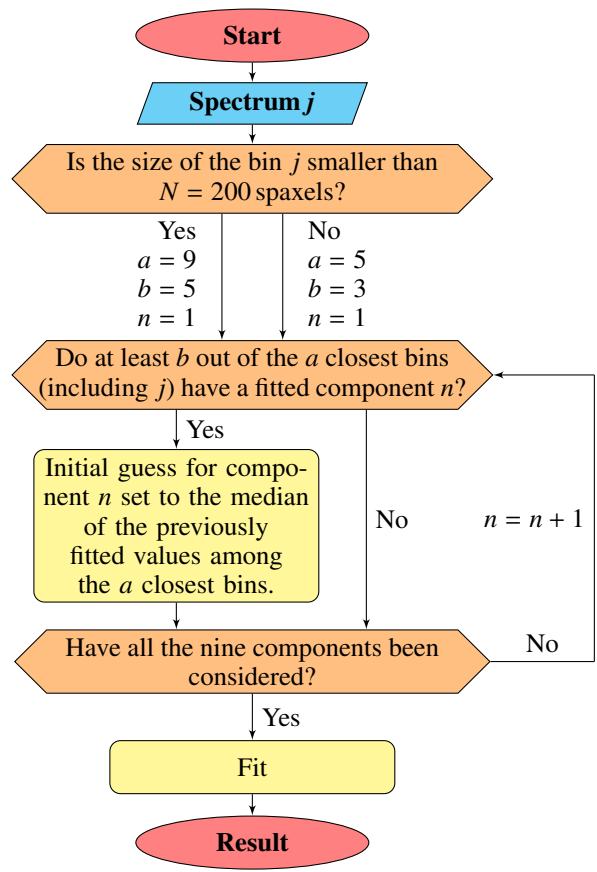

\begin{figure*}
   \centering
   \includegraphics[scale=0.30]{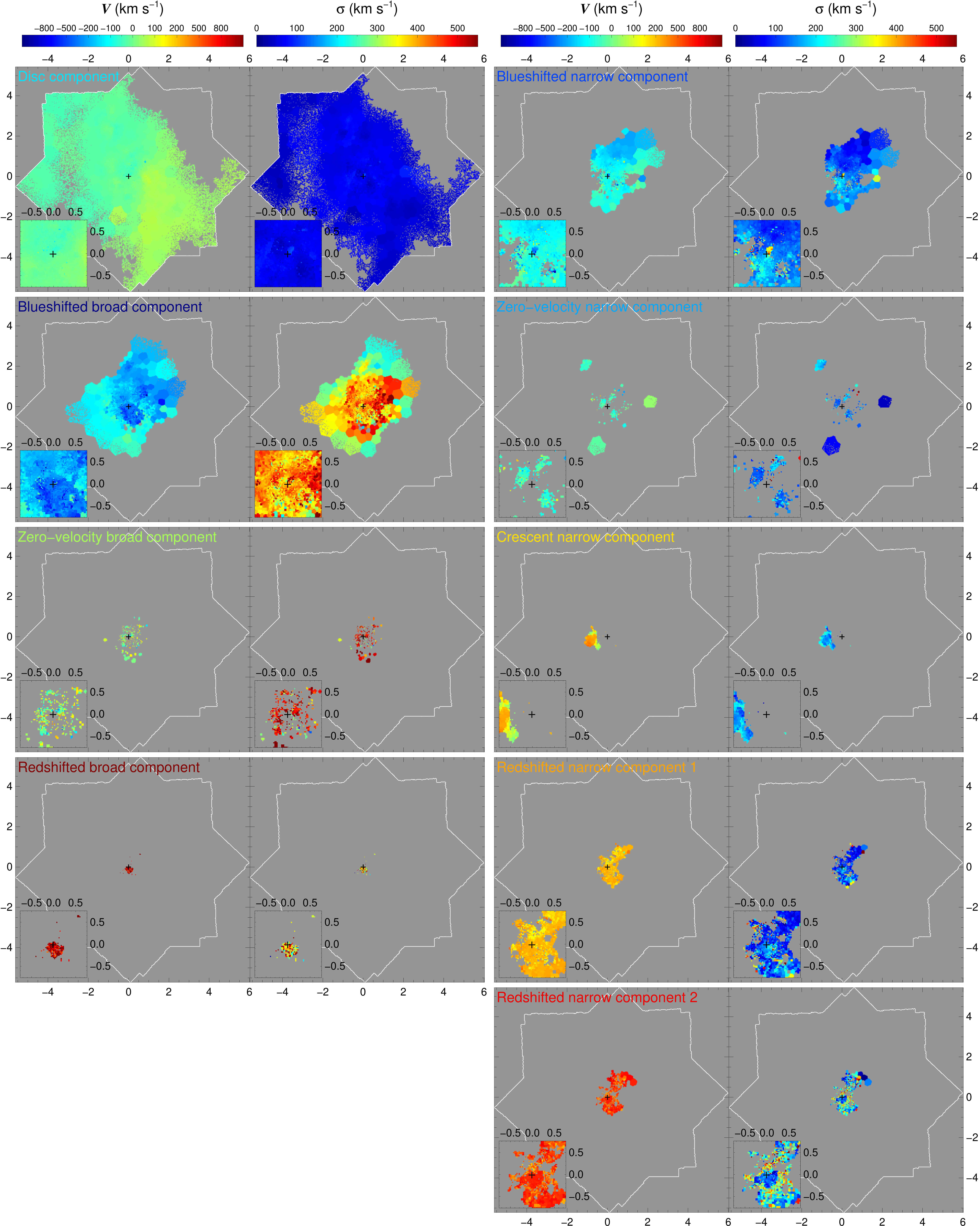}
   \caption{\label{gas_kinematics}{Kinematic maps of kinematically defined components. They are obtained from fits made accounting for all the emission lines of interest (Sect.~\ref{kinematics}) simultaneously. Nine sets of panels are shown, one for each of the kinematic components identified in the circumnuclear medium of NGC~7130. For each set, the {\it \emph{left-hand panel}} corresponds to the velocity map, and the {\it \emph{right-hand panel}} corresponds to the velocity dispersion map. The velocity colour bar is set so the spacing between velocity steps for $\left|V_{\rm gas}\right|>200\,{\rm km\,s^{-1}}$ is three times larger than for $\left|V_{\rm gas}\right|<200\,{\rm km\,s^{-1}}$. The zero in velocity corresponds to $z=0.016221$ ($cz=4863\,{\rm km\,s^{-1}}$; see Sect.~\ref{systemic}). The labels identifying components are given colours that are used consistently in all the plots in this article.}}
\end{figure*}

The velocity and velocity dispersion maps of the kinematic components obtained after the first run were not smooth (they showed large jumps in $V$ and $\sigma$ between neighbouring bins). Inspired by \citet{Davis2012} and \citet{Ho2016}, we refitted the spectra using the
median of the previously fitted values for the closest bins (including the
bin of interest itself) as initial guesses for $V$ and $\sigma$ . The distance between bins was calculated as that between geometric centres. The refitting algorithm is described in the flow diagram in Fig.~\ref{flow2}. This procedure was done three times to obtain the final maps (Fig.~\ref{gas_kinematics}). All the results, the discussion, and the conclusions are based on those refined fits.

The median absolute differences in $V$ and $\sigma$ between the third and the fourth iteration are considered to be representative of the fitting uncertainties. For line fluxes, we estimated median relative uncertainties as
\begin{equation}
 \frac{\delta f_n}{f_n}\equiv\text{median}\,\left({\frac{\left|f_{n,i,4}-f_{n,i,3}\right|}{f_{n,i,4}}}\right)
,\end{equation}
where the sub-indices 3 and 4 refer to the third and fourth iterations of the fits, respectively, and $i$ refers to the bins where the kinematic component $n$ is defined. The estimated uncertainties are listed in Table~\ref{uncertainties}.

\begin{table*}
  \setlength{\tabcolsep}{4.8pt}
 \caption{Estimated uncertainties of the fitted parameters.}
 \label{uncertainties}
 \centering
 \begin{tabular}{l c c c c c c c c c c}
 \hline\hline
  Component& $\delta V$ & $\delta\sigma$ & $\frac{\delta \text{H}\upbeta}{\text{H}\upbeta}$ &$\frac{\delta{[}\text{O\,{\sc iii}{]}}\,\uplambda5007}{{[}\text{O\,{\sc iii}}{]}\,\uplambda5007}$&$\frac{\delta{[}\text{O\,{\sc i}{]}}\,\uplambda6300}{{[}\text{O\,{\sc i}}{]}\,\uplambda6300}$&$\frac{\delta \text{H}\upalpha}{\text{H}\upalpha}$&$\frac{\delta{[}\text{N\,{\sc ii}{]}}\,\uplambda6583}{{[}\text{N\,{\sc ii}}{]}\,\uplambda6583}$&$\frac{\delta{[}\text{S\,{\sc ii}{]}}\,\uplambda6716}{{[}\text{S\,{\sc ii}}{]}\,\uplambda6716}$&$\frac{\delta{[}\text{S\,{\sc ii}{]}}\,\uplambda6731}{{[}\text{S\,{\sc ii}}{]}\,\uplambda6731}$&$\frac{\delta{[}\text{S\,{\sc iii}{]}}\,\uplambda9069}{{[}\text{S\,{\sc iii}}{]}\,\uplambda9069}$\\
  & \multicolumn{2}{c}{$\left({\rm km\,s^{-1}}\right)$}\\
  \hline
 Disc                 & 0.19  & 0.35 & 0.0089 & 0.056 & 0.017 & 0.0071 & 0.011 & 0.012 & 0.012 & 0.015\\
 \hline
 Blueshifted narrow   & 2.3   & 1.8  & 0.026  & 0.031 & 0.036 & 0.032  & 0.026 & 0.036 & 0.047 & 0.039\\
 Zero-velocity narrow & 7.7   & 6.3  & 0.089  & 0.20  & 0.22  & 0.092  & 0.11  & 0.12  & 0.15  & 0.15 \\
 Crescent narrow      & 0.72  & 0.82 & 0.011  & 0.011 & 0.0048& 0.012  & 0.0087& 0.0089& 0.0085& 0.015\\
 Redshifted narrow 1  & 4.4   & 5.7  & 0.11   & 0.081 & 0.15  & 0.11   & 0.15  & 0.23  & 0.16  & 0.082\\
 Redshifted narrow 2  & 13    & 8.9  & 0.069  & 0.068 & 0.074 & 0.11   & 0.099 & 0.22  & 0.16  & 0.13 \\
 \hline
 Blueshifted broad    & 5.9   & 3.1  & 0.024  & 0.020 & 0.021 & 0.023  & 0.017 & 0.026 & 0.061 & 0.033\\
 Zero-velocity broad  & 29    & 14   & 0.19   & 0.086 & 0.15  & 0.27   & 0.26  & 0.28  & 0.72  & 0.24 \\
 Redshifted broad     & 11    & 6.2  & 0.041  & 0.034 & 0.032 & 0.38   & 0.032 & 0.27  & 0.10  & 0.055\\
 \hline
 \end{tabular}
 \tablefoot{The uncertainties in $V$ and $\sigma$ are absolute, whereas those for the line fluxes are relative to the total line flux. Lines whose flux is tied to that of other lines (Table~\ref{ratios}) are not shown.}
\end{table*}

\begin{figure}
   \centering
   \includegraphics[scale=0.30]{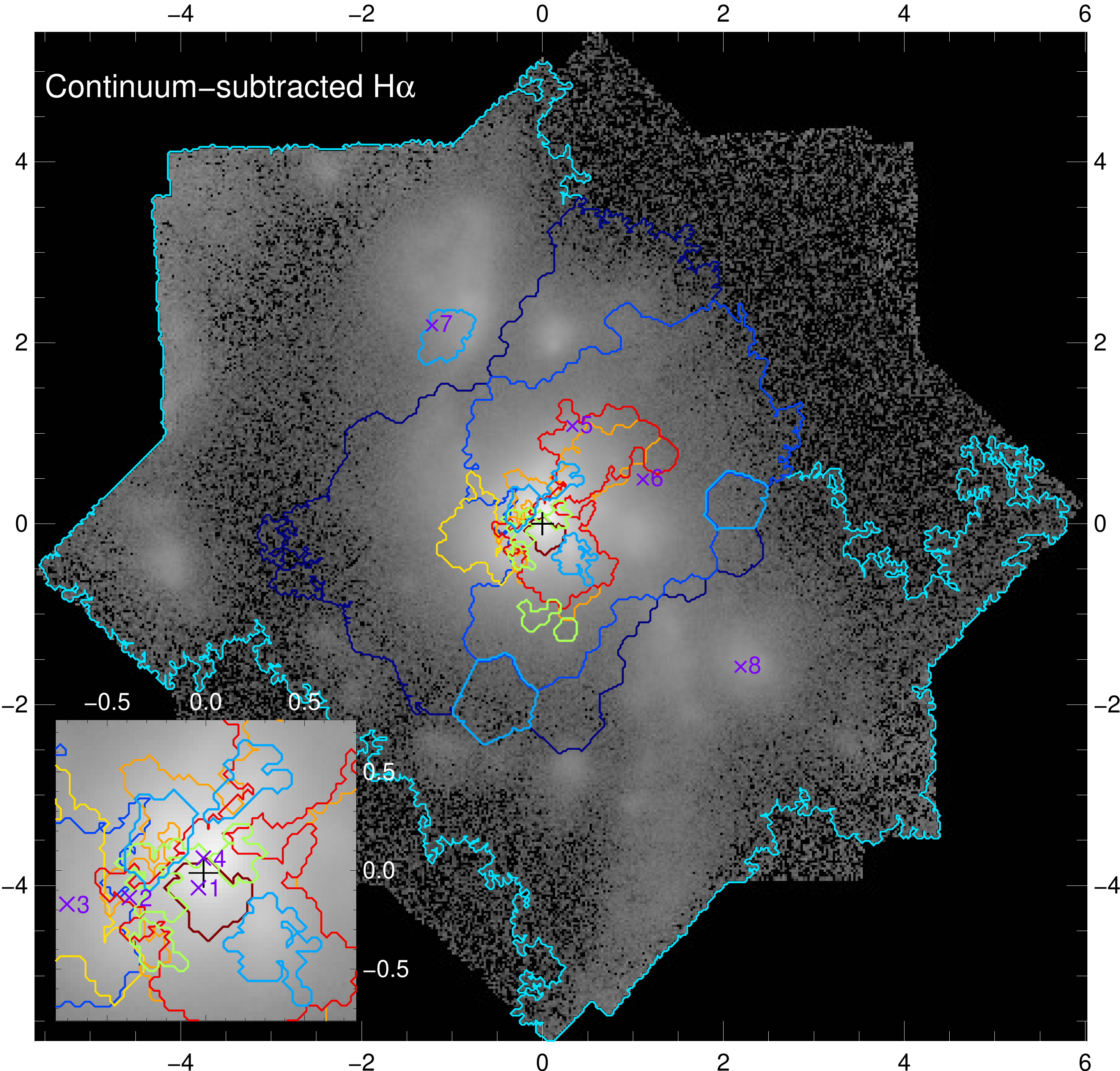}
   \caption{\label{locations}{Outlines of the nine kinematic components shown in Fig.~\ref{gas_kinematics} overlaid on the continuum-subtracted H$\upalpha$ image from Fig.~\ref{images}. The contours are colour-coded according to the labels identifying each component in Fig.~\ref{gas_kinematics}. To ease the readability of the figure, we omitted holes within components and regions smaller than $N=50$\,spaxels disconnected from the main body of the component. The purple x symbols and numbers indicate the bins whose spectra are discussed in Appendix~\ref{examples}. Those corresponding to bins in the inner $1\farcs5\times1\farcs5$ are indicated only in the inset to avoid clutter.}}
\end{figure}

\begin{figure}
   \centering
   \includegraphics[scale=0.30]{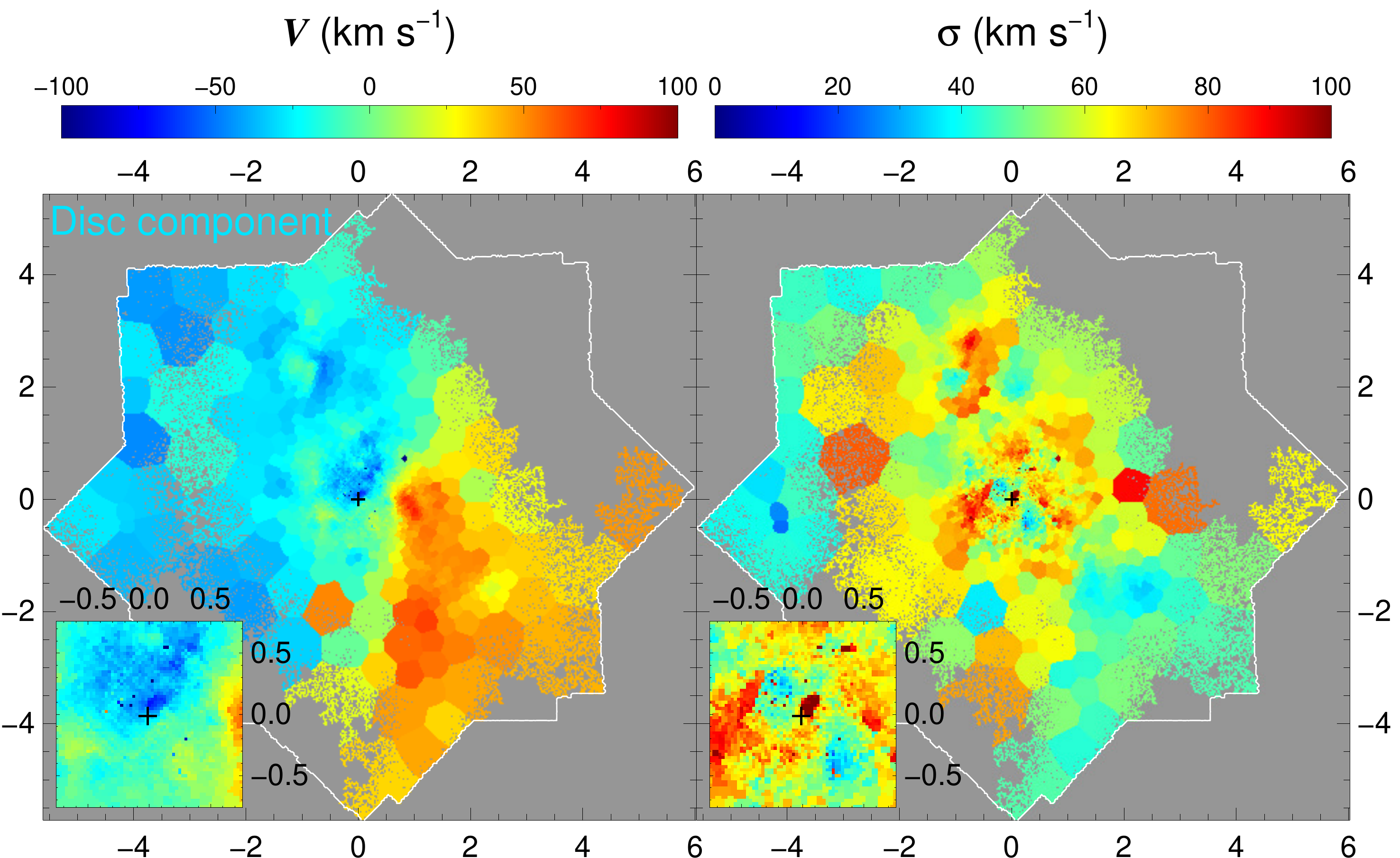}
   \caption{\label{disckin}{Velocity ({\it left}) and velocity dispersion ({\it right}) maps of disc component. The data are the same as in the {\it \emph{top-left}} panels in Fig.~\ref{gas_kinematics}, but the colour scales are adapted to highlight details.}}
\end{figure}

In Appendix~\ref{examples}, we show the spectra of a few representative bins indicated by the purple numbers and x signs in Fig.~\ref{locations}.  We also discuss their fits.

\subsection{The {[}Fe\,{\sc vii}{]}\,$\uplambda$6087 coronal line}

\label{sectcoronal}

We also fitted the {[}Fe\,{\sc vii}{]}\,$\uplambda$6087 coronal line kinematics. This line has a high ionisation potential \citep[99.1\,eV;][]{DeRobertis1984}, which unequivocally associates it with AGN photoionisation. Because of its faintness, we treated this line differently from the emission lines discussed above, by giving it its own binning constructed with a methodology similar to that employed for the H$\upalpha$ binning. Here, we used a 20\,\AA\ window centred at $\uplambda=6087$\,\AA\ to characterise the emission, and an equally wide window 50\,\AA\ bluewards for the continuum. We required the bins to have $\text{S/N}=10$. This resulted in 137 bins.

\begin{figure}
   \centering
   \includegraphics[scale=0.30]{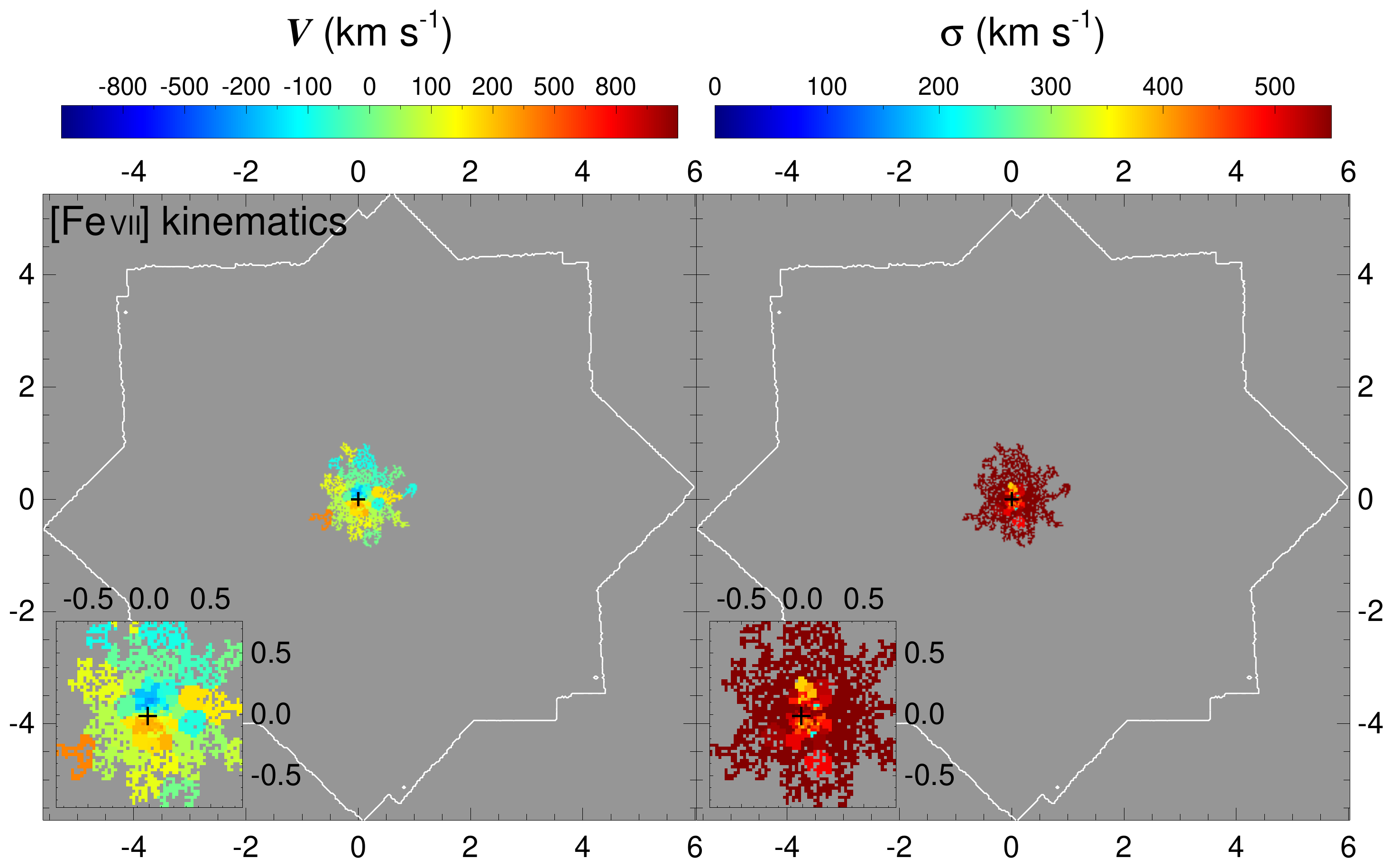}
   \caption{\label{coronal}NGC~7130 velocity map ({\it left panel}) and velocity dispersion map ({\it right panel}) derived from the {[}Fe\,{\sc vii}{]}\,$\uplambda$6087 line. The zero in velocity corresponds to $z=0.016221$ ($cz=4863\,{\rm km\,s^{-1}}$; see Sect.~\ref{systemic}). The colour scales are as in Fig.~\ref{gas_kinematics}.}
\end{figure}

The {[}Fe\,{\sc vii}{]} line was fitted with a single Gaussian component using \texttt{GIST} applied over the spectral range $6000\,\text{\AA}-6200\,\text{\AA}$.  The stellar continuum shape was modelled with a second-degree multiplicative Legendre polynomial. The $V$ and $\sigma$ maps are shown in Fig.~\ref{coronal}.

Within a radius of 0\farcs2, we find a bipolar structure centred at the nucleus and oriented in the north-south direction. The northern lobe is blueshifted, whereas the southern one is redshifted. The bipolar structure coincides with that in the single-component fits in \citet{Knapen2019}. Beyond the central $0\farcs2$, the maps are rather noisy, but there are also signs of a blueshifted northern component and a redshifted southern one, possibly corresponding to the blueshifted narrow component and redshifted narrow component~1, respectively.

\section{Results}

\label{results}

\subsection{The orientation of NGC~7130}

\label{orientation}

Disc orientation determinations for NGC~7130 are prone to uncertainties due to the distorted shape of its outer isophotes. The position angle $\text{PA}=160\degr$ obtained from 2MASS photometry \citep{Skrutskie2006} is clearly not compatible with the kinematic axes of the stellar and gas discs as seen in Figs.~\ref{stellar}, \ref{gas_kinematics}, and \ref{disckin}.

\begin{figure}
   \centering
   \includegraphics[scale=0.30]{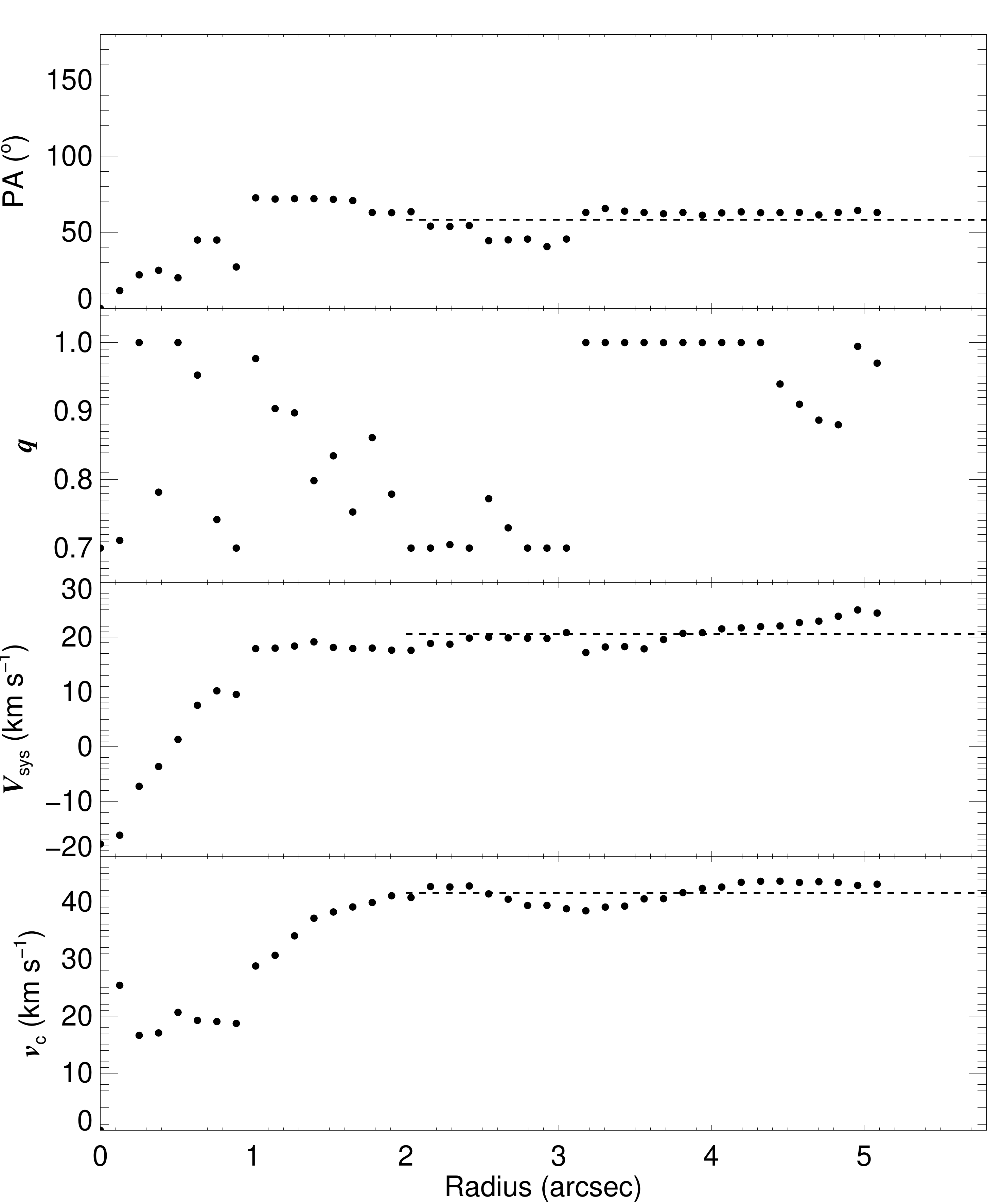}
   \caption{\label{kinemetry}{From the upper to the lower panel: Position angle, axis ratio, systemic velocity (with respect to $z=0.016151$), and circular velocity as functions of radius} for the disc component of NGC~7130, as obtained with \texttt{kinemetry}. The dashed horizontal lines show the values averaged over radii larger than $2^{\prime\prime}$.}
\end{figure}

We first attempted to estimate the orientation of NGC~7130 with \texttt{kinemetry} \citep{Krajnovic2006}, performing a tilted-ring fit of the velocity map of the disc component (Figs.~\ref{gas_kinematics} and \ref{disckin}). To constrain the parameter space, we limited the axis ratio range to $0.7<q<1.0$. We found that \texttt{kinemetry} yields a stable position angle at $\text{PA}=58\pm8\degr$ (obtained from averaging the fitted orientations for annuli with a semi-axis major axis larger than $2^{\prime\prime}$; uncertainty from the standard deviation), but not a stable axis ratio, which does not allow us to estimate the disc inclination (Fig.~\ref{kinemetry}).

To obtain an estimate of the inclination of NGC~7130, we assumed that the plateau in velocities seen in Fig.~\ref{kinemetry} corresponds to the maximum circular velocity. By averaging over radii larger than $2^{\prime\prime}$, we found the projected circular velocity to be $42\pm2\,{\rm km\,s^{-1}}$. Combining this information with the intrinsic luminosity of the galaxy, the Tully-Fisher relation can be used to estimate the inclination.

Because NGC~7130 is rather dusty, we measured its luminosity in the mid-infrared using $3.6\,\upmu{\rm m}$ science-ready images from the {\it Spitzer} Heritage Archive\footnote{\url{https://sha.ipac.caltech.edu/applications/Spitzer/SHA/}} (Programme ID 90031; PI O.~D.~Fox). We calculated the total magnitude of the galaxy using the curve of growth technique, following \citet{MunozMateos2015}. For this, we hand-masked stars and galaxies overlapping NGC~7130, then we measured the curve of growth as the cumulative sum of the flux of the galaxy using circular apertures that increase in size by 2\arcsec steps. The magnitude at infinite aperture is where the slope of the curve of growth reaches exactly zero, hence we calculated the gradient of the curve of growth as a function of radius, then fit a line between this gradient and the enclosed magnitude in the outer regions of the curve \citep[for an example, see Fig.~7 of][]{MunozMateos2015}. The intercept of this line is, by definition, the asymptotic magnitude.

We calculated the uncertainty as both the Poisson error, and the systematic uncertainty due to the local sky subtraction. We estimated the latter by refitting the asymptotic magnitude after subtracting randomised local sky values drawn from a Gaussian distribution with a mean and standard deviation matching those of the original sky measurement. The total magnitude we calculated for this galaxy is $m_{3.6\,\upmu{\rm m}} = 10.82 \pm 0.013$, where the error is the quadrature sum of the aforementioned random (0.009 mag) and systematic (0.01 mag) errors. Additional uncertainty (not included) comes from our flux interpolation across the relatively large masks, and we also included the AGN flux in this measurement, making this magnitude likely an overestimate.

Assuming a luminosity distance modulus $m-M=34.09$ we computed an absolute magnitude $M_{3.6\,\upmu{\rm m}}(\text{AB})=-23.27$. Using the Tully-Fisher relation from Eq.~1 in \citet{Sorce2014}, this corresponds to a circular velocity of rotation of $v_{\rm c}=318\pm32\,{\rm km\,s^{-1}}$, which is probably an upper limit to the real velocity due to the bias introduced by the AGN. The error bars were calculated using the scatter of 0.43\,mag found for the galaxies in \citet{Sorce2014}. If 50\% of the light came from the AGN, the circular velocity would be $v_{\rm c}=265\,{\rm km\,s^{-1}}$. As a sanity check, we estimated $v_{\rm c}$ from the Tully-Fisher relations for Cousins $B$ and $R$ in \citet{Pierce1992} and adopting the luminosity estimates from \citet{Lauberts1989} of $m_B=12.86$ and $m_R=11.57$. We obtain $v_{\rm c}=265\,{\rm km\,s^{-1}}$ and $v_{\rm c}=281\,{\rm km\,s^{-1}}$, respectively.

If we adopt a round number of $v_{\rm c}\approx300\,{\rm km\,s^{-1}}$, this would imply an inclination of $i\approx8\degr$. Even if the AGN were strongly biasing the Tully-Fisher results and the actual velocity were $v_{\rm c}=250\,{\rm km\,s^{-1}}$, the resulting inclination would be $i\approx10\degr$. We can therefore conclude that NGC~7130 is almost face-on.

The galaxy NGC~7130 has an inner pseudo-ring made of clockwise outward-winding spiral fragments encircling the bar \citep[see $HST$ images in][]{EliasRosa2018}. Assuming that the spiral fragments are trailing, the near side of NGC~7130 is the southern one.

\subsection{The systemic velocity of NGC~7130}

\label{systemic}

The third panel in Fig.~\ref{kinemetry} shows the systemic velocity as a function of radius for NGC~7130. The inner arcsecond of the galaxy has a velocity of $\approx-40\,{\rm km\,s^{-1}}$ with respect to the outer regions. Those outer regions have an average systemic velocity that is $21\pm2\,{\rm km\,s^{-1}}$ more than what is reported in the NED ($cz=4842\,{\rm km\,s^{-1}}$). We therefore reevaluate the recession velocity to be $cz=4863\,{\rm km\,s^{-1}}$ with respect to the barycentre of the Solar System, which corresponds to a redshift of $z=0.016221$. This systemic velocity has been used as the zero in velocity in the velocity maps in Figs.~\ref{stellar}, \ref{gas_kinematics}, \ref{disckin}, and \ref{coronal}.

\subsection{Morphology}

\label{morphology}

The inner regions of NGC~7130 host two nuclear rings (Fig.~\ref{images}), the shape of which has been obtained as described in \cite{Comeron2014}. The first one is the tiny elongated UCNR reported by \citet{GonzalezDelgado1998} and then confirmed by \citet{Knapen2019}. We remeasured it using our H$\upalpha$ continuum-subtracted image and determined a semi-major axis of $0\farcs68$ (210\,pc), an axis ratio of $q=0.59$, and a position angle of $\text{PA}=163\degr$. The continuum-subtracted H$\upalpha$ image reveals a larger and rounder ring that is not seen in the UV images in \citet{GonzalezDelgado1998}. This ring has a projected semi-major axis of $3\farcs0$ (930\,pc), an axis ratio of $q=0.77$, and a position angle of $\text{PA}=119\degr$. The largest of the rings is crossed by H$\upalpha$ emission and dusty spiral arms that reach the innermost region where the UCNR is found. Most of the western side of the ring is missing, which makes it hard to constrain its shape.

We characterised the bar with the $3.6\,{\rm \upmu m}$ image that we used to estimate the luminosity of NGC~7130 (Sect.~\ref{orientation}). By running Python's implementation of \texttt{ellipse} \citep{Jedrzejewski1987}, we found that the bar has its smallest axis ratio, $q=0.48$, at a radius of $7\farcs6$ (2.4\,kpc), where the position angle is $\text{PA}=6\degr$, which is representative of the whole bar.

We decided against deprojecting the ring parameters, because NGC~7130 is nearly face-on (Sect.~\ref{orientation}). We found that the difference in position angle between the largest of the rings and the bar is $\left|\Delta\text{PA}\right|=68\degr$. This does not quite match the theoretical picture of rings made of $x_2$ orbits perpendicular to the $x_1$ orbits constituting the backbone of the bar \citep[see e.g.~Sect.~4.2.1 in][]{Knapen1995}. We note that this expectation may not hold in statistical samples of nuclear rings \citep[][although the conclusions are also critically dependent on the reliability of the deprojection parameters of the galaxies]{Comeron2010}.

The outermost nuclear ring in NGC~7130 is exceptionally large: its semi-major axis is only 2.5 times smaller than that of the bar (as measured by the radius where it has its maximum ellipticity). This violates the empirical law in \citet{Comeron2010}: `for barred galaxies, the maximum radius that a nuclear ring can reach is a quarter of the bar radius'. To the best of our knowledge, the only other similar ring is that in ESO~565-11, where the peculiarities have been suggested to be caused by the ring being in an early and fast-evolving stage before it settles into a configuration with a smaller radius \citep{Buta1999}. We note, however, that the other peculiarity of the nuclear ring of ESO~565-11, its large ellipticity, is not shared by that in NGC~7130.

\subsection{The {[}O\,{\sc iii}{]}\,$\uplambda$5007 surface brightness of the kinematical components}

\label{intensity}

\begin{figure}[!h]
   \centering
   \includegraphics[scale=0.30]{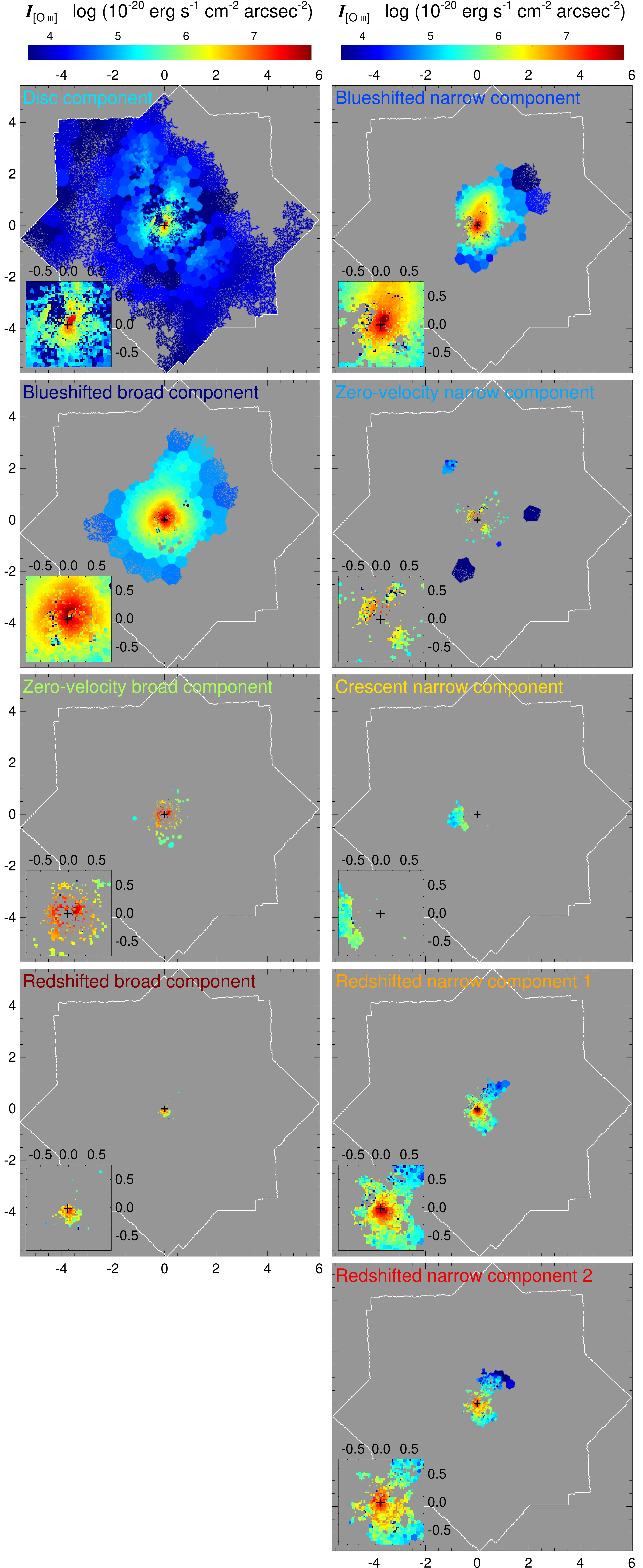}
   \caption{\label{OIII}{Surface brightness of {[}O\,{\sc iii}{]}\,$\uplambda$5007 emission of the nine kinematically identified components.}}
\end{figure}

Figure~\ref{OIII} shows the {[}O\,{\sc iii}{]}\,$\uplambda$5007 surface brightness maps for each of the nine kinematic components. We display {[}O\,{\sc iii}{]} because it is a tracer of AGN ionisation.

The disc component map is very noisy. This is because the {[}O\,{\sc iii}{]}\,$\uplambda$5007 disc emission is, in many regions, sub-dominant compared to that from the outflow as seen, for example, in the spectra in Figs.~\ref{fit91}, \ref{fit443}, \ref{fit2306}, and \ref{fit2221}. Hence, it is poorly constrained by the fits. H$\upalpha$ images of the disc component (not shown) display many details of the spiral arms and the star formation knots in the UCNR.

The {[}O\,{\sc iii}{]} emission of the blueshifted broad component is rather featureless. Its distribution has a nearly circular symmetry in the inner arcsecond (although the maximum emission is not at the galaxy centre, but $0\farcs1$ or 30\,pc to its north). At larger radii, the emission is enhanced in the region north-west of the nucleus where the narrow blueshifted component emission is also enhanced.

The blueshifted narrow component emission is less smooth than that of its broad counterpart. Both share the location of the maximum in emission. The blueshifted narrow component exhibits a well-collimated feature running to the north of the nucleus. This feature coincides with the regions where the component has its largest negative velocities (Sect.~\ref{kinematics} and Fig.~\ref{gas_kinematics}) and is also seen in the continuum-subtracted {[}O\,{\sc iii}{]}\,$\uplambda$5007 image (Fig.~\ref{images}).

In Fig.~\ref{OIII}, we see further evidence for the separation between the crescent narrow component and the redshifted narrow component~1. Indeed, whereas the redshifted narrow component~1 is centrally concentrated and fades as we move away from the central regions, we see that the crescent narrow component has a sharp border on its western side, where it is the closest to redshifted narrow component~1.

The redshifted narrow components~1 and 2 and the redshifted broad component have their peak in emissions slightly less than $0\farcs1$ south of the nucleus (roughly symmetric with the peak in the blueshifted components). The maxima in emission of the blue- and redshifted components coincide with the bipolar structure revealed by the coronal gas kinematics (Fig.~\ref{coronal}). The redshifted narrow components~1 and 2 were fitted both south and north of the nucleus. We see, however, that the extension of those components to the north-west has a very low surface brightness. Therefore, most of the light in redshifted components comes from regions south of the nucleus (as opposed to the blueshifted light, especially for the narrow component, that comes mostly from north of the nucleus).

\subsection{The radio jet}

\begin{figure}
   \centering
   \includegraphics[scale=0.30]{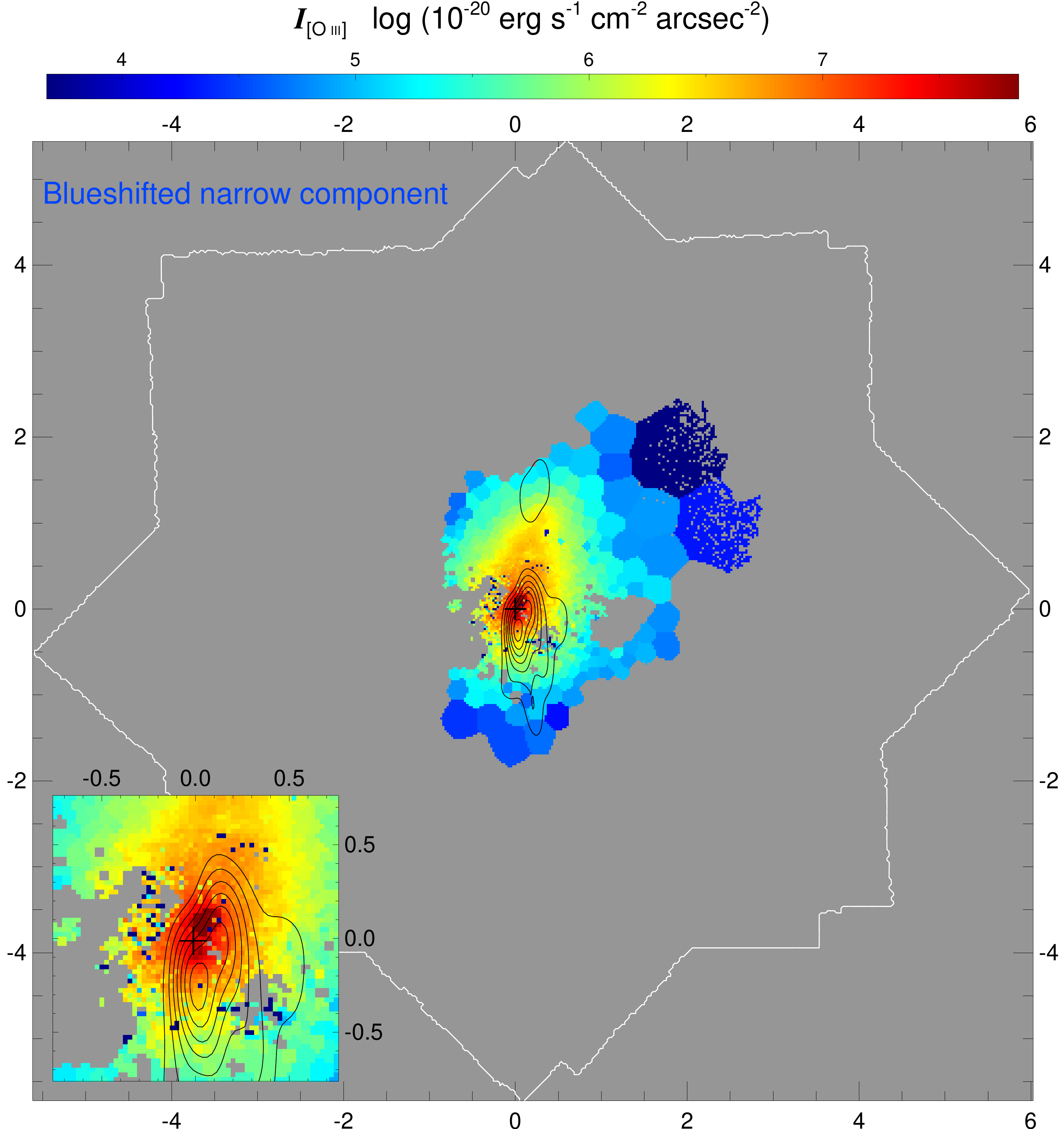}
   \caption{\label{vla}{{[}O\,{\sc iii}{]}\,$\uplambda$5007 surface brightness of blueshifted narrow component (same map as in the corresponding panel in Fig.~\ref{OIII}). The black line overlay shows the VLA 8.4\,GHz continuum data obtained for \citet{Thean2000} as processed by \citet{Zhao2016}. The contour levels cover a range from 0.001 to 0.007\,mJy/beam in steps of 0.001\,mJy/beam. The angular resolution in radio is $0\farcs60\times0\farcs19$ and the position angle of the point spread function ellipse is $\sim10\degr$ \citep{Zhao2016}.}}
\end{figure}

A radio jet as traced by 8.4\,GHz continuum emission was observed with the VLA by \citet{Thean2000}. Y.~Zhao kindly provided us with the re-reduced data presented in \citet{Zhao2016}. The radio emission south of the nucleus roughly coincides with the position of the redshifted narrow components~1 and 2. The detached blob of radio emission north of the nucleus is, strikingly, located just north of the collimated blueshifted feature seen in {[}O\,{\sc iii}{]} (Fig.~\ref{vla}).

\subsection{Resolved BPT diagrams: The ionisation mechanisms of the kinematical components}

\label{bpt}

To study the ionisation mechanisms of the kinematic components, we used three different types of the BPT line diagnostics \citep{Baldwin1981} and produced resolved BPT maps. The main ionisation mechanisms of the components are listed in Table~\ref{physics}.

A caveat of the diagnostics is that they rely on the H$\upalpha$ line. Unfortunately, H$\upalpha$ is often blended with {[}N\,{\sc ii}{]}, which makes it very hard to constrain. For example, in many bins the H$\upalpha$ amplitude of outflow components (in particular for the redshifted ones) is fitted as zero, which results in missing bins in the BPT maps. Even when a non-zero H$\upalpha$ amplitude is fitted, its value for components with a low relative H$\upalpha$ surface brightness is prone to large uncertainties (Table~\ref{uncertainties}), which increases the scatter in the horizontal axis in the line diagnostics shown in Figs.~\ref{bpts1} and \ref{bpts2}.

\begin{table*}
   \setlength{\tabcolsep}{5.5pt}
  \caption{Average physical properties of kinematic components of the ionised gas.}
 \label{physics}
 \centering
 \begin{tabular}{l c c c c c c c}
 \hline\hline  Component& BPT & $n_e$ &$E(B-V)$ & $\text{{[}S\,{\sc iii}{]}}/\text{{[}S\,{\sc ii}{]}}$ & $L\left({\rm H}\upbeta\right)_{\rm corr}$&$\dot{M}$ & $\dot{E}_{\rm kin}$\\
  & classification& $\left(\text{cm}^{-3}\right)$& & &$\left(10^{39}\,\text{erg\,s}^{-1}\right)$ &$\left(M_{\odot}\,\text{yr}^{-1}\right)$ &$\left(10^{39}\,\text{erg\,s}^{-1}\right)$\\
 \hline
 Disc & SF                                    &$90\pm20$       &$0.60\pm0.01$     &$0.385\pm0.008$   &$235\pm5$     & --           & --          \\
 \hline
 Blueshifted narrow                  & Seyfert&$500\pm100$     &$0.34\pm0.04$     &$1.20\pm0.08$     &$27\pm1.9$      & $0.05\pm0.03$& $1.2\pm1.1$ \\
 Zero-velocity narrow& SF + AGN               &$500\pm400$     &$0.53\pm0.11$     &$0.68\pm0.15$     &$15\pm3$      & --           & --          \\
 Crescent narrow& LINER                       &$180\pm20$      &$0.78\pm0.01$     &$0.134\pm0.003$   &$3.9\pm0.1$   & $0.09\pm0.04$& $3.0\pm2.1$ \\
 Redshifted narrow 1 & Seyfert                &$500\pm100^\ast$&$0.56\pm0.13$     &$2.7\pm0.7$       &$8\pm2$      & $0.09\pm0.05$& $2.9\pm3.0$ \\
 Redshifted narrow 2 & Seyfert                &$500\pm100^\ast$&$0.57\pm0.11$     &$0.9\pm0.3$       &$6.2\pm1.4$   & $0.16\pm0.10$& $25\pm19$   \\ 
 \hline
 Blueshifted broad & Seyfert                  &$300\pm100$     &$0.46\pm0.03$     &$1.27\pm0.07$     &$57\pm3$      & $0.39\pm0.24$& $69\pm47$   \\
 Zero-velocity broad& Seyfert                 &$300\pm100^\ast$&$0.80\pm0.28$     &$0.9\pm0.5$       &$23\pm13$     & --           & --          \\
 Redshifted broad & Seyfert                   &$300\pm100^\ast$&$0.57\pm0.11^\ast$&$1.27\pm0.07^\ast$&$1.8\pm0.4$   & $0.41\pm0.26$& $170\pm130$ \\
 \hline
 \end{tabular}
  \tablefoot{Error estimates are calculated by propagating the uncertainties quoted in Table~\ref{uncertainties} and Sects.~\ref{secSIII} and \ref{estimates} in quadrature. Values indicated by asterisks ($^*$) do not come from measurements, but are instead educated guesses used for our calculations (see Sects.~\ref{secSIII} and \ref{estimates}).}
\end{table*}

\subsubsection{The $\text{{[}O\,{\sc iii}{]}\,$\uplambda$5007}/\text{H$\upbeta$}$ versus $\text{{[}N\,{\sc ii}{]}\,$\uplambda$6563}/\text{H$\upalpha$}$ line diagnostic}

\begin{figure*}
   \centering
   \includegraphics[scale=0.30]{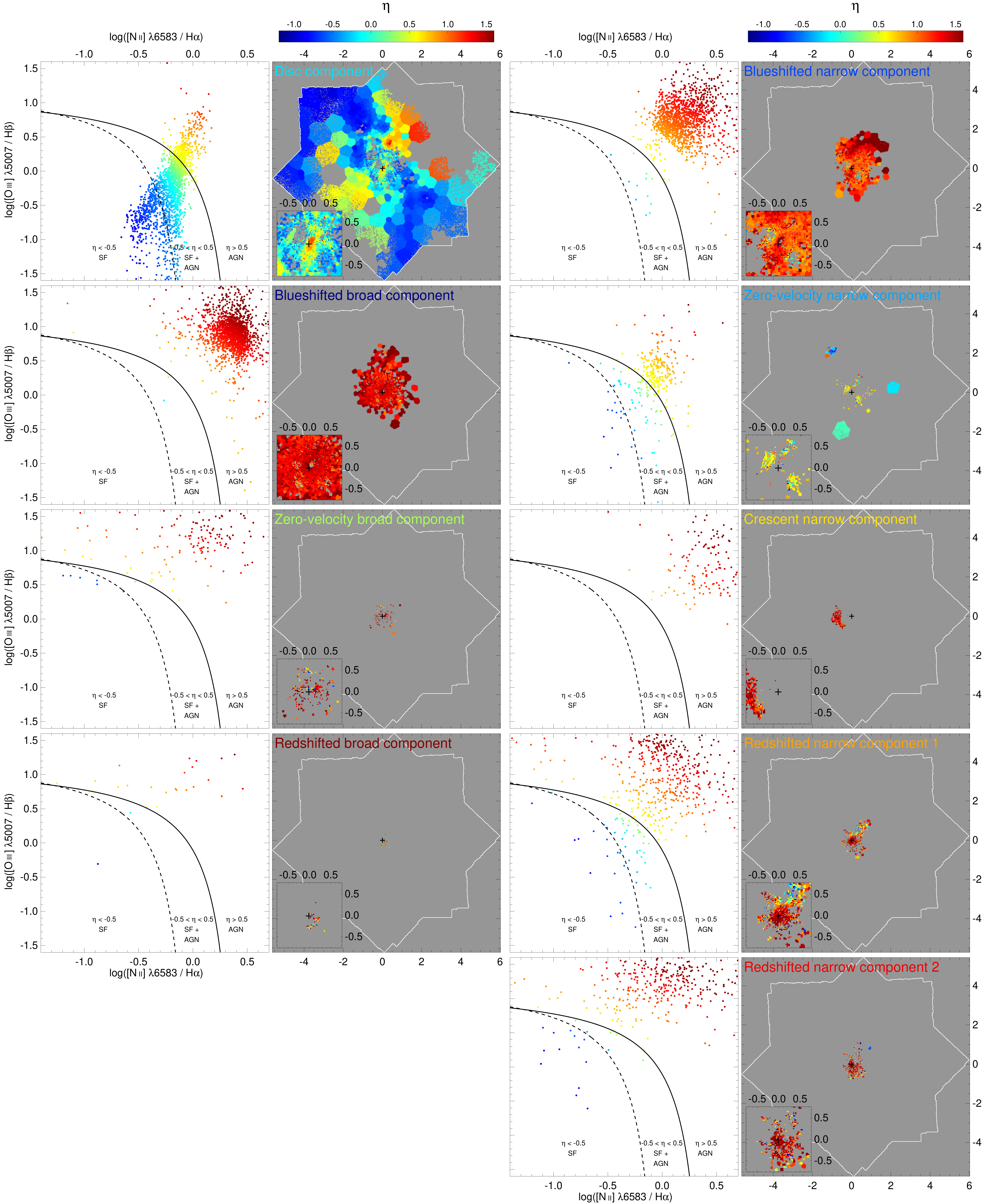}
   \caption{\label{bpts1}{{Nine sets of two panels are shown, one for each of the kinematic components. For each set, the {\it \emph{left-hand panel}} corresponds to the $\text{{[}O\,{\sc iii}{]}\,$\uplambda$5007}/\text{H$\upbeta$}$ versus $\text{{[}N\,{\sc ii}{]}\,$\uplambda$6563}/\text{H$\upalpha$}$ BPT diagram used to build the resolved BPT map in the {\it right-hand panel}. The colour coding indicates the $\eta$ parameter defined in \citet{ErrozFerrer2019}. The lines delimiting the regions in the BPT diagram are from \citet[][solid line]{Kewley2001} and \citet[][dashed line]{Kauffmann2003}.}}}
\end{figure*}

We first discuss the BPT diagnostic that uses the $\text{{[}O\,{\sc iii}{]}\,$\uplambda$5007}/\text{H$\upbeta$}$ and the $\text{{[}N\,{\sc ii}{]}\,$\uplambda$6563}/\text{H$\upalpha$}$ line ratios (Fig.~\ref{bpts1}). We follow the division of the $\text{{[}O\,{\sc iii}{]}\,$\uplambda$5007}/\text{H$\upbeta$}-\text{{[}N\,{\sc ii}{]}\,$\uplambda$6563}/\text{H$\upalpha$}$ plane by \citet{Kewley2006}, where areas below the dashed line \citep[derived semi-empirically by][]{Kauffmann2003} are considered to be ionised by star formation, and areas above the solid line \citep[derived theoretically by][]{Kewley2001} are considered to be ionised by the AGN. The region between the two lines is considered to be ionised by a mix of the two. A quantification of the strength of each mechanism is obtained through the parameter $\eta$ defined by \citet{ErrozFerrer2019}, so $\eta=-0.5$ at the transition between the star-forming and the mixed regimes, and $\eta=0.5$ at the limit between the mixed and the AGN regimes. In the star-forming regime, it is defined to be $\eta<0.5$ and measures the orthogonal distance to the \citet{Kauffmann2003} line. In the AGN regime, $\eta$ is defined to be $\eta>0.5$ and measures the orthogonal distance to the \citet{Kewley2001} line. In the mixed region, it measures the orthogonal distance to the bisector of the two above-mentioned lines, where $\eta=0$. The concrete formulae to compute $\eta$ are detailed in \citet{ErrozFerrer2019}. As indicated in Sect.~\ref{kinematics}, the parameter $\eta$ was used to decide whether a fitted component corresponded to the blueshifted narrow component ($\eta>0.8$) or the zero-velocity narrow component ($\eta<0.8$). In some bins, both components were present. There, we assigned the component with the largest negative velocity to the blueshifted narrow component, irrespective of $\eta$.

We show the BPT maps for the kinematic components in Fig.~\ref{bpts1}. As in \citet{Knapen2019}, the colour-coding traces $\eta$, with blue denoting star formation and red indicating AGN ionisation.

The disc is mostly ionised by star formation. In the south-east to north-west axis, there are regions with $\eta>-0.5$, which might indicate the effect of the AGN. We cannot, however, discard the effects of some confusion, especially between the disc and the blueshifted narrow components. Indeed, one of the disc regions with the largest $\eta$ corresponds to the position of the collimated feature seen in {[}O\,{\sc iii}{]} for the blueshifted narrow component (Sect.~\ref{intensity}).

The zero-velocity narrow component is also partially ionised by star formation (due partly to how we defined the component). The fraction of a spiral arm delineated by this component (Sect.~\ref{kinematics}) has a very low $\eta$, mostly compatible with a pure star-forming ionisation. The innermost clumps corresponding to the UCNR are probably contaminated by the blueshifted narrow component, which explains why $\eta>-0.5$. The remaining components have been ionised by the AGN, as indicated by their orange and red hues in Fig.~\ref{bpts1}.

\subsubsection{The $\text{{[}O\,{\sc iii}{]}\,$\uplambda$5007}/\text{H$\upbeta$}$ versus $(\text{{[}S\,{\sc ii}{]}\,$\uplambda$6716}+\text{{[}S\,{\sc ii}{]}\,$\uplambda$6731})/\text{H$\upalpha$}$ line diagnostic}

\begin{figure*}[!h]
   \centering
   \includegraphics[scale=0.30]{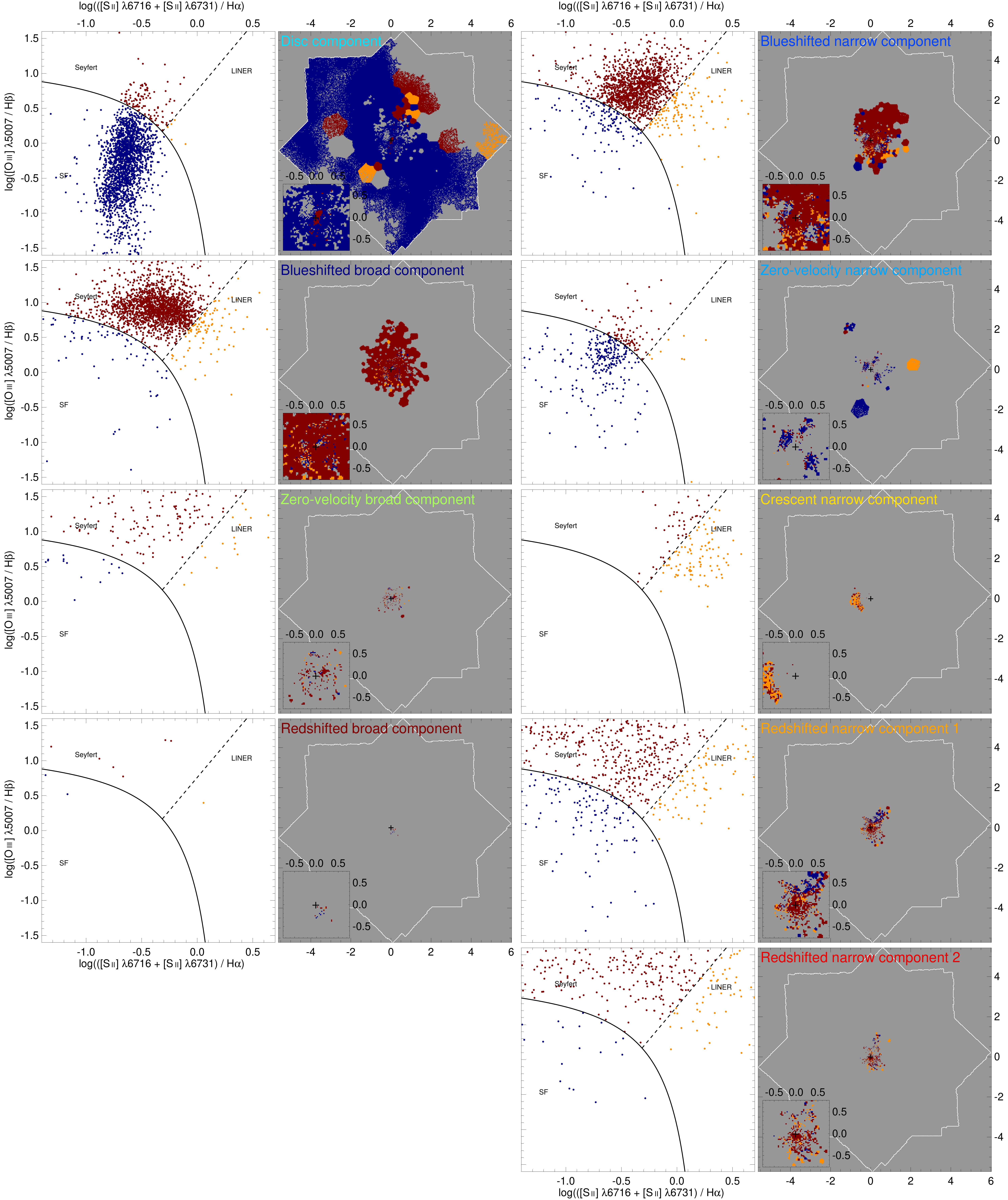}
   \caption{\label{bpts2}{Nine sets of two panels are shown, one for each of the kinematic components. For each set, the {\it \emph{left-hand panel}} corresponds to the $\text{{[}O\,{\sc iii}{]}\,$\uplambda$5007}/\text{H$\upbeta$}$ versus $(\text{{[}S\,{\sc ii}{]}\,$\uplambda$6716}+\text{{[}S\,{\sc ii}{]}\,$\uplambda$6731})/\text{H$\upalpha$}$ BPT diagram used to build the resolved BPT map in the {\it \emph{right-hand panel}}. The lines delimiting the regions in the BPT diagram are from \citet[][]{Kewley2006}.}}
\end{figure*}

For this line diagnostic, we used the criteria of \citet{Kewley2006}. The results are displayed in Fig.~\ref{bpts2} and are very similar to those found with the above BPT diagnostic (Fig.~\ref{bpts1}), namely that the disc and the zero-velocity narrow components are ionised by star formation, whereas the other components are ionised by the AGN. We find that the crescent narrow component straddles the Seyfert and LINER regimes, albeit weighted more strongly towards the LINER side.

We also studied the $\text{{[}O\,{\sc iii}{]}\,$\uplambda$5007}/\text{H$\upbeta$}$ versus $\text{{[}O\,{\sc i}{]}\,$\uplambda$6300}/\text{H$\upalpha$}$ line diagnostic. Because the results are nearly identical to those in Fig.~\ref{bpts2}, we do not show them here.

\subsection{The electron density, the reddening, and the $\text{{[}S\,{\sc iii}{]}}/\text{{[}S\,{\sc ii}{]}}$ line ratio}

\label{secSIII}

We estimated the mass outflow rates and kinetic powers (Sect.~\ref{estimates}). To calculate these values the electron density, $n_e$, and the reddening, $E(B-V)$, of the components must be measured. In our discussion (Sect.~\ref{discussion}), we use the $\text{{[}S\,{\sc iii}{]}}/\text{{[}S\,{\sc ii}{]}}$ line ratio as a proxy for the ionisation parameter \citep[e.g.][]{Diaz2000}. We attempted to produce maps of these three magnitudes for the nine kinematic components. Unfortunately, these maps were very noisy, so we opted to present a single integrated value for each component obtained from the total H$\upalpha$ and H$\upbeta$ fluxes (for the reddening), the total {[}S\,{\sc ii}{]}\,$\uplambda$6716 and {[}S\,{\sc ii}{]}\,$\uplambda$6731 fluxes (for the electron density), and the total {[}S\,{\sc ii}{]} and {[}S\,{\sc iii}{]} fluxes (for the ionisation parameter). The uncertainties in the integrated line fluxes are adopted to be those in Table~\ref{uncertainties}. Our estimated $n_e$, $E(B-V)$, and $\text{{[}S\,{\sc iii}{]}}/\text{{[}S\,{\sc ii}{]}}$ values are presented in Table~\ref{physics}.

We estimated $n_e$ using the {[}S\,{\sc ii}{]}\,$\uplambda$6716/{[}S\,{\sc ii}{]}\,$\uplambda$6731 ratio and the parametrisation from \citet{Sanders2016}. Unfortunately, the redshifted narrow and broad components are barely detected for these low-amplitude lines (see example spectra in Appendix~\ref{examples}). We therefore assumed that they have the same $n_e$ as their blueshifted counterparts.

Extinctions were estimated using the Balmer decrement and assuming a \citet{Calzetti2000} extinction law. Unfortunately, H$\upalpha$ is often blended with the {[}N\,{\sc ii}{]} lines, which makes it very hard to measure its flux, especially for the components with the lowest surface brightness. Thus, for the redshifted broad component, we assumed the extinction to be the same as for the redshifted narrow component~2, with which it overlaps in projection. As expected, the outflow components seen through the disc (the redshifted ones) are more extincted than the ones in front of the disc.

When estimating the $\text{{[}S\,{\sc iii}{]}}/\text{{[}S\,{\sc ii}{]}}$ line ratio, the flux in {[}S\,{\sc iii}{]} is defined to be the sum of the $\uplambda=9069$\,\AA\ and the $\uplambda=9532$\,\AA\ lines, whereas that in {[}S\,{\sc ii}{]} is that of the sum of the $\uplambda=6717$\,\AA\ and the $\uplambda=6731$\,\AA\ lines. Since MUSE does not cover $\text{{[}S\,{\sc iii}{]}}\,\uplambda9532$, we followed \citet{Mingozzi2019} and assumed $\text{{[}S\,{\sc iii}{]}}\,\uplambda9532/\text{{[}S\,{\sc iii}{]}}\,\uplambda9069=2.5$ as theoretically determined by \citet{Vilchez1996}. This value is similar to the line ratios found by \citet{RamosAlmeida2009} for a sample of five Seyfert galaxies (four of their galaxies have ratios in the 2.2--2.7 range). Because of the large separation in wavelength between the lines, we corrected their fluxes for extinction. We were not able to detect the redshifted broad component in {[}S\,{\sc iii}{]}, so we assume that its $\text{{[}S\,{\sc iii}{]}}/\text{{[}S\,{\sc ii}{]}}$ line ratio is similar to that of its blueshifted counterpart.

In the outflow components, the ionisation parameter is usually larger than in the disc. We find that the crescent narrow component has an extremely low ionisation parameter when compared to the other outflow components ($\text{{[}S\,{\sc iii}{]}}/\text{{[}S\,{\sc ii}{]}}\approx0.1$ versus at least 0.9).

\subsection{The mass outflow rate and the kinetic power}

\label{estimates}

Here, we followed the procedure from \citet{Rose2018} to calculate the mass outflow rate and the kinetic power of the outflow. We first estimated the mass in each of the kinematic components:
\begin{equation}
\label{mass}
 M=\frac{L({\rm H}\upbeta)_{\rm corr}m_p}{\alpha_{{\rm H}\upbeta}^{\rm eff}h\nu_{{\rm H}\upbeta}n_e},
\end{equation}
where $L({\rm H}\upbeta)$ is the H$\upbeta$ luminosity corrected for extinction (Table~\ref{physics}), $\alpha_{{\rm H}\upbeta}$ is the effective Case~B coefficient \citep[we adopted $3.03\times10^{-14}\,{\rm cm^3\,s^{-1}}$, which corresponds to an electron temperature $T_e=10^4$\,K;][]{Osterbrock2006}, $m_p$ is the mass of the proton, and $h\nu_{{\rm H}\upbeta}$ is the energy of an H$\upbeta$ photon. To obtain the mass-loss rate, we needed the timescale of the outflow, which can be estimated as
\begin{equation}
\label{tau}
 \tau_{\rm o}\sim\frac{R}{V_{\rm o}}
,\end{equation}
where $R$ and $V_{\rm o}$ are the size and the velocity of the outflow, respectively. For the error propagation, we estimated the uncertainty in $\tau_{\rm o}$ to be of 50\%.

If we divide Eq.~\ref{mass} by Eq.~\ref{tau}, we obtain the mass loss rate
\begin{equation}
 \label{mdot}
 \dot{M}=\frac{L({\rm H}\upbeta)_{\rm corr}m_pV_{\rm o}}{\alpha_{{\rm H}\upbeta}^{\rm eff}h\nu_{{\rm H}\upbeta}n_eR}.
\end{equation}
For $V_{\rm o}$, we adopted the velocity of each kinematic component averaged over all bins. The values of $R$ were obtained from the size of the components in Fig.~\ref{locations}. Both are listed in Table~\ref{adopted}.

We estimated the kinetic power by accounting for both the net velocity of the outflow, $V_{\rm o}$, and its velocity dispersion, $\sigma_{\rm o}$:
\begin{equation}
 \dot{E}_{\rm kin}=\frac{\dot{M}}{2}\left(V_{\rm o}^2+3\sigma_{\rm o}^2\right).
\end{equation}
For $\sigma_{\rm o}$, we adopted the velocity dispersion of each kinematic component averaged over all bins. To gauge the uncertainties in the adopted $V_{\rm o}$ and $\sigma_{\rm o}$, we estimated the inhomogeneities in each outflow component by finding the dispersion of the velocity and the velocity dispersion, $\varsigma(V)$ and $\varsigma(\sigma)$, over a kinematic component (Table~\ref{adopted}).

\begin{table}
 \caption{Values adopted as representative of the kinematics and the size of the kinematic components.}
 \label{adopted}
 \centering
 \begin{tabular}{l c c c c c}
 \hline\hline  Component& $\overline{\left|V\right|}$& $\varsigma(V)$& $\overline{\sigma}$&$\varsigma(\sigma)$& $R$\\
  &\multicolumn{2}{c}{$\left({\rm km\,s^{-1}}\right)$}&\multicolumn{2}{c}{$\left({\rm km\,s^{-1}}\right)$}&$\left(\text{pc}\right)$\\
 \hline
 Blueshifted narrow & 113 &  69 & 145 &  61 & 930\\
 Crescent narrow    & 174 & 112 & 163 &  40 & 310\\
 Redshifted narrow 1& 215 &  54 & 141 & 104 & 310\\
 Redshifted narrow 2& 559 & 148 & 235 & 123 & 310\\ 
 \hline
 Blueshifted nroad  & 266 & 136 & 401 & 65  & 930\\
 Redshifted nroad   & 918 & 231 & 391 & 128 &  93\\
 \hline
 \end{tabular}
 \tablefoot{$\varsigma$ denotes the standard deviation.}
\end{table}

The mass-loss rates and kinetic power estimates for each of the outflow components are shown in Table~\ref{physics}. We find a total mass loss rate of $\dot{M}=0.44\pm0.27\,M_{\odot}\,{\rm yr^{-1}}$ and a total kinetic power of $\dot{E}_{\rm kin}=\left(7.0\pm4.8\right)\times10^{40}\,{\rm erg\,s^{-1}}$ for the blueshifted components. If we also account for the more poorly constrained redshifted components, these values are $\dot{M}=1.2\pm0.7\,M_{\odot}\,{\rm yr^{-1}}$ and $\dot{E}_{\rm kin}=\left(2.7\pm2.0\right)\times10^{41}\,{\rm erg\,s^{-1}}$, respectively. The latter values have to be taken with caution because a third of the mass-loss rate and two thirds of the kinetic power come from the hard-to-characterise redshifted broad component. Indeed, we might only be detecting it over a small fraction of its true extent, which can cause a large overestimate in its associated mass-loss rate and kinetic power (through Eq.~\ref{mdot}). Assuming a bolometric AGN luminosity of $L_{\rm bol}=2.2\times10^{44}\,{\rm erg\,s^{-1}}$ \citep[][we corrected the luminosity for the different distance estimates]{Esquej2014}, the fraction of power emitted in kinetic energy is $F_{\rm kin}=0.032\pm0.022\%$ for the blueshifted components and $F_{\rm kin}=0.12\pm0.09\%$ if accounting for all components.

\section{Discussion}

\label{discussion}

\begin{figure*}
\begin{center}
\begin{tikzpicture}[]

\definecolor{disc}{RGB}{0,227,255};
\definecolor{broadblue}{RGB}{0,0,131};
\definecolor{narrowblue}{RGB}{0,67,255};
\definecolor{narrowred1}{RGB}{255,163,0};
\definecolor{narrowred2}{RGB}{237,0,0};
\definecolor{crescent}{RGB}{255,233,0};
\definecolor{zeronarrow}{RGB}{0,167,255};
\definecolor{broadred}{RGB}{131,0,0};

\pgfmathsetmacro{\ls}{4};
\pgfmathsetmacro{\angle}{20};
\pgfmathsetmacro{\xs}{\ls*cos(\angle)};
\pgfmathsetmacro{\ys}{\ls*sin(\angle)};

\pgfmathsetmacro{\lbn}{5.6};
\pgfmathsetmacro{\anglelbn}{\angle-10};
\pgfmathsetmacro{\anglehbn}{\angle+10};
\pgfmathsetmacro{\xlbn}{\lbn*cos(\anglelbn)};
\pgfmathsetmacro{\xhbn}{\lbn*cos(\anglehbn)};
\pgfmathsetmacro{\ylbn}{\lbn*sin(\anglelbn)};
\pgfmathsetmacro{\yhbn}{\lbn*sin(\anglehbn)};
\pgfmathsetmacro{\xtbn}{0.95*\lbn*cos(\angle)};
\pgfmathsetmacro{\ytbn}{0.95*\lbn*sin(\angle)};

\pgfmathsetmacro{\lbb}{6};
\pgfmathsetmacro{\anglelbb}{\angle-35};
\pgfmathsetmacro{\anglehbb}{\angle+35};
\pgfmathsetmacro{\xlbb}{\lbn*cos(\anglelbb)};
\pgfmathsetmacro{\xhbb}{\lbn*cos(\anglehbb)};
\pgfmathsetmacro{\ylbb}{\lbn*sin(\anglelbb)};
\pgfmathsetmacro{\yhbb}{\lbn*sin(\anglehbb)};
\pgfmathsetmacro{\anglelbbt}{\angle-35+180};
\pgfmathsetmacro{\anglehbbt}{\angle+35+180};
\pgfmathsetmacro{\xlbbt}{\lbn*cos(\anglelbbt)};
\pgfmathsetmacro{\xhbbt}{\lbn*cos(\anglehbbt)};
\pgfmathsetmacro{\ylbbt}{\lbn*sin(\anglelbbt)};
\pgfmathsetmacro{\yhbbt}{\lbn*sin(\anglehbbt)};
\pgfmathsetmacro{\xhbbts}{\lbn*cos(\anglehbbt)/10};
\pgfmathsetmacro{\yhbbts}{\lbn*sin(\anglehbbt)/10};
\pgfmathsetmacro{\xtbb}{0.1+0.75*\lbn*cos((\anglehbb+\anglehbn)/2)};
\pgfmathsetmacro{\ytbb}{0.75*\lbn*sin((\anglehbb+\anglehbn)/2)};
\pgfmathsetmacro{\xtrn}{-1.01*\lbn*cos(\angle)};
\pgfmathsetmacro{\ytrn}{-1.01*\lbn*sin(\angle)};

\pgfmathsetmacro{\lrno}{3};
\pgfmathsetmacro{\anglelrno}{\angle+180-10};
\pgfmathsetmacro{\anglehrno}{\angle+180+10};
\pgfmathsetmacro{\xlrno}{\lrno*cos(\anglelrno)};
\pgfmathsetmacro{\xhrno}{\lrno*cos(\anglehrno)};
\pgfmathsetmacro{\ylrno}{\lrno*sin(\anglelrno)};
\pgfmathsetmacro{\yhrno}{\lrno*sin(\anglehrno)};

\pgfmathsetmacro{\lrnt}{3};
\pgfmathsetmacro{\anglelrnt}{\angle+180-9};
\pgfmathsetmacro{\anglehrnt}{\angle+180+9};
\pgfmathsetmacro{\xlrnt}{\lrno*cos(\anglelrnt)};
\pgfmathsetmacro{\xhrnt}{\lrno*cos(\anglehrnt)};
\pgfmathsetmacro{\ylrnt}{\lrno*sin(\anglelrnt)};
\pgfmathsetmacro{\yhrnt}{\lrno*sin(\anglehrnt)};

\pgfmathsetmacro{\lc}{2};
\pgfmathsetmacro{\xc}{\lc*cos(\anglehrnt+8)};
\pgfmathsetmacro{\yc}{\lc*sin(\anglehrnt+8)};
\pgfmathsetmacro{\ae}{0.5}
\pgfmathsetmacro{\be}{0.25}

\pgfmathsetmacro{\eyey}{\yhbb*1.15}

\draw [line width=1mm,disc] (-8,0) -- (8,0);
\fill [color=zeronarrow] (-1,0) ellipse (0.3 and 0.1);
\fill [color=zeronarrow] (2.5,0) ellipse (0.45 and 0.1);
\fill [color=zeronarrow] (-3,0) ellipse (0.6 and 0.1);
\draw [thick,decoration={amplitude=4.pt,segment length=5.25pt, aspect=0.4,ncoil},decorate] (0,0) -- (\xs,\ys);
\draw [thick,decoration={amplitude=4.pt,segment length=5.25pt, aspect=0.4,ncoil},decorate] (0,0) -- (-\xs,-\ys);
\draw [line width=0.5mm,narrowblue] (0,0) -- (\xlbn,\ylbn);
\draw [line width=0.5mm,narrowblue] (0,0) -- (\xhbn,\yhbn);
\draw [line width=0.5mm,dashed,broadred] (0,0) -- (\xlbb,\ylbb);
\draw [line width=0.5mm,broadblue] (0,0) -- (\xhbb,\yhbb);
\draw [line width=0.5mm,broadblue] (0,0) -- (\xlbbt,\ylbbt);
\draw [line width=0.5mm,dashed,broadred] (\xhbbts,\yhbbts) -- (\xhbbt,\yhbbt);
\draw [line width=0.5mm,broadred] (0,0) -- (\xhbbts,\yhbbts);
\draw [line width=0.5mm,narrowred2] (0,0) -- (\xlrnt,\ylrnt);
\draw [line width=0.5mm,narrowred2] (0,0) -- (\xhrnt,\yhrnt);
\draw [line width=0.5mm,dashed,color=narrowred2] (\xlrnt,\ylrnt) -- (2*\xlrnt,2*\ylrnt);
\draw [line width=0.5mm,dashed,narrowred2] (\xhrnt,\yhrnt) -- (2*\xhrnt,2*\yhrnt);
\draw [line width=0.5mm,narrowred1] (0,0) -- (\xlrno,\ylrno);
\draw [line width=0.5mm,narrowred1] (0,0) -- (\xhrno,\yhrno);
\draw [line width=0.5mm,dashed,color=narrowred1] (\xlrno,\ylrno) -- (2*\xlrno,2*\ylrno);
\draw [line width=0.5mm,dashed,color=narrowred1] (\xhrno,\yhrno) -- (2*\xhrno,2*\yhrno);
\draw [line width=0.5mm,color=crescent,rotate around={\anglelrno+8:(\xc,\yc)}] (\xc,\yc) ellipse ({\ae} and {\be});
\draw [->] (\xc,\yc-0.4) -- (\xc,\yc-2.3);
\draw [->] (\xs/2+0.5,\ys/2) -- (\xs/2+3.2,\ys/2);
\draw [->] (-5,0.2) -- (-5,1.7);
\draw [->] (-3,0.2) -- (-3,1.7);
\node [align=center] at (\xtbn,\ytbn){Blueshifted\\narrow\\component};
\node [align=center] at (\xtbb,\ytbb){Blueshifted\\ broad\\component};
\node [align=center] at (\xtrn,\ytrn){Redshifted\\ narrow\\ components};
\node [align=center] at (\xc,\yc-3.0){Crescent\\ narrow\\ component};
\node [align=center] at (\xs/2+4,\ys/2){Radio jet};
\node [align=center] at (-5,2.0){Disc};
\node [align=center] at (-3,2.95){Star-forming\\regions\\causing the\\zero-velocity\\narrow\\component};
\node [align=center] at (0.2,-0.6){Torus};
\eye[radius=0.5,x=0,y=\eyey,rotation=-90]
\filldraw [line width=0.2mm,fill=white] (3,-3) -- (3,-3.3) -- (5,-3) -- cycle;
\filldraw [line width=0.2mm,fill=gray] (3,-3) -- (3,-2.7) -- (5,-3) -- cycle;
\filldraw [line width=0.2mm,fill=gray] (3,-3) -- (3,-3.3) -- (1,-3) -- cycle;
\filldraw [line width=0.2mm,fill=white] (3,-3) -- (3,-2.7) -- (1,-3) -- cycle;
\node [align=center] at (5.2,-3){\large{N}};
\node [align=center] at (0.8,-3){\large{S}};
\fill[rounded corners,black,rotate around={\angle+90:(0,0)}] (-0.4,-0.1) rectangle (0.4,0.1);

\end{tikzpicture}
\end{center}
\caption{\label{toy} Visual representation of our toy model for the ionised circumnuclear gas in NGC~7130. The colours of the lines follow the general colour scheme used throughout this article. Dashed lines indicate features that we hypothesise are obscured by disc dust. The torus is not drawn to scale. The observer's point of view is indicated by the eye graphic at the top of the image. North is to the right of the cartoon and south to the left.}
\end{figure*}
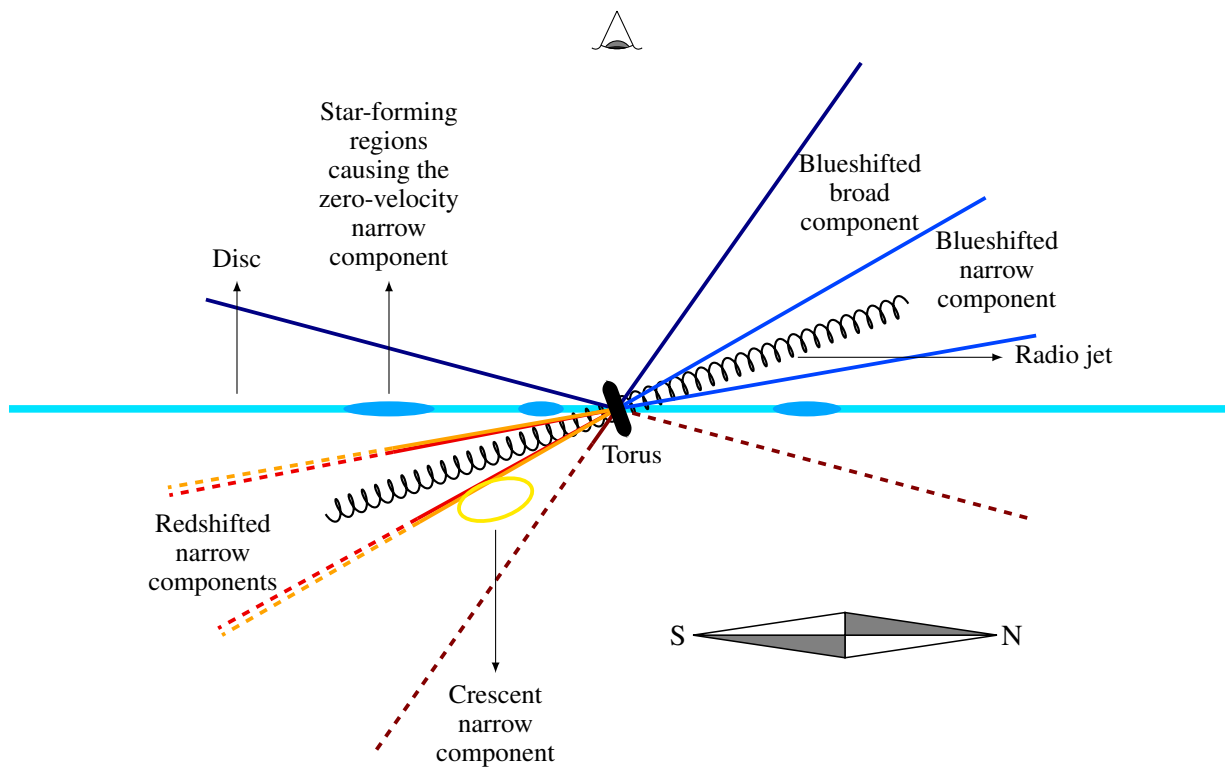

In this section, we discuss a plausible interpretation of the complex circumnuclear medium in NGC~7130, in terms of an AGN torus almost perpendicular to the galaxy disc, various red- and blueshifted components of AGN-driven outflow, and star-forming regions. A cartoon depicting our model is presented in Fig.~\ref{toy}.

Whereas NGC~7130 is nearly face-on (Sect.~\ref{orientation}), the torus of the AGN is probably close to edge-on, as indicated by its Seyfert~1.9 \citep{VeronCetty2010} or Seyfert~2 \citep{Phillips1983} classification. In fact, the spectrum of the central spaxel does not show any signature of a broad component associated with the broad-line region in the H$\upalpha$ line, which implies that the broad components are associated with the outflow. We therefore favour the Seyfert~2 classification.

The north-eastern arm of the disc is more blueshifted than its surroundings, and the south-western arm is more redshifted than its surroundings (see Fig.~\ref{images} for the location of the arms and Fig.~\ref{disckin} for the kinematics). Given that the southern side of the galaxy is the closest one (Sect.~\ref{orientation}), this indicates inward streaming motions bringing the gas to the inner few hundreds of parsecs. The disc component velocity map resulting from gas inflow through spiral arms is reminiscent of that for NGC~7213 \citep{SchnorrMueller2014}.

We find that both the outflow (Fig.~\ref{OIII}) and the synchrotron emission (Fig.~\ref{vla}) follow an almost north--south axis. An outflowing bicone close to the plane of the disc and perpendicular to the torus would explain the presence of blueshifted gas on both sides of the nucleus. This would imply that the AGN-driven wind is sweeping material from the disc outwards, including molecular gas. This could be potentially detected with ALMA observations aiming for a CO transition-tracing molecular gas with a lower density than the CO(6--5) line used in \citet{Zhao2016}.

The BPT line diagnostics of the outflow components (Figs.~\ref{bpts1} and \ref{bpts2}) indicate a Seyfert-like ionisation (except for the crescent narrow component, see below). Their typical velocities and velocity dispersions in a single-component fit \citep[$V\approx100\,{\rm km\,s^{-1}}$ and $\sigma\approx150\,{\rm km\,s^{-1}}$, respectively;][]{Knapen2019} are comparable to those in other Seyfert galaxies \citep[e.g.][]{MuellerSanchez2011, Mingozzi2019} and LIRGs \citep{Arribas2014, Cazzoli2014}. These pieces of evidence, together with the mass-loss rate and kinetic power estimates discussed below, strongly point to an AGN-driven origin of the outflow. 

We find the bipolar emission of the outflow to be asymmetric, with more emission coming from the blueshifted components (Fig.~\ref{OIII}). The asymmetry can be either intrinsic or, most likely, a result of absorption by disc dust. The fact that the redshifted outflow components are the most extincted (Table~\ref{physics}) is consistent with the latter possibility. This is in line with previous studies, where AGN line asymmetries have long been interpreted as a sign of an extincted redshifted outflow \citep{Grandi1977, Heckman1981}. More recently, \citet{Crenshaw2010} and \citet{Bae2014} established statistically that face-on Seyfert~2 galaxies more often have blueshifted outflow signatures, which was again linked to extinction. This is represented in Fig.~\ref{toy} with dashed lines indicating plausible locations for completely obscured components.

The blueshifted broad component is seen both north and south of the nucleus. The southern side is less blueshifted than the northern one (Fig.~\ref{gas_kinematics}). We also observe that, on its northern side, the blueshifted broad component covers a larger solid angle than the blueshifted narrow component. This is represented in Fig.~\ref{toy} by a larger aperture for the blueshifted broad component. In this model, the southern side is less blueshifted because the axis of the cone is pointing away from the observer. The redshifted broad component would be the (mostly obscured) receding counterpart of the blueshifted broad component.

Why do the narrow components cover a smaller solid angle than the blueshifted broad component? A possibility is that they correspond to different outflows emitted in separate episodes. Another possibility is that the mechanism powering the broad and the narrow components is different. Indeed, the alignment between the narrow components and the synchrotron emission suggests the possibility of a jet-driven component, whereas the broad components could be powered by another mechanism, such as AGN radiation. A third (and complementary) possibility, is that the two components are tracing gas with different properties. The broad components would correspond to lower-density gas, whereas the narrow ones would trace denser gas where efficient energy dissipation would keep the velocity dispersion low. We indeed find that $n_e=300\pm100\,{\rm cm^{-3}}$ for the blueshifted broad component, and $n_e=500\pm100\,{\rm cm^{-3}}$ for the blueshifted narrow component. Hence, we can say that there is tentative evidence that the blueshifted broad component has a smaller density than the narrow one.

In \citet{Knapen2019}, we found a `small kinematically decoupled core $0\farcs2$ (60\,pc) in radius \ldots [that] indicates a tiny inner disc'. This small bipolar structure seen in single-component fits is caused by the combined effects of the two blueshifted components and the redshifted narrow component~1, which dominate the emission in the two blobs just north and just south of the nucleus (Fig.~\ref{OIII}), which are also seen in coronal lines (Sect.~\ref{sectcoronal}). We find that the outflow components that contribute to these knots have Seyfert-like line ratios (Figs.~\ref{bpts1} and \ref{bpts2}). This, and the fact that coronal lines have large ionisation potentials \citep[$\gtrsim100\,{\rm eV}$;][]{Oke1968}, makes it hard to believe that the feature is a nuclear disc. Instead, we propose that we are observing the innermost parts of the outflow.

We estimated the mass outflow rates and kinetic power for each of the outflow components separately. We find that, although both narrow and broad outflow components have virtually the same flux in H$\upbeta$ ($L(\text{H}\upbeta)_{\rm corr}\sim(40-60)\times10^{39}\,{\rm erg\,s^{-1}}$), the broad components carry $\sim2/3$ of the mass outflow and maybe as much as $\sim90\%$ of the kinetic power. If we consider only the blueshifted components, which are better constrained due to the reduced extinction, we find that the luminosities of the narrow and the broad components are comparable ($L(\text{H}\upbeta)_{\rm corr}\approx30\times10^{39}\,{\rm erg\,s^{-1}}$ and $L(\text{H}\upbeta)_{\rm corr}\approx60\times10^{39}\,{\rm erg\,s^{-1}}$, respectively). In this case, the broad component carries 90\% of the mass loss and almost all the kinetic power (98\%). The relatively modest energy output of the ionised gas outflow, compared to the bolometric luminosity of the AGN ($F_{\rm kin}\approx0.12\%$ when accounting for all the components, and $F_{\rm kin}\approx0.03\%$ when accounting only for the blueshifted components), makes it unlikely that it has a significant impact at a galaxy-wide scale \citep[see discussion in][]{VillarMartin2016}. The mass-loss rate ($\dot{M}=1.2\pm0.7\,M_{\odot}\,{\rm yr^{-1}}$ for all the components and $\dot{M}=0.44\pm0.27\,M_{\odot}\,{\rm yr^{-1}}$ for the blueshifted components) is also low compared to the star formation rate, which is estimated to be $\dot{M}_{\rm SFR}=20.93\pm0.05\,M_{\odot}\,{\rm yr^{-1}}$ \citep{Gruppioni2016} or $\dot{M}_{\rm SFR}=6.7\,M_{\odot}\,{\rm yr^{-1}}$ \citep{DiamondStanic2012} for the galaxy as a whole, and $\dot{M}_{\rm SFR}=4.3\,M_{\odot}\,{\rm yr^{-1}}$ for the innermost kpc \citep{DiamondStanic2012}.

The estimated mass-outflow rate and the kinetic power are comparable to those computed in several recent studies of AGN by, for example, \citet{VillarMartin2016} and \citet{Rose2018}. The outflow in NGC~7130 appears to be a regular one in the $\dot{M}$ versus $L_{\rm bol}$ and $\dot{E}_{\rm kin}$ versus $L_{\rm bol}$ plots in Fig.~1 of \citet{Fiore2017}. We note, however, that just as in \citet{VillarMartin2016} and \citet{Rose2018}, we find the mass-outflow rates to be significantly lower than others derived in the literature \citep[e.g.,][]{Liu2013, Harrison2014, McElroy2015}. Part of the differences come from the fact that, as discussed in \citet{Rose2018}, these authors assumed low electron densities on the order of $n_e=100\,{\rm cm^{-3}}$, which results in an increase of the derived outflow mass (Eq.~\ref{mass}). Our estimates are also lower than those in \citet{MuellerSanchez2011}, who established their measurements based on coronal lines. On the other hand, they are larger than in \citet{Davies2020}, but this is likely because they used a different indicator to measure $n_e$, which results in larger estimated densities that lower the mass-loss-rate estimate by up to two dex in some cases.

All of the outflow components have Seyfert line ratios, except the crescent narrow component, whose line ratios are compatible with a LINER excitation (Fig.~\ref{bpts2}). This is demonstrated by the {[}N\,{\sc ii}{]}/H$\upalpha$ and {[}S\,{\sc ii}{]}/H$\upalpha$ line ratios (low-ionisation line ratios; hereafter LILrs) that are larger in the crescent narrow component than elsewhere in the outflow. Assuming a north--south axis for the outflow, the crescent narrow component would be found at the edge of the southern lobe of the ionisation cone (Fig.~\ref{locations}) and may be physically disconnected from the redshifted narrow components~1 and 2 (Sect.~\ref{intensity}).

The presence of high LILrs at the edges of ionisation cones has been reported by \citet{Mingozzi2019}. They also found a correlation between high LILrs and the velocity dispersion of {[}O\,{\sc iii}{]}, although this behaviour is not universal. We do not find that the crescent narrow component has a velocity dispersion that is significantly larger than that of the other narrow components (Fig.~\ref{gas_kinematics}), that is to say, the crescent narrow component average velocity dispersion is $\sigma=163\,{\rm km\,s^{-1}}$, whereas it is $145\,{\rm km\,s^{-1}}$, $141\,{\rm km\,s^{-1}}$, and $235\,{\rm km\,s^{-1}}$ for the blueshifted narrow component, redshifted narrow component~1, and redshifted narrow component~2, respectively (Table~\ref{adopted}). \citet{Mingozzi2019} also found that high LILrs correlate with a low-ionisation parameter, which agrees with the observed $\text{{[}S\,{\sc iii}{]}}/\text{{[}S\,{\sc ii}{]}}$ line ratio for the crescent narrow component (Sect.~\ref{secSIII}).

Since the velocity dispersion of the crescent narrow component is not particularly large, and because reasonable shock models by \citet{Allen2008} fail to explain line ratios with $\text{log}\,(\text{{[}N\,{\sc ii}{]}}/\text{H$\upalpha$})>0.3$ \citep[see discussion in][]{Mingozzi2019}, shock ionisation seems unlikely to explain the peculiar line ratios. Thus, in the case of NGC~7130, the high LILrs could be caused by the radiation field being different from that in other cone regions \citep{Mingozzi2019}. The light reaching the high LILr regions could have been filtered by clumpy ionised clouds between the SMBH and the crescent narrow component.

\section{Summary and conclusions}

\label{conclusion}

We present a detailed analysis of the circumnuclear medium of NGC~7130, a composite Seyfert~2 galaxy located at a proper distance of 64.8\,Mpc. Our work is a follow-up of \citet{Knapen2019}. We have used data obtained with MUSE + GALACSI (laser guide star adaptive optics) in narrow-field mode. We achieve an exquisite angular resolution of $0\farcs17$ which corresponds to $\sim50\,{\rm pc}$. Figure~\ref{images} and the comparison with $HST$ imaging in \citet{Knapen2019} bear witness to the extraordinary quality of the data.

In addition to the well-known UCNR, we found a nuclear ring with a major axis of $3^{\prime\prime}$, which is only 2.5\,times smaller than the major axis of the bar (Fig.~\ref{images}). This nuclear ring is, relative to the bar, exceptionally large. 

We analysed the data with adapted tools based on the \texttt{GIST} pipeline. The stellar velocity map (Fig.~\ref{stellar}) displays a butterfly pattern with an amplitude of a few tens of ${\rm km\,s^{-1}}$. We used the Tully-Fisher relation to estimate that NGC~7130 is nearly face-on (Sect.~\ref{orientation}).

We found that for several regions the line profiles are very complex, which is a clear indication of multiple kinematic components in the same line of sight. We have devised an algorithm to automatically fit any number of components between one and six (Figs.~\ref{flow} and \ref{flow2}). Such multi-component fits were produced for several relevant spectral lines simultaneously (Sect.~\ref{resolved}). Examples of the fits can be found in Appendix~\ref{examples}. We also produced single-component fits of the very weak {[}Fe\,{\sc vii}{]}\,$\uplambda$6087 coronal line (kinematic maps in Fig.~\ref{coronal}).

We identified nine distinct kinematic components for the circumnuclear ionised gas. For each of these components we produced kinematic maps (Fig.~\ref{gas_kinematics}), {[}O\,{\sc iii}{]} maps (Fig.~\ref{OIII}), resolved BPT line diagnostic maps (Figs.~\ref{bpts1} and \ref{bpts2}), the electron density $n_e$, the extinction $E(B-V)$, and the $\text{{[}S\,{\sc iii}{]}}/\text{{[}S\,{\sc ii}{]}}$ line ratio (Table~\ref{physics}).
Our interpretation of the circumnuclear medium of NGC~7130, in terms of a disc and an AGN-powered outflow, is illustrated in Fig.~\ref{toy}.

The disc of NGC~7130 displays signs of inward streaming motions through the spiral arms (Fig.~\ref{disckin}) bringing material to the inner few-hundred parsecs. We also distinguish a zero-velocity narrow component that we interpret as an outflow associated with intense star formation in the UCNR and the spiral arm. The above two components have BPT line ratios compatible with star-forming, or star-forming + AGN, ionisation.

The torus of the AGN is not in the same plane as the disc. The Seyfert~2 spectrum of the central region (at our resolution of $\sim50$\,pc) indicates that the torus is close to edge-on, and therefore almost perpendicular to the plane of the disc. We identify a north--south orientation in the signs of AGN activity: a radio jet and a biconic ionised gas outflow. The lobe length is at least 3\arcsec ($\sim900$\,pc).

Outflow components on the northern side of the galaxy are dominated by a blueshifted narrow and a blueshifted broad component. The narrow component is much more collimated than the broad one, as indicated by the solid angle covered by each of them. Also, while the narrow component has some well-defined substructure (Fig.~\ref{OIII}), the broad one is much smoother. The blueshifted narrow component is probably at least partially jet-powered as indicated by the alignment between its collimated substructure and the radio synchrotron emission (Fig.~\ref{vla}).

To the south of the nucleus, we observe two redshifted narrow components that occupy the same region in projection. A redshifted broad component is also seen in a small blob near the nucleus. The redshifted components are not as extended or bright as the blueshifted ones. Since they are seen through the host's disc, this is likely due to dust obscuration.

In \citet{Knapen2019}, we suggested that the tiny bipolar structure seen in the single-component velocity maps was caused by a nuclear disc. However, its north–south alignment, Seyfert line ratios, and coronal line emission indicate that this feature corresponds to the innermost region of the outflow.

We have measured the ionised gas mass outflow rate and the kinetic power of the outflow (Sect.~\ref{estimates}). Accounting for all the (the blueshifted only) components, we find them to be $\dot{M}=1.2\pm0.7\,M_{\odot}\,{\rm yr^{-1}}$ ($\dot{M}=0.44\pm0.27\,M_{\odot}\,{\rm yr^{-1}}$) and $\dot{E}_{\rm kin}=\left(2.7\pm2.0\right)\times10^{41}\,{\rm erg\,s^{-1}}$ ($\dot{E}_{\rm kin}=\left(7.0\pm4.8\right)\times10^{40}\,{\rm erg\,s^{-1}}$), respectively. The kinetic power is $F_{\rm kin}=0.12\pm0.09\%$ ($F_{\rm kin}=0.032\pm0.022\%$) of the bolometric AGN output. These values are comparable to those of other AGN \citep{VillarMartin2016, Rose2018} and are roughly a factor of ten lower than the star formation rate. They are probably too low for the outflow to have a galaxy-wide effect. The broad outflow components are responsible for $\sim2/3$ ($\sim90\%$) of the mass outflow rate and about 90\% (98\%) of the kinetic power output.

All of the outflow components have Seyfert line ratios, except for the crescent narrow component, which has LINER line ratios. This component is located off-axis (presumably at the border of the ionisation cone). We hypothesise that the most likely scenario for its peculiar line ratios is that it is ionised by light that has been filtered by clouds found between the central engine and the crescent narrow component region \citep[see][]{Mingozzi2019}.

Our study has once again proven the extraordinary quality of MUSE data, the finesse of which is such that it fully reveals the complexity of the multi-component outflow. Although multi-component ionised outflows have already been observed \citep[e.g.][]{McElroy2015}, none have, to our knowledge, required the many kinematic components that we have used in our description. This is partly due to the great care taken at scrutinising the data manually before devising an optimal procedure for the automatic multi-Gaussian line fit. It is also because of the great angular resolution provided by the NFM mode of MUSE, which prevents details from being lost due to beam smearing. We conclude that high $S/N$, high-angular-resolution data, and careful analysis are required to unearth many of the details of AGN outflows. It is to be expected that further AO-assisted IFU observations will show similarly complex ionised outflows in other active galaxies.

\begin{acknowledgements} 
We thank the anonymous referee for useful comments. We thank Dr.~Marja K.~Seidel for her contributions to the Very Large Telescope proposal \href{http://archive.eso.org/wdb/wdb/eso/sched_rep_arc/query?progid=60.A-9493(A)}{60.A-9493(A)}, Dr.~Yinghe Zhao for kindly sharing the reduced 8.4\,GHz image used in \citet{Zhao2016}, Dr.~Lodovico Coccato for his help at processing the MUSE data cube, Adrian Bittner for his prompt responses on \texttt{GIST}, Dr.~Jes\'us Falc\'on-Barroso for pointing to useful references, and Prof.~Tom Oosterloo and Dr.~Sim\'on D\'iaz-Garc\'ia for discussions. J.H.K.~acknowledges financial support from the European Union's Horizon 2020 research and innovation programme under Marie Sk{\l}odowska-Curie grant agreement No~721463 to the SUNDIAL ITN network, from the State Research Agency (AEI-MCINN) of the Spanish Ministry of Science and Innovation under the grant `The structure and evolution of galaxies and their central regions' with reference PID2019-105602GB-I00/10.13039/501100011033, and from IAC project P/300724, financed by the Ministry of Science and Innovation, through the State Budget and by the Canary Islands Department of Economy, Knowledge and Employment, through the Regional Budget of the Autonomous Community. C.R.A.~acknowledges financial support from the Spanish Ministry of Science, Innovation and Universities (MCIU) under grant with reference RYC-2014-15779, from the European Union's Horizon 2020 research and innovation programme under Marie Sk{\l}odowska-Curie grant agreement No~860744 (BID4BEST), from the State Research Agency (AEI-MCINN) of the Spanish MCIU under grants `Feeding and feedback in active galaxies' with reference PID2019-106027GB-C42, `Feeding, feedback and obscuration in active galaxies'
with reference AYA2016-76682-C3-2-P, and from IAC project P/301404, financed by the Ministry of Science and Innovation, through the State Budget and by the Canary Islands Department of Economy, Knowledge and Employment, through the Regional Budget of the Autonomous Community. Part of the results are based on public data released from the MUSE NFM-AO commissioning observations at the VLT Yepun (UT4) telescope under Programme \href{http://archive.eso.org/wdb/wdb/eso/sched_rep_arc/query?progid=60.A-9100(K)}{ID~60.A-9100, run K}. This research has made use of the NASA/IPAC Extragalactic Database (NED) which is operated by the Jet Propulsion Laboratory, California Institute of Technology, under contract with the National Aeronautics and Space Administration. This research has made use of the NASA/IPAC Infrared Science Archive, which is funded by the National Aeronautics and Space Administration and operated by the California Institute of Technology.
\end{acknowledgements}

   \bibliographystyle{aa} 
   \bibliography{/home/sebastien/JabRef/main_library.bib} 

\begin{thebibliography}{135}
\expandafter\ifx\csname natexlab\endcsname\relax\def\natexlab#1{#1}\fi

\bibitem[{Allen {et~al.}(2008)Allen, Groves, Dopita, Sutherland, \&
  Kewley}]{Allen2008}
Allen, M.~G., Groves, B.~A., Dopita, M.~A., Sutherland, R.~S., \& Kewley, L.~J.
  2008, \apjs, 178, 20

\bibitem[{Arribas {et~al.}(2014)Arribas, Colina, Bellocchi, Maiolino, \&
  Villar-Mart{\'\i}n}]{Arribas2014}
Arribas, S., Colina, L., Bellocchi, E., Maiolino, R., \& Villar-Mart{\'\i}n, M.
  2014, \aap, 568, A14

\bibitem[{Athanassoula(1992)}]{Athanassoula1992}
Athanassoula, E. 1992, \mnras, 259, 345

\bibitem[{Bacon {et~al.}(2010)Bacon, Accardo, Adjali, Anwand, Bauer, Biswas,
  Blaizot, Boudon, Brau-Nogue, Brinchmann, Caillier, Capoani, Carollo, Contini,
  Couderc, Daguis{\'e}, Deiries, Delabre, Dreizler, Dubois, Dupieux, Dupuy,
  Emsellem, Fechner, Fleischmann, Fran{\c{c}}ois, Gallou, Gharsa, Glindemann,
  Gojak, Guiderdoni, Hansali, Hahn, Jarno, Kelz, Koehler, Kosmalski, Laurent,
  Le~Floch, Lilly, Lizon, Loupias, Manescau, Monstein, Nicklas, Olaya, Pares,
  Pasquini, P{\'e}contal-Rousset, Pell{\'o}, Petit, Popow, Reiss, Remillieux,
  Renault, Roth, Rupprecht, Serre, Schaye, Soucail, Steinmetz, Streicher,
  Stuik, Valentin, Vernet, Weilbacher, Wisotzki, \& Yerle}]{Bacon2010}
Bacon, R., Accardo, M., Adjali, L., {et~al.} 2010, Society of Photo-Optical
  Instrumentation Engineers (SPIE) Conference Series, Vol. 7735, The MUSE
  second-generation VLT instrument, 773508

\bibitem[{Bacon {et~al.}(2017)Bacon, Conseil, Mary, Brinchmann, Shepherd,
  Akhlaghi, Weilbacher, Piqueras, Wisotzki, Lagattuta, Epinat, Guerou, Inami,
  Cantalupo, Courbot, Contini, Richard, Maseda, Bouwens, Bouch{\'e},
  Kollatschny, Schaye, Marino, Pello, Herenz, Guiderdoni, \&
  Carollo}]{Bacon2017}
Bacon, R., Conseil, S., Mary, D., {et~al.} 2017, \aap, 608, A1

\bibitem[{Bae \& Woo(2014)}]{Bae2014}
Bae, H.-J. \& Woo, J.-H. 2014, \apj, 795, 30

\bibitem[{Baldwin {et~al.}(1981)Baldwin, Phillips, \& Terlevich}]{Baldwin1981}
Baldwin, J.~A., Phillips, M.~M., \& Terlevich, R. 1981, \pasp, 93, 5

\bibitem[{Bellocchi {et~al.}(2012)Bellocchi, Arribas, \&
  Colina}]{Bellocchi2012}
Bellocchi, E., Arribas, S., \& Colina, L. 2012, \aap, 542, A54

\bibitem[{Bittner {et~al.}(2019)Bittner, Falc{\'o}n-Barroso, Nedelchev, Dorta,
  Gadotti, Sarzi, Molaeinezhad, Iodice, Rosado-Belza, de~Lorenzo-C{\'a}ceres,
  Fragkoudi, Gal{\'a}n-de Anta, Husemann, M{\'e}ndez-Abreu, Neumann, Pinna,
  Querejeta, S{\'a}nchez-Bl{\'a}zquez, \& Seidel}]{Bittner2019}
Bittner, A., Falc{\'o}n-Barroso, J., Nedelchev, B., {et~al.} 2019, \aap, 628,
  A117

\bibitem[{Busko \& Steiner(1988)}]{Busko1988}
Busko, I.~C. \& Steiner, J.~E. 1988, \mnras, 232, 525

\bibitem[{Buta {et~al.}(1999)Buta, Crocker, \& Byrd}]{Buta1999}
Buta, R., Crocker, D.~A., \& Byrd, G.~G. 1999, \aj, 118, 2071

\bibitem[{Buton {et~al.}(2013)Buton, Copin, Aldering, Antilogus, Aragon,
  Bailey, Baltay, Bongard, Canto, Cellier-Holzem, Childress, Chotard, Fakhouri,
  Gangler, Guy, Hsiao, Kerschhaggl, Kowalski, Loken, Nugent, Paech, Pain,
  P{\'e}contal, Pereira, Perlmutter, Rabinowitz, Rigault, Runge, Scalzo,
  Smadja, Tao, Thomas, Weaver, Wu, \& Factory}]{Buton2013}
Buton, C., Copin, Y., Aldering, G., {et~al.} 2013, \aap, 549, A8

\bibitem[{Calzetti {et~al.}(2000)Calzetti, Armus, Bohlin, Kinney, Koornneef, \&
  Storchi-Bergmann}]{Calzetti2000}
Calzetti, D., Armus, L., Bohlin, R.~C., {et~al.} 2000, \apj, 533, 682

\bibitem[{Cappellari \& Copin(2003)}]{Cappellari2003}
Cappellari, M. \& Copin, Y. 2003, \mnras, 342, 345

\bibitem[{Cazzoli {et~al.}(2014)Cazzoli, Arribas, Colina, Piqueras-L{\'o}pez,
  Bellocchi, Emonts, \& Maiolino}]{Cazzoli2014}
Cazzoli, S., Arribas, S., Colina, L., {et~al.} 2014, \aap, 569, A14

\bibitem[{Combes {et~al.}(2014)Combes, Garc{\'\i}a-Burillo, Casasola, Hunt,
  Krips, Baker, Boone, Eckart, Marquez, Neri, Schinnerer, \&
  Tacconi}]{Combes2014}
Combes, F., Garc{\'\i}a-Burillo, S., Casasola, V., {et~al.} 2014, \aap, 565,
  A97

\bibitem[{Comer{\'o}n {et~al.}(2010)Comer{\'o}n, Knapen, Beckman, Laurikainen,
  Salo, Mart{\'\i}nez-Valpuesta, \& Buta}]{Comeron2010}
Comer{\'o}n, S., Knapen, J.~H., Beckman, J.~E., {et~al.} 2010, \mnras, 402,
  2462

\bibitem[{Comer{\'o}n {et~al.}(2008)Comer{\'o}n, Knapen, Beckman, \&
  Shlosman}]{Comeron2008}
Comer{\'o}n, S., Knapen, J.~H., Beckman, J.~E., \& Shlosman, I. 2008, \aap,
  478, 403

\bibitem[{Comer{\'o}n {et~al.}(2014)Comer{\'o}n, Salo, Laurikainen, Knapen,
  Buta, Herrera-Endoqui, Laine, Holwerda, Sheth, Regan, Hinz, Mu{\~n}oz-Mateos,
  Gil~de Paz, Men{\'e}ndez-Delmestre, Seibert, Mizusawa, Kim, Erroz-Ferrer,
  Gadotti, Athanassoula, Bosma, \& Ho}]{Comeron2014}
Comer{\'o}n, S., Salo, H., Laurikainen, E., {et~al.} 2014, \aap, 562, A121

\bibitem[{Contini {et~al.}(2002)Contini, Radovich, Rafanelli, \&
  Richter}]{Contini2002}
Contini, M., Radovich, M., Rafanelli, P., \& Richter, G.~M. 2002, \apj, 572,
  124

\bibitem[{Crenshaw {et~al.}(2010)Crenshaw, Schmitt, Kraemer, Mushotzky, \&
  Dunn}]{Crenshaw2010}
Crenshaw, D.~M., Schmitt, H.~R., Kraemer, S.~B., Mushotzky, R.~F., \& Dunn,
  J.~P. 2010, \apj, 708, 419

\bibitem[{Davies {et~al.}(2020)Davies, Baron, Shimizu, Netzer, Burtscher,
  de~Zeeuw, Genzel, Hicks, Koss, Lin, Lutz, Maciejewski,
  M{\"u}ller-S{\'a}nchez, de~Xivry, Ricci, Riffel, Riffel, Rosario, Schartmann,
  Schnorr-M{\"u}ller, Shangguan, Sternberg, Sturm, Storchi-Bergmann, Tacconi,
  \& Veilleux}]{Davies2020}
Davies, R., Baron, D., Shimizu, T., {et~al.} 2020, \mnras
  [\eprint[arXiv]{2003.06153}]

\bibitem[{Davies {et~al.}(2009)Davies, Maciejewski, Hicks, Tacconi, Genzel, \&
  Engel}]{Davies2009}
Davies, R.~I., Maciejewski, W., Hicks, E. K.~S., {et~al.} 2009, \apj, 702, 114

\bibitem[{Davies {et~al.}(2014)Davies, Rich, Kewley, \& Dopita}]{Davies2014}
Davies, R.~L., Rich, J.~A., Kewley, L.~J., \& Dopita, M.~A. 2014, \mnras, 439,
  3835

\bibitem[{Davis {et~al.}(2012)Davis, Krajnovi{\'c}, McDermid, Bureau, Sarzi,
  Nyland, Alatalo, Bayet, Blitz, Bois, Bournaud, Cappellari, Crocker, Davies,
  de~Zeeuw, Duc, Emsellem, Khochfar, Kuntschner, Lablanche, Morganti, Naab,
  Oosterloo, Scott, Serra, Weijmans, \& Young}]{Davis2012}
Davis, T.~A., Krajnovi{\'c}, D., McDermid, R.~M., {et~al.} 2012, \mnras, 426,
  1574

\bibitem[{De~Robertis \& Osterbrock(1984)}]{DeRobertis1984}
De~Robertis, M.~M. \& Osterbrock, D.~E. 1984, \apj, 286, 171

\bibitem[{de~Vaucouleurs {et~al.}(1991)de~Vaucouleurs, de~Vaucouleurs, Corwin,
  Buta, Paturel, \& Fouque}]{Vaucouleurs1991}
de~Vaucouleurs, G., de~Vaucouleurs, A., Corwin, Herold~G., J., {et~al.} 1991,
  Third Reference Catalogue of Bright Galaxies

\bibitem[{de~Vaucouleurs {et~al.}(1976)de~Vaucouleurs, de~Vaucouleurs, \&
  Corwin}]{Vaucouleurs1976}
de~Vaucouleurs, G., de~Vaucouleurs, A., \& Corwin, J.~R. 1976, Second reference
  catalogue of bright galaxies, 1976, 0

\bibitem[{de~Vaucouleurs {et~al.}(1964)de~Vaucouleurs, de~Vaucouleurs, \&
  Shapley}]{Vaucouleurs1964}
de~Vaucouleurs, G.~H., de~Vaucouleurs, A., \& Shapley, H. 1964, Reference
  catalogue of bright galaxies

\bibitem[{Devereux(1987)}]{Devereux1987}
Devereux, N. 1987, \apj, 323, 91

\bibitem[{Diamond-Stanic \& Rieke(2012)}]{DiamondStanic2012}
Diamond-Stanic, A.~M. \& Rieke, G.~H. 2012, \apj, 746, 168

\bibitem[{D{\'\i}az {et~al.}(2000)D{\'\i}az, Castellanos, Terlevich, \& Luisa
  Garc{\'\i}a-Vargas}]{Diaz2000}
D{\'\i}az, A.~I., Castellanos, M., Terlevich, E., \& Luisa Garc{\'\i}a-Vargas,
  M. 2000, \mnras, 318, 462

\bibitem[{D{\'\i}az-Garc{\'\i}a {et~al.}(2020)D{\'\i}az-Garc{\'\i}a, Moyano,
  Comer{\'o}n, Knapen, Salo, \& Bouquin}]{DiazGarcia2020}
D{\'\i}az-Garc{\'\i}a, S., Moyano, F.~D., Comer{\'o}n, S., {et~al.} 2020, arXiv
  e-prints, arXiv:2009.00962

\bibitem[{D{\'\i}az-Garc{\'\i}a {et~al.}(2016)D{\'\i}az-Garc{\'\i}a, Salo, \&
  Laurikainen}]{DiazGarcia2016}
D{\'\i}az-Garc{\'\i}a, S., Salo, H., \& Laurikainen, E. 2016, \aap, 596, A84

\bibitem[{Dopita {et~al.}(2002)Dopita, Pereira, Kewley, \&
  Capaccioli}]{Dopita2002}
Dopita, M.~A., Pereira, M., Kewley, L.~J., \& Capaccioli, M. 2002, \apjs, 143,
  47

\bibitem[{Elias-Rosa {et~al.}(2018)Elias-Rosa, Van~Dyk, Benetti, Cappellaro,
  Smith, Kotak, Turatto, Filippenko, Pignata, Fox, Galbany,
  Gonz{\'a}lez-Gait{\'a}n, Miluzio, Monard, \& Ergon}]{EliasRosa2018}
Elias-Rosa, N., Van~Dyk, S.~D., Benetti, S., {et~al.} 2018, \apj, 860, 68

\bibitem[{Erroz-Ferrer {et~al.}(2019)Erroz-Ferrer, Carollo, den Brok, Onodera,
  Brinchmann, Marino, Monreal-Ibero, Schaye, Woo, Cibinel, Debattista, Inami,
  Maseda, Richard, Tacchella, \& Wisotzki}]{ErrozFerrer2019}
Erroz-Ferrer, S., Carollo, C.~M., den Brok, M., {et~al.} 2019, \mnras, 484,
  5009

\bibitem[{Esquej {et~al.}(2014)Esquej, Alonso-Herrero,
  Gonz{\'a}lez-Mart{\'\i}n, H{\"o}nig, Hern{\'a}n-Caballero, Roche,
  Ramos~Almeida, Mason, D{\'\i}az-Santos, Levenson, Aretxaga,
  Rodr{\'\i}guez~Espinosa, \& Packham}]{Esquej2014}
Esquej, P., Alonso-Herrero, A., Gonz{\'a}lez-Mart{\'\i}n, O., {et~al.} 2014,
  \apj, 780, 86

\bibitem[{Falc{\'o}n-Barroso {et~al.}(2006)Falc{\'o}n-Barroso, Bacon, Bureau,
  Cappellari, Davies, de~Zeeuw, Emsellem, Fathi, Krajnovi{\'c}, Kuntschner,
  McDermid, Peletier, \& Sarzi}]{FalconBarroso2006}
Falc{\'o}n-Barroso, J., Bacon, R., Bureau, M., {et~al.} 2006, \mnras, 369, 529

\bibitem[{Ferrarese \& Merritt(2000)}]{Ferrarese2000}
Ferrarese, L. \& Merritt, D. 2000, \apjl, 539, L9

\bibitem[{Feruglio {et~al.}(2010)Feruglio, Maiolino, Piconcelli, Menci, Aussel,
  Lamastra, \& Fiore}]{Feruglio2010}
Feruglio, C., Maiolino, R., Piconcelli, E., {et~al.} 2010, \aap, 518, L155

\bibitem[{Fiore {et~al.}(2017)Fiore, Feruglio, Shankar, Bischetti, Bongiorno,
  Brusa, Carniani, Cicone, Duras, Lamastra, Mainieri, Marconi, Menci, Maiolino,
  Piconcelli, Vietri, \& Zappacosta}]{Fiore2017}
Fiore, F., Feruglio, C., Shankar, F., {et~al.} 2017, \aap, 601, A143

\bibitem[{Freudling {et~al.}(2013)Freudling, Romaniello, Bramich, Ballester,
  Forchi, Garc{\'\i}a-Dabl{\'o}, Moehler, \& Neeser}]{Freudling2013}
Freudling, W., Romaniello, M., Bramich, D.~M., {et~al.} 2013, \aap, 559, A96

\bibitem[{Gebhardt {et~al.}(2000)Gebhardt, Bender, Bower, Dressler, Faber,
  Filippenko, Green, Grillmair, Ho, Kormendy, Lauer, Magorrian, Pinkney,
  Richstone, \& Tremaine}]{Gebhardt2000}
Gebhardt, K., Bender, R., Bower, G., {et~al.} 2000, \apjl, 539, L13

\bibitem[{Gonz{\'a}lez~Delgado {et~al.}(1998)Gonz{\'a}lez~Delgado, Heckman,
  Leitherer, Meurer, Krolik, Wilson, Kinney, \& Koratkar}]{GonzalezDelgado1998}
Gonz{\'a}lez~Delgado, R.~M., Heckman, T., Leitherer, C., {et~al.} 1998, \apj,
  505, 174

\bibitem[{Grandi(1977)}]{Grandi1977}
Grandi, S.~A. 1977, \apj, 215, 446

\bibitem[{Gruppioni {et~al.}(2016)Gruppioni, Berta, Spinoglio,
  Pereira-Santaella, Pozzi, Andreani, Bonato, De~Zotti, Malkan, Negrello,
  Vallini, \& Vignali}]{Gruppioni2016}
Gruppioni, C., Berta, S., Spinoglio, L., {et~al.} 2016, \mnras, 458, 4297

\bibitem[{Hanuschik(2003)}]{Hanuschik2003}
Hanuschik, R.~W. 2003, \aap, 407, 1157

\bibitem[{Harrison(2017)}]{Harrison2017}
Harrison, C.~M. 2017, Nature Astronomy, 1, 0165

\bibitem[{Harrison {et~al.}(2014)Harrison, Alexander, Mullaney, \&
  Swinbank}]{Harrison2014}
Harrison, C.~M., Alexander, D.~M., Mullaney, J.~R., \& Swinbank, A.~M. 2014,
  \mnras, 441, 3306

\bibitem[{Hawarden {et~al.}(1986)Hawarden, Mountain, Leggett, \&
  Puxley}]{Hawarden1986}
Hawarden, T.~G., Mountain, C.~M., Leggett, S.~K., \& Puxley, P.~J. 1986,
  \mnras, 221, 41P

\bibitem[{Heckman(1980)}]{Heckman1980}
Heckman, T.~M. 1980, \aap, 88, 365

\bibitem[{Heckman {et~al.}(1981)Heckman, Miley, van Breugel, \&
  Butcher}]{Heckman1981}
Heckman, T.~M., Miley, G.~K., van Breugel, W. J.~M., \& Butcher, H.~R. 1981,
  \apj, 247, 403

\bibitem[{Hinshaw {et~al.}(2009)Hinshaw, Weiland, Hill, Odegard, Larson,
  Bennett, Dunkley, Gold, Greason, Jarosik, Komatsu, Nolta, Page, Spergel,
  Wollack, Halpern, Kogut, Limon, Meyer, Tucker, \& Wright}]{Hinshaw2009}
Hinshaw, G., Weiland, J.~L., Hill, R.~S., {et~al.} 2009, The Astrophysical
  Journal Supplement Series, 180, 225

\bibitem[{Ho {et~al.}(2016)Ho, Medling, Groves, Rich, Rupke, Hampton, Kewley,
  Bland-Hawthorn, Croom, Richards, Schaefer, Sharp, \& Sweet}]{Ho2016}
Ho, I.~T., Medling, A.~M., Groves, B., {et~al.} 2016, \apss, 361, 280

\bibitem[{H{\"o}nig(2019)}]{Hoenig2019}
H{\"o}nig, S.~F. 2019, \apj, 884, 171

\bibitem[{Hummel(1981)}]{Hummel1981}
Hummel, E. 1981, \aap, 93, 93

\bibitem[{Hunt {et~al.}(2008)Hunt, Combes, Garc{\'\i}a-Burillo, Schinnerer,
  Krips, Baker, Boone, Eckart, L{\'e}on, Neri, \& Tacconi}]{Hunt2008}
Hunt, L.~K., Combes, F., Garc{\'\i}a-Burillo, S., {et~al.} 2008, \aap, 482, 133

\bibitem[{Jedrzejewski(1987)}]{Jedrzejewski1987}
Jedrzejewski, R.~I. 1987, \mnras, 226, 747

\bibitem[{Kauffmann {et~al.}(2003)Kauffmann, Heckman, Tremonti, Brinchmann,
  Charlot, White, Ridgway, Brinkmann, Fukugita, Hall, Ivezi{\'c}, Richards, \&
  Schneider}]{Kauffmann2003}
Kauffmann, G., Heckman, T.~M., Tremonti, C., {et~al.} 2003, \mnras, 346, 1055

\bibitem[{Kewley {et~al.}(2001)Kewley, Dopita, Sutherland, Heisler, \&
  Trevena}]{Kewley2001}
Kewley, L.~J., Dopita, M.~A., Sutherland, R.~S., Heisler, C.~A., \& Trevena, J.
  2001, \apj, 556, 121

\bibitem[{Kewley {et~al.}(2006)Kewley, Groves, Kauffmann, \&
  Heckman}]{Kewley2006}
Kewley, L.~J., Groves, B., Kauffmann, G., \& Heckman, T. 2006, \mnras, 372, 961

\bibitem[{Kim \& Elmegreen(2017)}]{Kim2017}
Kim, W.-T. \& Elmegreen, B.~G. 2017, \apjl, 841, L4

\bibitem[{Kim \& Kim(2014)}]{Kim2014}
Kim, Y. \& Kim, W.-T. 2014, \mnras, 440, 208

\bibitem[{Knapen {et~al.}(1995)Knapen, Beckman, Heller, Shlosman, \&
  de~Jong}]{Knapen1995}
Knapen, J.~H., Beckman, J.~E., Heller, C.~H., Shlosman, I., \& de~Jong, R.~S.
  1995, \apj, 454, 623

\bibitem[{Knapen {et~al.}(2019)Knapen, Comer{\'o}n, \& Seidel}]{Knapen2019}
Knapen, J.~H., Comer{\'o}n, S., \& Seidel, M.~K. 2019, \aap, 621, L5

\bibitem[{Koski(1978)}]{Koski1978}
Koski, A.~T. 1978, \apj, 223, 56

\bibitem[{Kraemer \& Crenshaw(2000)}]{Kraemer2000}
Kraemer, S.~B. \& Crenshaw, D.~M. 2000, \apj, 532, 256

\bibitem[{Krajnovi{\'c} {et~al.}(2006)Krajnovi{\'c}, Cappellari, de~Zeeuw, \&
  Copin}]{Krajnovic2006}
Krajnovi{\'c}, D., Cappellari, M., de~Zeeuw, P.~T., \& Copin, Y. 2006, \mnras,
  366, 787

\bibitem[{Kroupa(2001)}]{Kroupa2001}
Kroupa, P. 2001, \mnras, 322, 231

\bibitem[{Lauberts \& Valentijn(1989)}]{Lauberts1989}
Lauberts, A. \& Valentijn, E.~A. 1989, The surface photometry catalogue of the
  ESO-Uppsala galaxies

\bibitem[{Lena {et~al.}(2015)Lena, Robinson, Storchi-Bergman,
  Schnorr-M{\"u}ller, Seelig, Riffel, Nagar, Couto, \& Shadler}]{Lena2015}
Lena, D., Robinson, A., Storchi-Bergman, T., {et~al.} 2015, \apj, 806, 84

\bibitem[{Levenson {et~al.}(2005)Levenson, Weaver, Heckman, Awaki, \&
  Terashima}]{Levenson2005}
Levenson, N.~A., Weaver, K.~A., Heckman, T.~M., Awaki, H., \& Terashima, Y.
  2005, \apj, 618, 167

\bibitem[{Lin {et~al.}(2017)Lin, Li, He, Xiao, \& Wang}]{Lin2017}
Lin, L., Li, C., He, Y., Xiao, T., \& Wang, E. 2017, \apj, 838, 105

\bibitem[{Liu {et~al.}(2013)Liu, Zakamska, Greene, Nesvadba, \& Liu}]{Liu2013}
Liu, G., Zakamska, N.~L., Greene, J.~E., Nesvadba, N. P.~H., \& Liu, X. 2013,
  \mnras, 436, 2576

\bibitem[{Lubow {et~al.}(1986)Lubow, Balbus, \& Cowie}]{Lubow1986}
Lubow, S.~H., Balbus, S.~A., \& Cowie, L.~L. 1986, \apj, 309, 496

\bibitem[{Lynden-Bell(1969)}]{LyndenBell1969}
Lynden-Bell, D. 1969, \nat, 223, 690

\bibitem[{Malkan {et~al.}(1998)Malkan, Gorjian, \& Tam}]{Malkan1998}
Malkan, M.~A., Gorjian, V., \& Tam, R. 1998, \apjs, 117, 25

\bibitem[{McElroy {et~al.}(2015)McElroy, Croom, Pracy, Sharp, Ho, \&
  Medling}]{McElroy2015}
McElroy, R., Croom, S.~M., Pracy, M., {et~al.} 2015, \mnras, 446, 2186

\bibitem[{Mingozzi {et~al.}(2019)Mingozzi, Cresci, Venturi, Marconi, Mannucci,
  Perna, Belfiore, Carniani, Balmaverde, Brusa, Cicone, Feruglio, Gallazzi,
  Mainieri, Maiolino, Nagao, Nardini, Sani, Tozzi, \& Zibetti}]{Mingozzi2019}
Mingozzi, M., Cresci, G., Venturi, G., {et~al.} 2019, \aap, 622, A146

\bibitem[{Morganti(2017)}]{Morganti2017}
Morganti, R. 2017, Frontiers in Astronomy and Space Sciences, 4, 42

\bibitem[{Morganti {et~al.}(2005)Morganti, Oosterloo, Tadhunter, van Moorsel,
  \& Emonts}]{Morganti2005}
Morganti, R., Oosterloo, T.~A., Tadhunter, C.~N., van Moorsel, G., \& Emonts,
  B. 2005, \aap, 439, 521

\bibitem[{Mulchaey {et~al.}(1997)Mulchaey, Regan, \& Kundu}]{Mulchaey1997}
Mulchaey, J.~S., Regan, M.~W., \& Kundu, A. 1997, \apjs, 110, 299

\bibitem[{M{\"u}ller-S{\'a}nchez {et~al.}(2011)M{\"u}ller-S{\'a}nchez, Prieto,
  Hicks, Vives-Arias, Davies, Malkan, Tacconi, \& Genzel}]{MuellerSanchez2011}
M{\"u}ller-S{\'a}nchez, F., Prieto, M.~A., Hicks, E. K.~S., {et~al.} 2011,
  \apj, 739, 69

\bibitem[{Mu{\~n}oz~Mar{\'\i}n {et~al.}(2007)Mu{\~n}oz~Mar{\'\i}n,
  Gonz{\'a}lez~Delgado, Schmitt, Cid~Fernandes, P{\'e}rez, Storchi-Bergmann,
  Heckman, \& Leitherer}]{MunozMarin2007}
Mu{\~n}oz~Mar{\'\i}n, V.~M., Gonz{\'a}lez~Delgado, R.~M., Schmitt, H.~R.,
  {et~al.} 2007, \aj, 134, 648

\bibitem[{Mu{\~n}oz-Mateos {et~al.}(2015)Mu{\~n}oz-Mateos, Sheth, Regan, Kim,
  Laine, Erroz-Ferrer, Gil~de Paz, Comeron, Hinz, Laurikainen, Salo,
  Athanassoula, Bosma, Bouquin, Schinnerer, Ho, Zaritsky, Gadotti, Madore,
  Holwerda, Men{\'e}ndez-Delmestre, Knapen, Meidt, Querejeta, Mizusawa,
  Seibert, Laine, \& Courtois}]{MunozMateos2015}
Mu{\~n}oz-Mateos, J.~C., Sheth, K., Regan, M., {et~al.} 2015, \apjs, 219, 3

\bibitem[{Negroponte \& White(1983)}]{Negroponte1983}
Negroponte, J. \& White, S. D.~M. 1983, \mnras, 205, 1009

\bibitem[{Norris {et~al.}(1990)Norris, Allen, Sramek, Kesteven, \&
  Troup}]{Norris1990}
Norris, R.~P., Allen, D.~A., Sramek, R.~A., Kesteven, M.~J., \& Troup, E.~R.
  1990, \apj, 359, 291

\bibitem[{Oke \& Sargent(1968)}]{Oke1968}
Oke, J.~B. \& Sargent, W. L.~W. 1968, \apj, 151, 807

\bibitem[{Osterbrock \& Ferland(2006)}]{Osterbrock2006}
Osterbrock, D.~E. \& Ferland, G.~J. 2006, Astrophysics of gaseous nebulae and
  active galactic nuclei

\bibitem[{Osterbrock \& Fulbright(1996)}]{Osterbrock1996}
Osterbrock, D.~E. \& Fulbright, J.~P. 1996, \pasp, 108, 183

\bibitem[{Padovani {et~al.}(2017)Padovani, Alexander, Assef, De~Marco, Giommi,
  Hickox, Richards, Smol{\v{c}}i{\'c}, Hatziminaoglou, Mainieri, \&
  Salvato}]{Padovani2017}
Padovani, P., Alexander, D.~M., Assef, R.~J., {et~al.} 2017, \aapr, 25, 2

\bibitem[{Peirani {et~al.}(2008)Peirani, Kay, \& Silk}]{Peirani2008}
Peirani, S., Kay, S., \& Silk, J. 2008, \aap, 479, 123

\bibitem[{Phillips {et~al.}(1983)Phillips, Charles, \& Baldwin}]{Phillips1983}
Phillips, M.~M., Charles, P.~A., \& Baldwin, J.~A. 1983, \apj, 266, 485

\bibitem[{Pierce \& Tully(1992)}]{Pierce1992}
Pierce, M.~J. \& Tully, R.~B. 1992, \apj, 387, 47

\bibitem[{Pietrinferni {et~al.}(2004)Pietrinferni, Cassisi, Salaris, \&
  Castelli}]{Pietrinferni2004}
Pietrinferni, A., Cassisi, S., Salaris, M., \& Castelli, F. 2004, \apj, 612,
  168

\bibitem[{Pozzi {et~al.}(2017)Pozzi, Vallini, Vignali, Talia, Gruppioni,
  Mingozzi, Massardi, \& Andreani}]{Pozzi2017}
Pozzi, F., Vallini, L., Vignali, C., {et~al.} 2017, \mnras, 470, L64

\bibitem[{Querejeta {et~al.}(2016)Querejeta, Meidt, Schinnerer,
  Garc{\'\i}a-Burillo, Dobbs, Colombo, Dumas, Hughes, Kramer, Leroy, Pety,
  Schuster, \& Thompson}]{Querejeta2016}
Querejeta, M., Meidt, S.~E., Schinnerer, E., {et~al.} 2016, \aap, 588, A33

\bibitem[{Radovich {et~al.}(1997)Radovich, Rafanelli, Birkle, \&
  Richter}]{Radovich1997}
Radovich, M., Rafanelli, P., Birkle, K., \& Richter, G.~M. 1997, Astronomische
  Nachrichten: A Journal on all Fields of Astronomy, 318, 229

\bibitem[{Ramos~Almeida {et~al.}(2009)Ramos~Almeida, P{\'e}rez~Garc{\'\i}a, \&
  Acosta-Pulido}]{RamosAlmeida2009}
Ramos~Almeida, C., P{\'e}rez~Garc{\'\i}a, A.~M., \& Acosta-Pulido, J.~A. 2009,
  \apj, 694, 1379

\bibitem[{Ramos~Almeida \& Ricci(2017)}]{RamosAlmeida2017}
Ramos~Almeida, C. \& Ricci, C. 2017, Nature Astronomy, 1, 679

\bibitem[{Riffel {et~al.}(2009)Riffel, Storchi-Bergmann, \&
  McGregor}]{Riffel2009}
Riffel, R.~A., Storchi-Bergmann, T., \& McGregor, P.~J. 2009, \apj, 698, 1767

\bibitem[{Riffel {et~al.}(2010)Riffel, Storchi-Bergmann, \& Nagar}]{Riffel2010}
Riffel, R.~A., Storchi-Bergmann, T., \& Nagar, N.~M. 2010, \mnras, 404, 166

\bibitem[{Rose {et~al.}(2018)Rose, Tadhunter, Ramos~Almeida,
  Rodr{\'\i}guez~Zaur{\'\i}n, Santoro, \& Spence}]{Rose2018}
Rose, M., Tadhunter, C., Ramos~Almeida, C., {et~al.} 2018, \mnras, 474, 128

\bibitem[{Sakamoto {et~al.}(1999)Sakamoto, Okumura, Ishizuki, \&
  Scoville}]{Sakamoto1999}
Sakamoto, K., Okumura, S.~K., Ishizuki, S., \& Scoville, N.~Z. 1999, \apj, 525,
  691

\bibitem[{Salpeter(1964)}]{Salpeter1964}
Salpeter, E.~E. 1964, \apj, 140, 796

\bibitem[{Sandage \& Bedke(1994)}]{Sandage1994}
Sandage, A. \& Bedke, J. 1994, The Carnegie atlas of galaxies, Vol. 638

\bibitem[{Sanders {et~al.}(2003)Sanders, Mazzarella, Kim, Surace, \&
  Soifer}]{Sanders2003}
Sanders, D.~B., Mazzarella, J.~M., Kim, D.~C., Surace, J.~A., \& Soifer, B.~T.
  2003, \aj, 126, 1607

\bibitem[{Sanders \& Mirabel(1996)}]{Sanders1996}
Sanders, D.~B. \& Mirabel, I.~F. 1996, \araa, 34, 749

\bibitem[{Sanders {et~al.}(2016)Sanders, Shapley, Kriek, Reddy, Freeman, Coil,
  Siana, Mobasher, Shivaei, Price, \& de~Groot}]{Sanders2016}
Sanders, R.~L., Shapley, A.~E., Kriek, M., {et~al.} 2016, \apj, 816, 23

\bibitem[{Sarzi {et~al.}(2006)Sarzi, Falc{\'o}n-Barroso, Davies, Bacon, Bureau,
  Cappellari, de~Zeeuw, Emsellem, Fathi, Krajnovi{\'c}, Kuntschner, McDermid,
  \& Peletier}]{Sarzi2006}
Sarzi, M., Falc{\'o}n-Barroso, J., Davies, R.~L., {et~al.} 2006, \mnras, 366,
  1151

\bibitem[{Schnorr-M{\"u}ller {et~al.}(2014)Schnorr-M{\"u}ller,
  Storchi-Bergmann, Nagar, \& Ferrari}]{SchnorrMueller2014}
Schnorr-M{\"u}ller, A., Storchi-Bergmann, T., Nagar, N.~M., \& Ferrari, F.
  2014, \mnras, 438, 3322

\bibitem[{Schwarz(1984)}]{Schwarz1984}
Schwarz, M.~P. 1984, \mnras, 209, 93

\bibitem[{Sheth {et~al.}(2005)Sheth, Vogel, Regan, Thornley, \&
  Teuben}]{Sheth2005}
Sheth, K., Vogel, S.~N., Regan, M.~W., Thornley, M.~D., \& Teuben, P.~J. 2005,
  \apj, 632, 217

\bibitem[{Shields \& Filippenko(1990)}]{Shields1990}
Shields, J.~C. \& Filippenko, A.~V. 1990, \aj, 100, 1034

\bibitem[{Shlosman {et~al.}(1989)Shlosman, Frank, \& Begelman}]{Shlosman1989}
Shlosman, I., Frank, J., \& Begelman, M.~C. 1989, \nat, 338, 45

\bibitem[{Skrutskie {et~al.}(2006)Skrutskie, Cutri, Stiening, Weinberg,
  Schneider, Carpenter, Beichman, Capps, Chester, Elias, Huchra, Liebert,
  Lonsdale, Monet, Price, Seitzer, Jarrett, Kirkpatrick, Gizis, Howard, Evans,
  Fowler, Fullmer, Hurt, Light, Kopan, Marsh, McCallon, Tam, Van~Dyk, \&
  Wheelock}]{Skrutskie2006}
Skrutskie, M.~F., Cutri, R.~M., Stiening, R., {et~al.} 2006, \aj, 131, 1163

\bibitem[{Sorce {et~al.}(2014)Sorce, Tully, Courtois, Jarrett, Neill, \&
  Shaya}]{Sorce2014}
Sorce, J.~G., Tully, R.~B., Courtois, H.~M., {et~al.} 2014, \mnras, 444, 527

\bibitem[{Storchi-Bergmann \& Schnorr-M{\"u}ller(2019)}]{StorchiBergmann2019}
Storchi-Bergmann, T. \& Schnorr-M{\"u}ller, A. 2019, Nature Astronomy, 3, 48

\bibitem[{Storey \& Zeippen(2000)}]{Storey2000}
Storey, P.~J. \& Zeippen, C.~J. 2000, \mnras, 312, 813

\bibitem[{Stuik {et~al.}(2006)Stuik, Bacon, Conzelmann, Delabre, Fedrigo,
  Hubin, Le~Louarn, \& Str{\"o}bele}]{Stuik2006}
Stuik, R., Bacon, R., Conzelmann, R., {et~al.} 2006, \nar, 49, 618

\bibitem[{Telesco(1988)}]{Telesco1988}
Telesco, C.~M. 1988, \araa, 26, 343

\bibitem[{Thean {et~al.}(2000)Thean, Pedlar, Kukula, Baum, \&
  O'Dea}]{Thean2000}
Thean, A., Pedlar, A., Kukula, M.~J., Baum, S.~A., \& O'Dea, C.~P. 2000,
  \mnras, 314, 573

\bibitem[{Thuan(1984)}]{Thuan1984}
Thuan, T.~X. 1984, \apj, 281, 126

\bibitem[{Tody(1986)}]{Tody1986}
Tody, D. 1986, Society of Photo-Optical Instrumentation Engineers (SPIE)
  Conference Series, Vol. 627, The IRAF Data Reduction and Analysis System, ed.
  D.~L. {Crawford}, 733

\bibitem[{Vazdekis {et~al.}(2016)Vazdekis, Koleva, Ricciardelli, R{\"o}ck, \&
  Falc{\'o}n-Barroso}]{Vazdekis2016}
Vazdekis, A., Koleva, M., Ricciardelli, E., R{\"o}ck, B., \&
  Falc{\'o}n-Barroso, J. 2016, \mnras, 463, 3409

\bibitem[{Veilleux {et~al.}(2005)Veilleux, Cecil, \&
  Bland-Hawthorn}]{Veilleux2005}
Veilleux, S., Cecil, G., \& Bland-Hawthorn, J. 2005, \araa, 43, 769

\bibitem[{Veilleux {et~al.}(2020)Veilleux, Maiolino, Bolatto, \&
  Aalto}]{Veilleux2020}
Veilleux, S., Maiolino, R., Bolatto, A.~D., \& Aalto, S. 2020, \aapr, 28, 2

\bibitem[{V{\'e}ron(1981)}]{Veron1981}
V{\'e}ron, M.~P. 1981, \aap, 100, 12

\bibitem[{V{\'e}ron-Cetty \& V{\'e}ron(2010)}]{VeronCetty2010}
V{\'e}ron-Cetty, M.~P. \& V{\'e}ron, P. 2010, \aap, 518, A10

\bibitem[{Vilchez \& Esteban(1996)}]{Vilchez1996}
Vilchez, J.~M. \& Esteban, C. 1996, \mnras, 280, 720

\bibitem[{Villar-Mart{\'\i}n {et~al.}(2016)Villar-Mart{\'\i}n, Arribas, Emonts,
  Humphrey, Tadhunter, Bessiere, Cabrera~Lavers, \&
  Ramos~Almeida}]{VillarMartin2016}
Villar-Mart{\'\i}n, M., Arribas, S., Emonts, B., {et~al.} 2016, \mnras, 460,
  130

\bibitem[{Weilbacher {et~al.}(2012)Weilbacher, Streicher, Urrutia, Jarno,
  P{\'e}contal-Rousset, Bacon, \& B{\"o}hm}]{Weilbacher2012}
Weilbacher, P.~M., Streicher, O., Urrutia, T., {et~al.} 2012, Society of
  Photo-Optical Instrumentation Engineers (SPIE) Conference Series, Vol. 8451,
  Design and capabilities of the MUSE data reduction software and pipeline,
  84510B

\bibitem[{Weilbacher {et~al.}(2014)Weilbacher, Streicher, Urrutia,
  P{\'e}contal-Rousset, Jarno, \& Bacon}]{Weilbacher2014}
Weilbacher, P.~M., Streicher, O., Urrutia, T., {et~al.} 2014, Astronomical
  Society of the Pacific Conference Series, Vol. 485, The MUSE Data Reduction
  Pipeline: Status after Preliminary Acceptance Europe, ed. N.~{Manset} \&
  P.~{Forshay}, 451

\bibitem[{Zhao {et~al.}(2016)Zhao, Lu, Xu, Gao, Barcos-Mun{\~o}z,
  D{\'\i}az-Santos, Appleton, Charmandaris, Armus, van~der Werf, Evans, Cao,
  Inami, \& Murphy}]{Zhao2016}
Zhao, Y., Lu, N., Xu, C.~K., {et~al.} 2016, \apj, 820, 118

\end{thebibliography}

\begin{appendix}   
   
\section{Selected examples of multi-component fits}

\label{examples}

\subsection{Example~1: A bin $0\mkern2.0mu.\mkern-6.5mu^{\prime\prime}1$ south of the nucleus with two redshifted narrow components}

\label{example1}

\begin{figure*}
   \centering
   \includegraphics[scale=0.30]{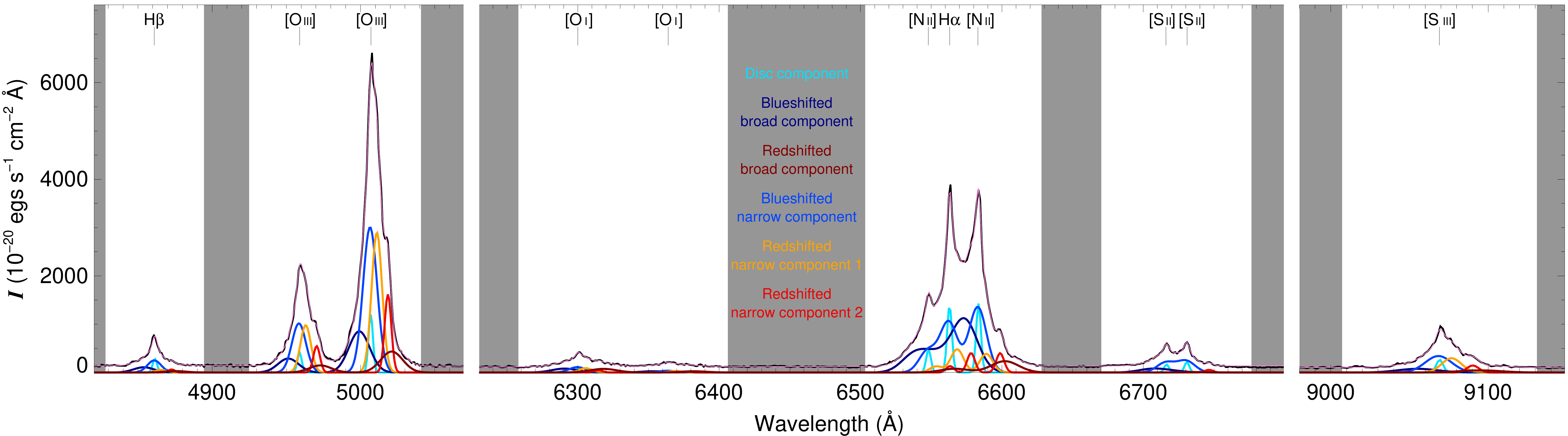}
   \caption{\label{fit91}{Spectrum of Voronoi bin discussed as Example~1 in Sect.~\ref{examples}, and indicated by a number `1' in Fig~\ref{locations}. The black line describes the spectrum and the almost coincident purple line describes the fit. The contribution of the six components in this fit are indicated by curves, colour-coded as in Figs.~\ref{gas_kinematics} and \ref{locations}. The shaded areas were masked away while using \texttt{pyGandALF}. The wavelengths are displayed at the rest frame of $z=0.016151$.}}
\end{figure*}

The bins in the central regions of NGC~7130 show very complex line profiles that require multiple components to be characterised. This is well illustrated in our Example~1 (Fig.~\ref{fit91}). The {[}O\,{\sc iii}{]}\,$\uplambda$5007 line profile is a good place to start a visual analysis, because it is isolated and has a high S/N. The {[}O\,{\sc iii}{]}\,$\uplambda$5007 line is asymmetric and shows a peak with two redward shoulders, corresponding to the two redshifted narrow components, and a blueshifted tail, corresponding to the blueshifted broad component.

The peak of the lines in {[}O\,{\sc iii}{]} is displaced blueward with respect to that in H$\upalpha$ and H$\upbeta$. This is because the peak is actually made of the superposition of two narrow components, namely the disc component and the blueshifted narrow component. In {[}O\,{\sc iii}{],} the blueshifted narrow component dominates over the disc component, causing the peak of the line to be blueshifted compared to that of the Balmer lines. Fitting the central peak with a single component instead of two slightly shifted ones causes the fit to miss a large fraction of light either in the peaks in the {[}O\,{\sc iii}{]} doublet or in the Balmer lines (i.e. the observed peaks would have a larger amplitude than in the fits).

The redshifted broad component appears as a low-amplitude red tail in both {[}O\,{\sc iii}{]}\,$\uplambda$5007 and {[}N\,{\sc ii}{]}\,$\uplambda$6583. While it has a small flux, not including it in the fit means that the kinks and shoulders on the red side of {[}N\,{\sc ii}{]}\,$\uplambda$6583 are not well described.

\subsection{Example~2: A bin at the periphery of the region with two redshifted narrow components}

\begin{figure*}
   \centering
   \includegraphics[scale=0.30]{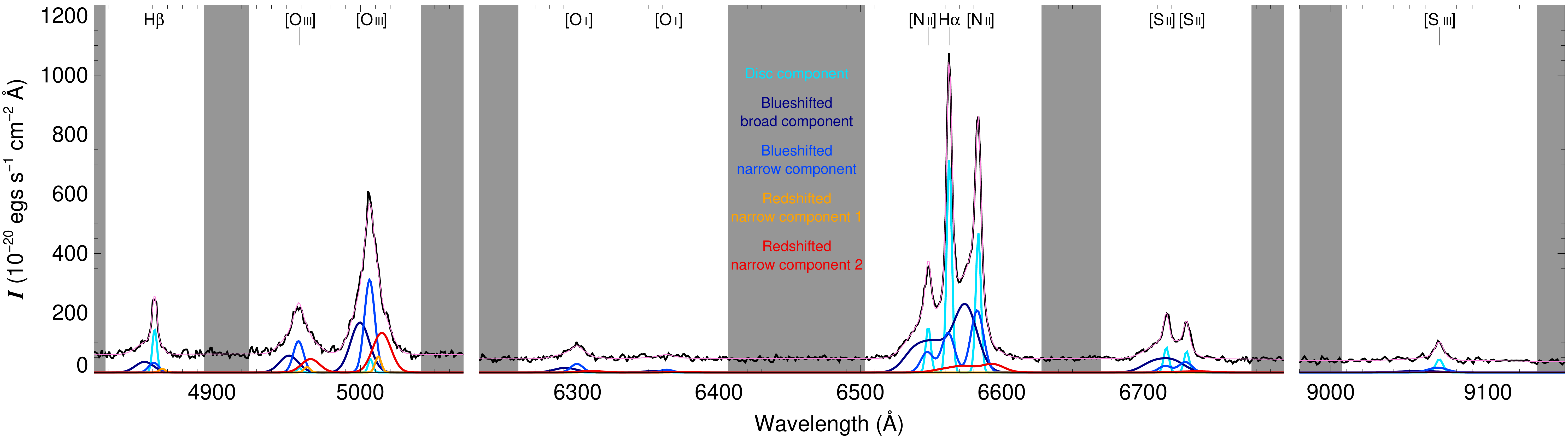}
   \caption{\label{fit443}{Same as Fig.~\ref{fit91} but for bin indicated by number `2' in Fig~\ref{locations}. This fit is discussed as Example~2 in Sect.~\ref{examples}.}}
\end{figure*}

\label{example2}

Just outside the region with two fitted redshifted narrow components, subtle signs of redshifted shoulders are seen under visual inspection but have not been fitted because of the $\chi^2$ limits that we imposed. This is an effect of both the surface brightness decreasing with radius, and also the increasing domination of the disc component with increasing distance from the centre, which ultimately causes the traces of redshifted components to be hidden within the disc component wings. Allowing for less stringent ratios of $\chi^2$ limits makes the region where these two components are fitted larger, at a cost of introducing spurious components everywhere in the field of view.

In Fig.~\ref{fit91}, we see that the kinks caused by the redshifted narrow components are observed at a higher surface brightness in {[}O\,{\sc iii}{]} than in {[}N\,{\sc ii}{]}\,$\uplambda$6583. As a consequence, in noisy bins, clear traces of one or two of these narrow components are seen in {[}O\,{\sc iii}{]} only. An example of that is the spectrum displayed in our Example~2 (Fig.~\ref{fit443}), where the red side of {[}N\,{\sc ii}{]}\,$\uplambda$6583 barely shows any signs of the two redshifted narrow components. A global $\chi^2$ criterion does, in this case, easily overlook the clear indications for the redshifted components seen in {[}O\,{\sc iii}{]}. Hence, this kind of spectrum was the motivation to introduce a $\chi^2$ criterion referring to the narrow spectral window around {[}O\,{\sc iii}{]}\,$\uplambda$5007 (see Sect.~\ref{resolved}).

\subsection{Example~3: A bin in the region with the crescent narrow component}

\label{example3}

\begin{figure*}
   \centering
   \includegraphics[scale=0.30]{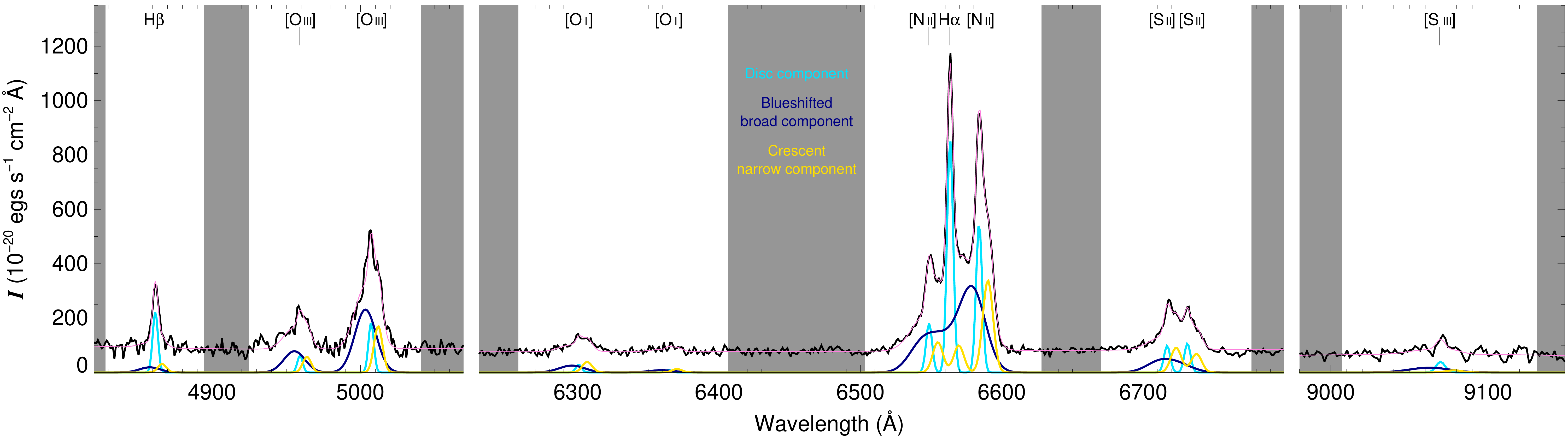}
   \caption{\label{fit1056}{Same as Fig.~\ref{fit91} but for the bin indicated by number `3' in Fig~\ref{locations} and with three components. This fit is discussed as Example~3 in Sect.~\ref{examples}.}}
\end{figure*}

The bins in this region have a conspicuous redshifted shoulder in their {[}N\,{\sc ii}{]}\,$\uplambda$6583 line that is also seen, albeit less clearly, in the {[}O\,{\sc iii}{]} lines. The example displayed in Fig.~\ref{fit1056} is an exceptionally good case, but at the periphery of the crescent, the redshifted shoulder of {[}N\,{\sc ii}{]}\,$\uplambda$6583 is also obvious. In these periphery bins, the global $\chi^2$ ratio criterion is not sensitive enough to capture that crescent narrow component. To solve that without loosening the $\chi^2$ ratio criteria too much, we included the $\chi^2$ ratio criteria for the wavelength range in the line complex including {[}N\,{\sc ii}{]} and H$\upalpha$ (Sect.~\ref{resolved}).

\subsection{Example~4: A bin $0\mkern2.0mu.\mkern-6.5mu^{\prime\prime}1$ north of the nucleus with a conspicuous blueshifted broad component}

\label{example4}

\begin{figure*}
   \centering
   \includegraphics[scale=0.30]{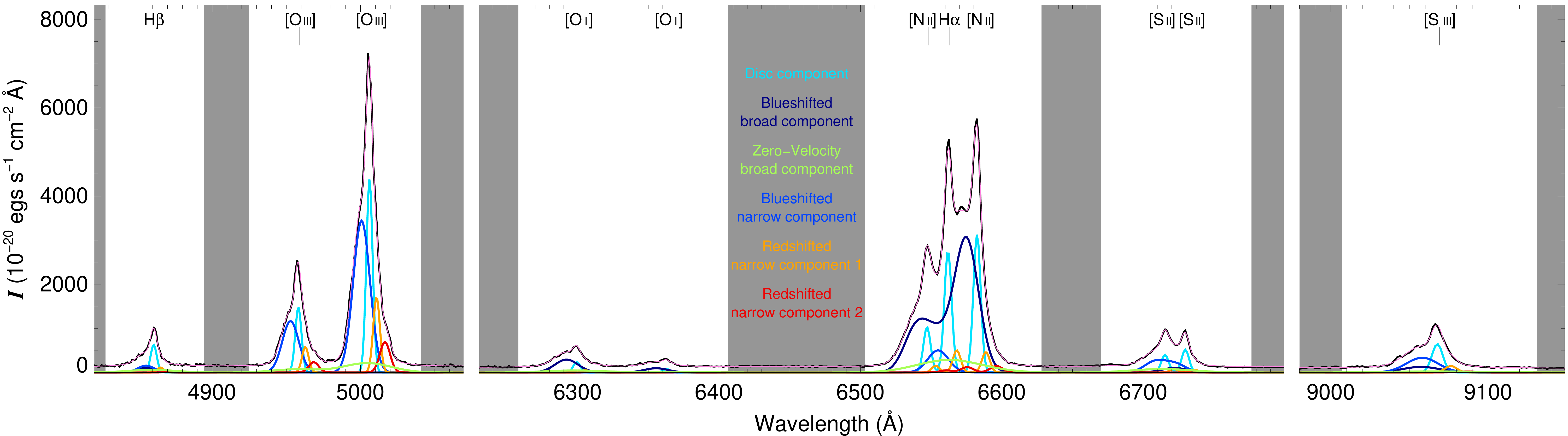}
   \caption{\label{fit1}{Same as Fig.~\ref{fit91} but for the bin indicated by number `4' in Fig~\ref{locations}. This fit is discussed as Example~4 in Sect.~\ref{examples}.}}
\end{figure*}

The innermost bins on the northern side of the nucleus show a clear blueshifted broad shoulder in the {[}O\,{\sc iii}{]} line profiles (Fig.~\ref{fit1}). Hints of the two redshifted narrow components observed south of the nucleus (Example~1) are also seen here, albeit with a smaller amplitude.

The line profile in the {[}N\,{\sc ii}{]} plus H$\upalpha$ complex is much more complicated to interpret because of the blending of lines. A small bump appears almost midway between H$\upalpha$ and {[}N\,{\sc ii}{]}\,$\uplambda$6583. The peak of this bump is found at $\uplambda=6571.5\,\text{\AA}$, which would correspond to a H$\upalpha$ line redshifted by roughly $400\,{\rm km\,s^{-1}}$, or to an {[}N\,{\sc ii}{]}\,$\uplambda$6583 line blueshifted by about $500\,{\rm km\,s^{-1}}$ (with quite some imprecision because of the low contrast of the line). In {[}O\,{\sc iii}{],} we see that both the blueshifted broad component and the redshifted narrow component 2 are shifted by several hundreds of kilometres per second with respect to the disc, hence both could contribute to the bump between H$\upalpha$ and {[}N\,{\sc ii}{]}\,$\uplambda$6583.

In practice, if the bump is an {[}N\,{\sc ii}{]}\,$\uplambda$6583 line, it is compatible with the blue wing in {[}N\,{\sc ii}{]}\,$\uplambda$6548, but a bit too blueshifted to be explained by the component that makes the blueshifted wing in {[}O\,{\sc iii}{]}. What \texttt{pyGandALF} does in this case is transform the component that is set according to the initial conditions as a blueshifted narrow component into the blueshifted broad component, and it takes the component with initial conditions of a blueshifted broad component and blueshifts it a bit more in order for its velocity to be compatible with that of the bump. Thus, the resulting fit has two blueshifted broad components, but the bluest one is too blue to be compatible with the shape of the {[}O\,{\sc iii}{]} lines, so its amplitude is fitted to have a very low value there. In Fig.~\ref{gas_kinematics}, the most blueshifted component is plotted as the blueshifted broad component, and the least blueshifted one is plotted as the narrow blueshifted component where it can be distinguished from the genuinely narrow component because of its yellow colour, corresponding to $\sigma\sim350\,{\rm km\,s^{-1}}$, in the velocity dispersion map. The patch where this occurs is very small (less than $0\farcs2$ across), and hence it corresponds to an angularly unresolved region.

An alternative to the solution with two blueshifted broad components proposed by \texttt{pyGandALF} would be the bump to be related to the redshifted narrow component~2. However in Figs.~\ref{fit91} and \ref{fit443}, the amplitude of this component in H$\upalpha$ is small compared to that of {[}O\,{\sc iii}{]}\,$\uplambda$5007. If this is true everywhere for this component, it would be insufficient to explain the bump. On the other hand, the solution offered by \texttt{pyGandALF} poses the problem of why the second blueshifted broad component is not seen in {[}O\,{\sc iii}{]}.

In some bins, such as the one discussed here, the redshifted broad component is replaced by a zero-velocity broad component. Even though its addition adheres to our number of component selection criteria, this component is the only one in our fits that does not seem justified upon visual inspection. Including it might be overfitting the data. However, making the criteria more stringent to eliminate the need for this component also causes components that are clearly identified elsewhere under visual inspection to be ignored, so we decided against it. In the particular case shown here, the zero-velocity broad component has a low amplitude. However, in some cases, it has a significant amplitude in {[}O\,{\sc iii}{]}. In these cases the zero-velocity broad component is fitting all the blueshifted and redshifted components with a single very broad Gaussian.

\subsection{Example~5: A bin $1^{\prime\prime}$ north-west of the nucleus with a clear blueshifted narrow component}

\label{example5}

\begin{figure*}
   \centering
   \includegraphics[scale=0.30]{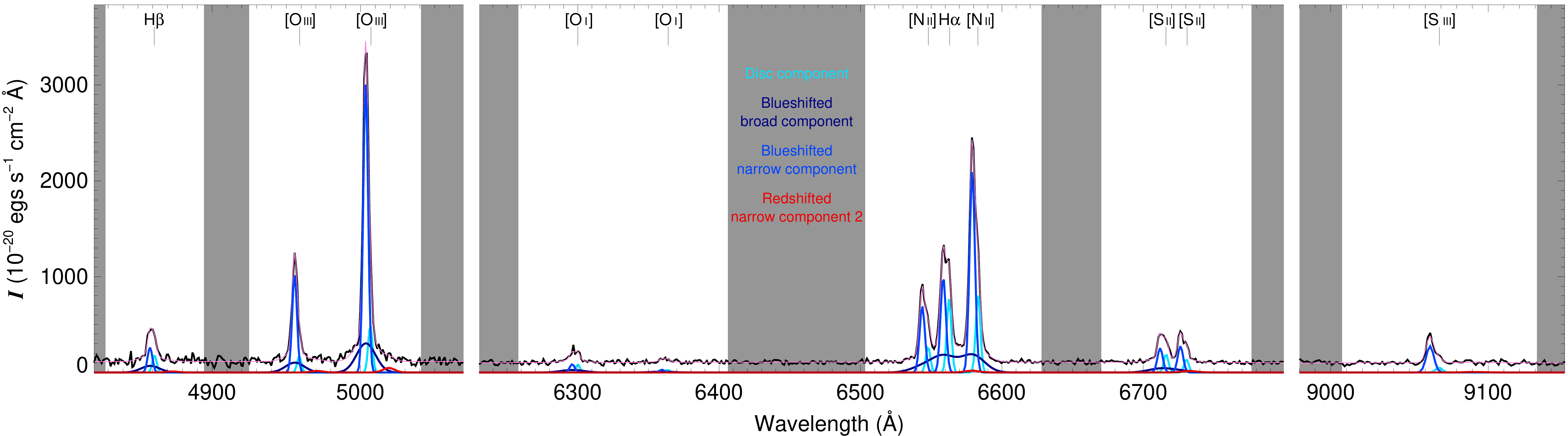}
   \caption{\label{fit2306}{Same as Fig.~\ref{fit91} but for the bin indicated by number `5' in Fig~\ref{locations} and with three components. This fit is discussed as Example~5 in Sect.~\ref{examples}.}}
\end{figure*}

The bin whose spectrum is shown in Fig.~\ref{fit2306} belongs to a region to the north-west of the nucleus where, except for the {[}O\,{\sc iii}{]} doublet, the disc component and the blueshifted narrow component have similar contributions to the line flux. Thus, the blueshifted narrow component does not appear as a blue wing of the disc component as it does in Examples~1, 2, and 3 (Figs.~\ref{fit91}, \ref{fit443}, and \ref{fit1056}), but instead the lines are double-peaked. In {[}O\,{\sc iii}{]}, the blueshifted narrow component clearly dominates, as seen by comparing the positions of the peaks with those of the vertical marks below the {[}O\,{\sc iii}{]} label that show the positions of lines at rest at $z=0.016151$.

In this bin, the difference in velocity between the narrow and blueshifted broad components is smaller than $50\,{\rm km\,s^{-1}}$, maybe denoting a single blueshifted component with a core and broad wings. This kind of relatively simple spectrum (only three obvious components) proved surprisingly challenging to fit. Indeed, moving from one to two components did not improve the $\chi^2$ substantially. A third component was required to obtain a significantly better fit than with a single component. This is the sort of spectrum that motivated us to consider both the $Y_{n+1,n}$ and $Y_{n+2,n+1}$ chi-squared ratios when deciding whether to add an extra component to an $n$-component fit (Sect.~\ref{resolved}).

Our algorithm has fitted the redward wings of the {[}O\,{\sc iii}{]} doublet with a low-amplitude redshifted narrow component, which is probably unnecessary. However, adopting more stringent thresholds to add a fourth component would result in visually identified components not being fitted elsewhere.

\subsection{Example~6: A bin $1^{\prime\prime}$ north-west of the nucleus with a subtler blueshifted narrow component}

\label{example6}

\begin{figure*}
   \centering
   \includegraphics[scale=0.30]{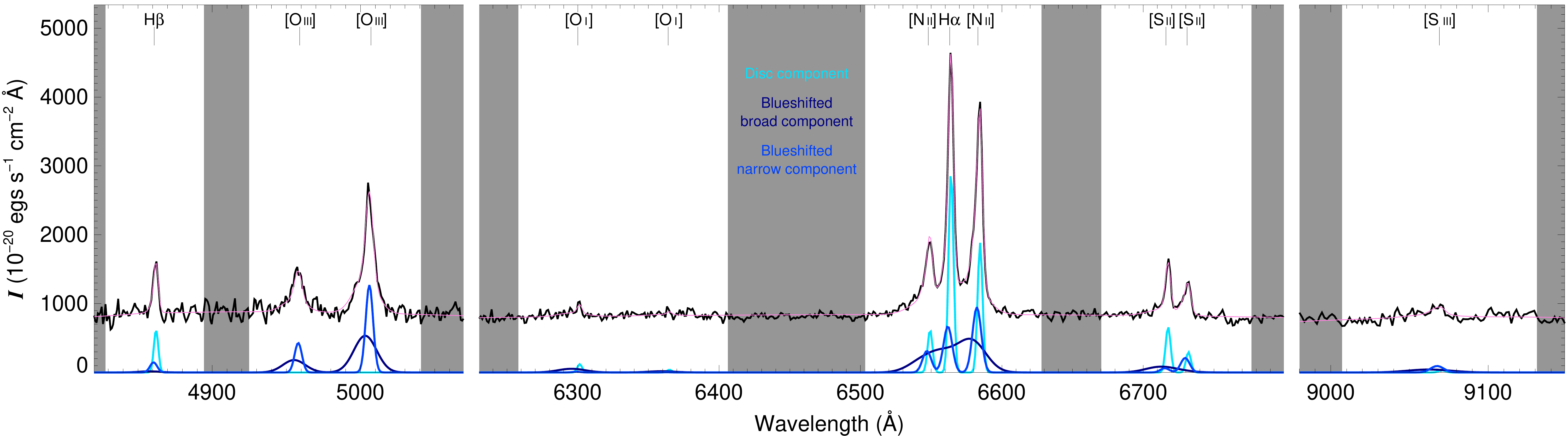}
   \caption{\label{fit2221}{Same as Fig.~\ref{fit91} but for the bin indicated by number `6' in Fig~\ref{locations} and with three components. This fit is discussed as Example~6 in Sect.~\ref{examples}.}}
\end{figure*}

The case shown in Fig.~\ref{fit2221} is similar to Example~5, but with a much subtler blueshifted narrow component, which could be confused with a blueward extension of an asymmetric disc component if {[}O\,{\sc iii}{]} were not considered. Indeed, the peak of {[}O\,{\sc iii}{]} is blueshifted compared to the restframe of the galaxy, which indicates that the blueshifted narrow component is genuine. On this occasion, the blueshifted broad component is clearly more blueshifted than the blueshifted narrow component.

\subsection{Example~7: A bin $2\mkern2.0mu.\mkern-6.5mu^{\prime\prime}5$ north-east of the nucleus with the zero-velocity narrow component}

\label{example7}

\begin{figure*}
   \centering
   \includegraphics[scale=0.30]{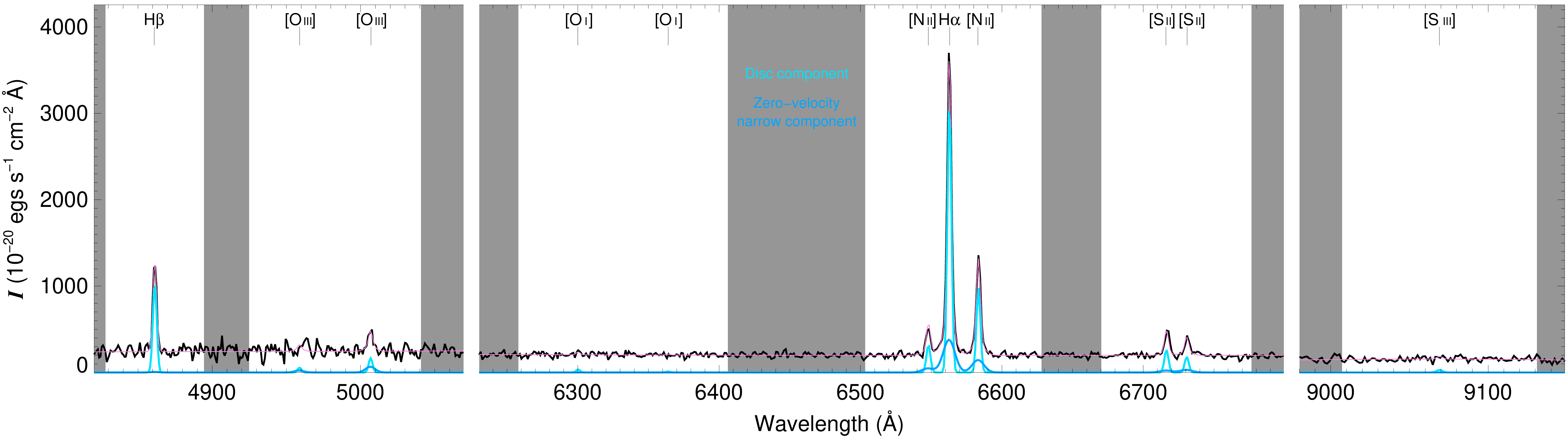}
   \caption{\label{fit2536}{Same as Fig.~\ref{fit91} but for the bin indicated by number `7' in Fig~\ref{locations} and with two components. This fit is discussed as Example~7 in Sect.~\ref{examples}.}}
\end{figure*}

In some regions, the disc component lines have low-amplitude wings that are not well fitted by a single Gaussian function. In this case, the H$\upalpha$ and the {[}N\,{\sc ii}{]}\,$\uplambda$6583 lines show the need for a second, slightly blueshifted component (Fig.~\ref{fit2536}). The existence of this component (the zero-velocity narrow component) cannot be confirmed in other lines, because, if present, it is buried in the noise.

\subsection{Example~8: A bin $3^{\prime\prime}$ south-west of the nucleus with only the disc component}

\label{example8}

\begin{figure*}
   \centering
   \includegraphics[scale=0.30]{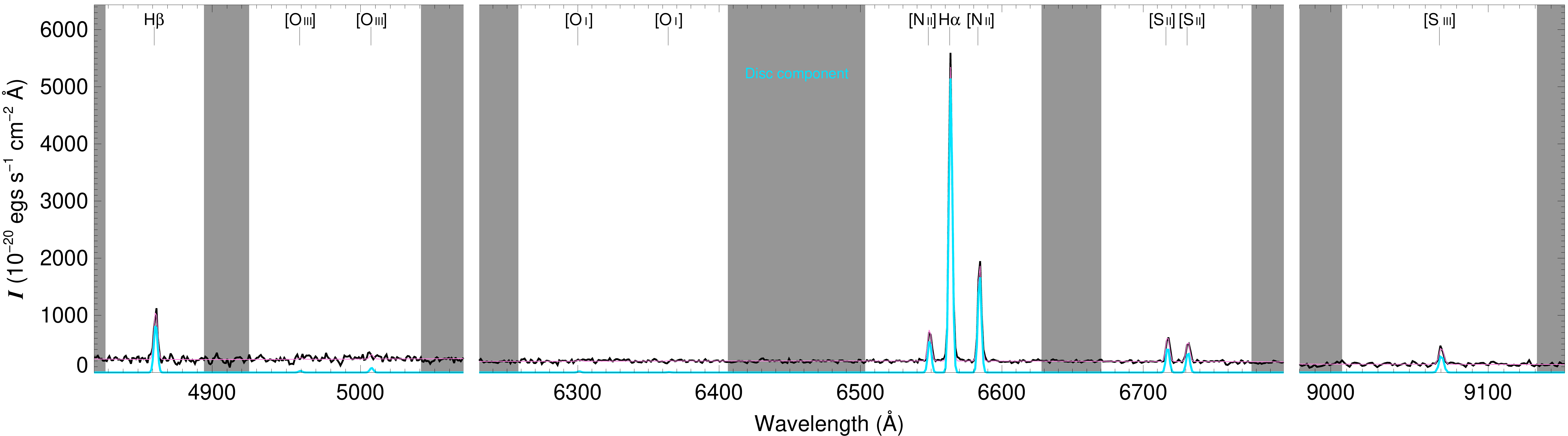}
   \caption{\label{fit2579}{Same as Fig.~\ref{fit91} but for the bin indicated by number `8' in Fig~\ref{locations} and with one component. This fit is discussed as Example~8 in Sect.~\ref{examples}.}}
\end{figure*}

Far from the nucleus, especially in regions to its north-east and its south-west, bins with a single component are observed (Fig.~\ref{fit2579}). While this lack of complexity might be partly caused by the low surface brightness of these areas, which prevents detections of low-amplitude components, this cannot fully explain the simplicity of these spectra. Indeed, if we use the area of the bins as a proxy for the surface brightness (the lower the surface brightness, the larger the Voronoi bin), then we find bins with a disc plus a blueshifted broad component with sizes comparable to those of the bins showing the disc component only.
\end{appendix}
   
\end{document}